%% file: z0_MasterFile.tex
\documentclass[%
reprint,
nobibnotes,   
 amsmath,amssymb,
 aps,
 prc,  
floatfix,
superscriptaddress
]{revtex4-2}

\usepackage[sectionbib]{bibunits} 

\usepackage{graphicx}
\usepackage{dcolumn}
\usepackage{bm}
\usepackage[colorlinks=true,bookmarks=true]{hyperref}

\usepackage[mathlines]{lineno}
\usepackage{comment}
\usepackage{xcolor}
\usepackage{listings}
\usepackage{orcidlink}
\usepackage{tabularx}

\definecolor{mGreen}{rgb}{0,0.6,0}
\definecolor{mGray}{rgb}{0.5,0.5,0.5}
\definecolor{mPurple}{rgb}{0.58,0,0.82}
\definecolor{backgroundColour}{rgb}{0.95,0.95,0.92}

\lstdefinestyle{C}{
    backgroundcolor=\color{backgroundColour},   
    commentstyle=\color{mGreen},
    keywordstyle=\color{magenta},
    numberstyle=\tiny\color{mGray},
    stringstyle=\color{mPurple},
    basicstyle=\footnotesize,
    breakatwhitespace=false,         
    breaklines=true,                 
    captionpos=b,                    
    keepspaces=true,                 
    numbers=left,                    
    numbersep=5pt,                  
    showspaces=false,                
    showstringspaces=false,
    showtabs=false,                  
    tabsize=2,
    language=C
}
\usepackage{dcolumn}
\usepackage{multirow}
\usepackage{booktabs}
\usepackage{xcolor}
\usepackage{ulem}
\usepackage{makecell}
\usepackage{tikz, pgfplots}
    \usetikzlibrary{arrows, intersections, math}
    \pgfplotsset{compat=1.15}
\usepackage{listings}
\usepackage{array}

\newcolumntype{k}[1]{>{\centering\arraybackslash}m{#1\linewidth}} 
\newcolumntype{d}[1]{D{.}{.}{#1}}  
\newcommand{\expt}[1]{\langle #1 \rangle}
\newcommand{\headerwrap}[2]{\multicolumn{1}{c}{\parbox{#1\linewidth}{\centering #2}}}

\graphicspath{{images/}}

\usepackage{lineno}
\usepackage{amsmath}
\usepackage{etoolbox} 

\newcommand*\linenomathpatch[1]{%
  \cspreto{#1}{\linenomath}%
  \cspreto{#1*}{\linenomath}%
  \csappto{end#1}{\endlinenomath}%
  \csappto{end#1*}{\endlinenomath}%
}
\linenomathpatch{equation}
\linenomathpatch{gather}
\linenomathpatch{multline}
\linenomathpatch{align}
\linenomathpatch{alignat}
\linenomathpatch{flalign}

\usepackage[T1]{fontenc}

\begin{document}

\title{Precision Neutron Skins of $^{208}$Pb and $^{48}$Ca from Parity-Violating Electron Scattering}

\collaboration{The PREX Collaboration}
\input{AuthorList.tex}

\date{September 6, 2026}

\begin{abstract}

We have measured the
parity-violating elastic electron scattering asymmetry in
the PREX and CREX experiments on ${}^{208}$Pb and ${}^{48}$Ca respectively;
these are both doubly-magic nuclei whose excited states can be
discriminated from the ground state by the high resolution spectrometers in Hall A at Jefferson Lab.
This asymmetry provides a
precise determination of the weak charge form factor at one $Q^2$ and pins
down the neutron radius in these two nuclei in a relatively clean and
model-independent way.  This is because the $Z^0$ boson of the weak
interaction couples primarily to neutrons.  
The heavier lead nucleus, with a neutron excess,
provides an interpretation of the neutron skin thickness in terms
of properties of bulk neutron matter.  For the lighter
${}^{48}$Ca nucleus, which is also rich in neutrons, comparisons to microscopic
nuclear theory calculations are sensitive to poorly
constrained 3-neutron forces. The weak neutral form factors $F_W(Q^2)$ were extracted to be $0.368 \pm 0.013$ at $Q = 0.3977 {\rm\ fm}^{-1}$ for $^{208}$Pb from PREX-2 and $0.1304 \pm 0.0055$ at $Q = 0.8733 {\rm\ fm}^{-1}$ for $^{48}$Ca. The form factor differences $(F_{ch}-F_W)(Q^2)$ were calculated to be $0.041 \pm 0.013$ at $Q = 0.3977 {\rm fm}^{-1}$ for $^{208}$Pb from PREX-2 and $0.0277 \pm 0.0055$ at $Q = 0.8733 {\rm fm}^{-1}$ for $^{48}$Ca. 
Correcting for Coulomb distortions and using nuclear model information, we
find the neutron skin thicknesses to be 
$R^{208}_{\rm skin} = 0.283 \pm 0.071$ fm combining PREX-1 and PREX-2 and
$R^{48}_{\rm skin} = 0.121 \pm 0.035$ fm.
This paper provides a full description of the special experimental and data analysis 
techniques employed for precisely measuring these small asymmetries.
\vskip 0.05in
{\hskip 0.02in 25.30.Bf} {Elastic Electron Scattering} 
     {21.65.Ef} {Symmetry Energy}   
      {21.10.Gv} {Nucleon Distributions} 

\end{abstract}


\maketitle

\input{z1_introV2}

\input{z2_expApparatus}
\input{z3_analysisV2}

\input{z4_resultsV2}

\section{Acknowledgments}
We thank the entire staff of JLab for their efforts to develop and maintain the polarized beam and the experimental apparatus and acknowledge the support of the U.S. Department of Energy, the National Science Foundation, and NSERC (Canada). 
We thank J.~Piekarewicz, P.~G.~Reinhard and X.~Roca-Maza for RPA calculations of $^{48}$Ca excited states and J.~Erler and M.~Gorchtein for calculations of $\gamma\ -\ Z$ radiative corrections.
This material is based upon work supported by the U.S. Department of Energy, Office of Science, Office of Nuclear Physics under Contract No. 89243126CSC000213.

\bibliography{masterBib}

\newpage
\section{Appendix I - models}
\input{z5_App}

\end{document}

%% file: AuthorList.tex

\author{D.~Adhikari\,\orcidlink{0000-0001-8514-5759}}\affiliation{Idaho State University, Pocatello, Idaho 83209, USA}
\affiliation{Virginia Tech, Blacksburg, Virginia 24061, USA}\affiliation{Hampton Roads Academy, Newport News, VA 23602}
\author{H.~Albataineh\,\orcidlink{0000-0002-9302-8453}}\affiliation{Texas  A  \&  M  University - Kingsville,  Kingsville,  Texas  78363,  USA}
\author{D.~Androi\'{c}\,\orcidlink{0000-0002-3921-5696}}\affiliation{University of Zagreb, Faculty of Science, Zagreb, HR 10002, Croatia}
\author{K.A.~Aniol\,\orcidlink{0000-0002-5294-9264}}\affiliation{California State University, Los Angeles, Los Angeles, California  90032, USA}
\author{D.S.~Armstrong\,\orcidlink{0000-0002-0859-3459}}\affiliation{William \& Mary, Williamsburg, Virginia 23185, USA}
\author{T.D.~Averett\,\orcidlink{0000-0002-0440-5836}}\affiliation{William \& Mary, Williamsburg, Virginia 23185, USA}
\author{\mbox{C. Ayerbe Gayoso}\,\orcidlink{0000-0001-8640-5380 }}\affiliation{William \& Mary, Williamsburg, Virginia 23185, USA}
\author{S.K.~Barcus\,\orcidlink{0000-0002-4470-5921}}\affiliation{Thomas Jefferson National Accelerator Facility, Newport News, Virginia 23606, USA} 
\author{V.~Bellini\,\orcidlink{0000-0001-6906-7463}}\affiliation{Instituto  Nazionale  di  Fisica  Nucleare,  Sezione  di  Catania,  95123  Catania,  Italy}
\author{R.S.~Beminiwattha\,\orcidlink{0000-0002-1473-1651}}\affiliation{Louisiana Tech University, Ruston, Louisiana 71272, USA}
% ...

%% file: z1_introV2.tex
\section{INTRODUCTION}
\label{part:overview}

The PREX-2 and CREX experiments precisely measured the weak nuclear form factors of $^{208}$Pb~\cite{ref:PrexII} and $^{48}$Ca~\cite{ref:crex} respectively. These results were used to extract the neutron skins of those nuclei. We note that PREX-1~\cite{ref:prexI} made the first measurement for $^{208}$Pb nine years before PREX-2.
While there are other methods~\cite{PhysRevLett.43.844,PhysRevC.21.1488,PhysRevC.29.1295,BARNETT1985172,PhysRevC.49.2118,PhysRevC.82.044611,PhysRevC.65.044306,PhysRevC.67.054605,KRASZNAHORKAY2004224,KRASZNAHORKAY1994521,PhysRevLett.87.082501,PhysRevC.76.014311,PhysRevLett.107.062502,PhysRevResearch.5.L022044,PhysRevC.85.041302,PhysRevC.92.064304,PhysRevLett.131.202302,PhysRevC.80.024316,DANIELEWICZ2003233,PhysRevC.86.015803} to determine the neutron radius of heavy nuclei, the use of parity-violating electron scattering (PVES) minimizes the contributions of strong interaction uncertainties leading to a cleaner theoretical interpretation.

Electron scattering has provided much of current knowledge of nuclear sizes, in particular charge radii~\cite{doi:10.1080/00018736600101254,DEVRIES1987495,ANGELI2004185} of nuclei.  By measuring the parity-violating asymmetry in the scattering of longitudinally polarized electrons from unpolarized solid targets, we can determine the weak charge radius of the nuclei~\cite{DONNELLY1989589}. These measurements are relevant for testing nuclear structure theory~\cite{FSUGold}, for corrections to measurements of atomic parity violation~\cite{Marie-Anne_Bouchiat_1997,PhysRevLett.65.2857,PhysRevC.46.2587,POLLOCK1999177,PhysRevC.48.1392}, for coherent neutrino
scattering~\cite{ref:COHERENT0,ref:COHERENT1,ref:COHERENT2}, for nuclear spectroscopy, and more~\cite{PhysRevC.81.051303}. Moreover, since a precise measurement of the weak charge radius provides information about isovector properties~\cite{PhysRevC.84.064302,PhysRevC.88.034325}, they are important at extremes of matter as found in neutron stars~\cite{doi:10.1126/science.1090720,doi:10.1126/science.1078070,PhysRevLett.85.5296}. 

PVES experiments make use of the fact that weak interaction is the only known interaction to violate parity, in order to extract the weak contribution to the scattering cross section.
In the Born approximation, the parity-violating cross-section asymmetry for longitudinally polarized electrons elastically scattered from an unpolarized nucleus, $A_{\rm PV}$,
is proportional to the weak form factor $F_W(Q^2)$,
\begin{equation}
\label{eq:apv}
A_{\rm PV}=\frac{\sigma_R-\sigma_L}{\sigma_R+\sigma_L} \approx \frac{G_FQ^2|Q_W|}{4\pi\alpha\sqrt{2}Z}
\frac{F_W(Q^2)}{F_{\rm ch}(Q^2)}
\end{equation}
where $Q$ is the four-momentum transfer, $\sigma_{R(L)}$ is the differential cross section for elastic scattering of right ($R$) or left ($L$) handed longitudinally polarized electrons, $G_F$ is the Fermi constant, $\alpha$ the fine structure constant, $Q_W$ the total weak charge, $Z$ the electric charge, and $F_{\rm ch}(Q^2)$ the charge form factor~\cite{DONNELLY1989589,Horowitz:1999fk}. 

This paper is organized as follows. In this section, we continue with a more detailed discussion of the motivation for these experiments together with the formalism required for the weak form factor extraction, followed by the experimental challenges.
Section~\ref{sec:Apparatus} describes the apparatus in 
detail, including previously unpublished information about the high-power lead and calcium targets.  
Section~\ref{sec:analysis} presents the analysis of the data. This section details all experimental corrections to the data, including the beam polarization, beam asymmetries, backgrounds, acceptance, and kinematics. The method for analyzing the beam asymmetries is provided for the first time in Sec.~\ref{sec:BeamAsymmetries}.
Section~\ref{sec:results} presents the model-independent experimental results. Section~\ref{sec:interpretation} shows the extraction of the weak radius and the neutron radius and their model dependence, followed by the implications to the nuclear equation of state.
Finally,  Sec.~\ref{sec:Conclusions} draws the conclusions of this work.

\subsection{Theoretical Motivation}
\label{sec:theory_motiv}

At forward angles, the parity-violating asymmetry can be interpreted as the ratio of form factors from $Z^0$ exchange and photon exchange. Both of these interactions are well known from electroweak theory. While photon exchange is primarily sensitive to the proton distribution, $Z^0$ exchange is almost entirely determined by the neutron distribution. Therefore, parity-violating elastic electron scattering from nuclei provides a way to determine the RMS radius of the distribution of neutrons,
$R_n$, which is relatively free of model uncertainties~\cite{DONNELLY1990179,Horowitz:1999fk,Erler:2014fqa}.

Neutron-rich doubly magic nuclei make ideal targets for this kind of investigation. In addition to being spherical $J^\pi=0^+$ nuclei, their nuclear structure is particularly well understood so that corrections to the naive picture can be made reliably. Only two such nuclei are stable, however, namely $^{208}$Pb and $^{48}$Ca, so these are the targets with the cleanest interpretation.
These two nuclei have been very well studied \cite{PhysRevLett.38.152,Emrich,wise,quint,PhysRevLett.58.195,DEVRIES1987495,Kelly:2002}. They are doubly-magic, i.e. have closed shells in both neutrons and protons, providing a simple nuclear structure. Both nuclei have a large separation of the first excited state (2.61 MeV for $^{208}$Pb and 3.83 MeV for $^{48}$Ca~\cite{nndc}), leading to small corrections arising from inelastic excited states.

The equation of state (EOS) of nuclear matter provides a bridge between the nuclei and neutron stars. On the one hand, it is critical to our understanding of the bulk properties of nuclei~\cite{PhysRevLett.85.5296}. On the other, it provides crucial input to calculating the properties of neutron stars, as well as their formation~\cite{doi:10.1126/science.1078070,doi:10.1126/science.1090720,LATTIMER2007109}. The EOS provides a description of the relation between the energy density and pressure in nuclear matter,
be it a heavy nucleus like $^{208}$Pb or the inside of a neutron star.

Unlike an ideal gas, nuclear matter behaves like a quantum gas where the density is very large and the temperature is very low, meaning that the pressure is mostly given by the degeneracy of the fermionic system. Moreover, the two flavors of particles in nuclear matter (protons and neutrons) provide an additional complication in describing the system. An established methodology~\cite{Yang:2019fvs} can connect symmetric nuclear matter (SNM) -- where the number of protons equals the number of neutrons -- to pure neutron matter -- where only neutrons exist in the system -- through the expansion
\begin{equation}
\frac{E}{A}-M\equiv{\cal E}(\rho,\alpha)
={\cal E}_{\rm SNM}(\rho)+\alpha^2S(\rho)+{\cal O}(\alpha^4)
\end{equation}
where
$\alpha\equiv\frac{\rho_n-\rho_p}{\rho_n+\rho_p}$ 
is the asymmetry in the density of the protons ($\rho_p$) and neutrons ($\rho_n$), $E$ is the energy of a system with mass number $A$ and $M$ is the nucleon mass.  
The density dependence is expanded in terms of the parameter
$x\equiv\frac{\rho-\rho_0}{3\rho_0}$
where $\rho_0$ is the density of nuclear matter at saturation. The energy of symmetric nuclear matter is 
\begin{equation}
{\cal E}_{\rm SNM}(\rho)=\epsilon_0+\frac{1}{2}K_0x^2+\cdots
\end{equation}
and the symmetry energy is given by
\begin{equation}
S(\rho)=J+Lx+\frac{1}{2}K_{\rm sym}x^2+\cdots
\end{equation}
where $J$ is the correction to the binding energy for symmetric nuclear matter, and $L$ describes how fast the symmetry energy increases with density.  The incompressibility of symmetric nuclear matter is given by $K_0$, while $K_{\rm sym}$ describes how the incompressibility increases with isospin asymmetry. 
Note that the pressure of symmetric nuclear matter vanishes at saturation density while the pressure of pure neutron matter is $P_0=\frac{1}{3}\rho_0L$.

Despite the wealth of available experimental data, there are precious few measurements that constrain the density dependence of the symmetry energy, $S$, typically parameterized as $L = 3\rho_0\frac{\partial S}{\partial \rho}|_{\rho_0}$. An accurate determination of $L$ (called the slope of the symmetry energy) allows an extrapolation from the nuclear EOS to that for pure neutron matter, and therefore has a direct bearing on the properties of neutron stars~\cite{Thiel:2019tkm,Drischler:2020}.

A number of nuclear model calculations~\cite{PhysRevLett.85.5296,PhysRevLett.106.252501,Reinhard:2021,PhysRevC.109.035803} show a strong correlation between $L$ and the difference between the proton and neutron distribution radii, $R_n-R_p$, that is, the ``neutron skin thickness.'' Although $R_p$ is well measured from elastic electron scattering and atomic properties, $R_n$ is much more difficult to determine because probes that are sensitive to $R_n$ typically suffer from large distortions due to their hadronic nature.

Lastly, measurements of the neutron radii of $^{208}$Pb and $^{48}$Ca provide complementary tests of theoretical models. While the large size of $^{208}$Pb makes it particularly amenable to bulk models of uniform nuclear matter, it is feasible to carry out microscopic calculations for the much smaller $^{48}$Ca nucleus~\cite{Hagen:2016}. 

\subsection{Weak form factor extraction}
\label{ssec:weakFF}
The parity-violating asymmetry arises from the interference term of the vector (Coulomb, $V(r)$) and axial $A(r)$ potentials between an electron and a nucleus.  
The potential may be written
\begin{equation}\hat V(r)=V(r)+\gamma_5A(r),\end{equation}

where the Coulomb potential is given by
\begin{equation}
V(r)=\int d^3r' \rho_{\rm ch}(r')/|\vec r - \vec r\, ' |.
\end{equation}

The charge and weak density distributions, $\rho_{\rm ch}(r)$ and $\rho_W(r)$ respectively, are approximately equal to the point proton (neutron) density $\rho_{p,n}(r)$ given by
\begin{equation}
\rho_{p,n}(r)=\sum_{p,n}\langle\bar{\psi}_{p,n}(r)\gamma^0\psi_{p,n}(r)\rangle\, ,
\end{equation}
folded with the charge (weak charge) distribution of a single proton (neutron).  
The weak charge distribution of the nucleus requires a small correction to account for the weak charge of the protons.  The axial potential is given by
\begin{equation}
A(r)={\frac{G_F}{2^{3/2}}}\rho_W(r).
\end{equation}
Since the weak charge of a proton is small, $A(r)$ depends mainly on the neutron density $\rho_n(r)$, and for scattering electrons with momentum transfer $Q$ the cross-section is given by
\begin{equation}
\frac{d\sigma}{d\Omega}=\frac{d\sigma}{d\Omega}_{\rm Mott}Z^2|F_{\rm ch}(Q^2)|^2
\end{equation}
where $\Omega$ is the solid angle, $\sigma_{\rm Mott}$ describes the scattering from a point-like charged particle, $Z$ is the charge of the nucleus, and the charge form factor is $F_{\rm ch}(Q^2)$.

The charge form factor $F_{\rm ch}(Q^2)$, given by
\begin{equation}
F_{\rm ch}(Q^2) = \frac{1}{4\pi Z}\int d^3r j_0(Qr)\rho_{ch}(r), 
\label{eq:ChargeFFdef}
\end{equation}
describes the distribution of the charge; it is the Fourier transform of the radial charge density distribution of the nucleus.  Similarly, there is also a weak form factor 
\begin{equation}
F_W(Q^2)=\frac{1}{4\pi Q_W}\int d^3r j_0(Qr)\rho_W(r)\, ,
\label{eq:WeakFFdef}
\end{equation} 
which is the Fourier transform of the radial weak charge density distribution of the nucleus.
These form factors are normalized such that the value of each at $Q^2 \rightarrow 0$ are equal to 1 and the total weak charge of a nucleus $Q_W=\int d^3r \rho_W(r)$.
Since $F_{\rm ch}$ is known, one can extract the weak density as well as the closely related neutron density from the measurement of the asymmetry.
$A_{\rm PV}$ is approximately proportional to the ratio of the weak to charge form factors as provided in Eq.~\ref{eq:apv}.

In order to extract $F_W$ from $A_{\rm PV}$ we rely on the predictions of density functional models. To accurately probe the parameter space, we use the predictions of several relativistic and non-relativistic models for both $^{208}{\rm Pb}$ and $^{48}{\rm Ca}$ that predict very different bulk properties of nuclear matter and nuclei. The non-relativistic models use the framework of Skyrme~\cite{Skyrme:1958} and relativistic models (RMF) use the framework of relativistic mean-field theory~\cite{FSUGold}. We list the models considered in this work in Table~\ref{tab:models1} (relativistic) and Table~\ref{tab:models2} (non-relativistic). 

For our calculation of $F_W$ we follow the procedure described in \cite{ref:crex}, summarized here. The value of $A_{\rm PV}$ depends both on the weak charge and electromagnetic charge distribution in the target nucleus. These two quantities are calculated from the single-nucleon form factors including a spin-orbit term which has been shown to be non-negligible for lighter nuclei such as $^{48}{\rm Ca}$~\cite{Horowitz:2012we}. 
The charge and weak form factors are calculated first from both the vector and tensor densities of the nucleus,
\begin{eqnarray}
    F_V(q^2) = \int \Bar{\psi}(\textbf{r})e^{i\textbf{q}\cdot\textbf{r}}\gamma^0\psi(\textbf{r})d^3r\\
    F_T(q^2) = \int \Bar{\psi}(\textbf{r})e^{i\textbf{q}\cdot\textbf{r}}\gamma^0\mathbf{\gamma}\cdot\mathbf{\hat{q}}\psi(\textbf{r})d^3r ,
\end{eqnarray}
where \textbf{q} is the spatial component of the four-momentum transfer. These integrals can be rewritten for our purposes into their neutron and proton components directly from solutions to the Hartree equations assuming spherical symmetry as
\begin{eqnarray}
    F_V^{p,n}(q^2)=\frac{4\pi}{(Z,N)}\int_0^\infty \rho^{p,n}_V(r)j_0(qr)r^2dr\\
    F_T^{p,n}(q^2)=\frac{4\pi}{(Z,N)}\int_0^\infty \rho^{p,n}_T(r)j_1(qr)r^2dr,
\end{eqnarray}
where $\rho_V$ and $\rho_T$ are the vector and tensor densities, $j_i$ are spherical Bessel functions, and ($Z,N$) is the total number of protons or neutrons in the nucleus. We note here that the exact form of $F_T$ presented here is different for our calculations of non-relativistic models; this will be elaborated on later. With these two in hand for each of the protons and neutrons, one may calculate the full charge and weak form factors as
\begin{eqnarray}
 \nonumber   ZF_{\rm ch}(Q) = &&\sum_{i=p,n}\Big(G_E^i(Q)F_V^i(Q)+\Big(\frac{G_M^i(Q)-G_E^i(Q)}{1+\tau}\Big)\\
    &&\times \Big[\tau F_V^i(Q)+\frac{Q}{2m}F_T^i(Q)\Big]\\
 \nonumber   Q_WF_W(Q) = &&\sum_{i=p,n}\Big(\widetilde{G}_E^i(Q)F_V^i(Q)+\Big(\frac{\widetilde{G}_M^i(Q)-\widetilde{G}_E^i(Q)}{1+\tau}\Big)\\
    &&\times \Big[\tau F_V^i(Q)+\frac{Q}{2m}F_T^i(Q)\Big],
\end{eqnarray}
where $Q_W$ is the total weak charge of the nucleus, $m=939$~MeV is the mass of the nucleon, $G_E$ and $G_M$ are the single-particle Sachs electric and magnetic form factors~\cite{Horowitz:2012tj}, respectively, tilde denotes the weak form factors, and $\tau=\frac{Q^2}{4m^2}$. The form of the single-nucleon form factors we take as the Galster parameterization \cite{Kelly:2002}. For the weak Sachs form factors, we calculate them from the electric and magnetic form factors according to
\begin{eqnarray}
    \widetilde{G}_{E,M}^p(Q)=q_pG^p_{E,M}(Q)+q_nG^n_{E,M}(Q)\\
    \widetilde{G}_{E,M}^n(Q)=q_pG^n_{E,M}(Q)+q_nG^p_{E,M}(Q),
\end{eqnarray}
where $q_p$ and $q_n$ are the weak charge of the proton and neutron, respectively. Note that we neglect the contribution of strange quarks. 

The above formalism is complete to calculate the weak and electromagnetic form factors for RMF models. However, non-relativistic models do not possess a tensor density and instead this additional term would contain contributions from the spin-orbit current density $\textbf{J}(r)$. These contributions are much smaller than in RMF models, so we approximate the spin-orbit contribution of non-relativistic models using the tensor density of the FSUGold~\cite{FSUGold} RMF interaction in the free-space relation
\begin{eqnarray}
F(r) \approx \frac{1}{2m}\Big(\frac{dG}{dr}+\frac{\kappa}{r}G\Big),
\end{eqnarray}
where $G(r)$ and $F(r)$ are the upper and lower components of the Dirac wavefunction respectively. The proton and neutron vector densities are directly calculable from the solution of the Skyrme-Hartree equations. Therefore, using the free-space relation to calculate the tensor density, we can calculate the weak and electromagnetic form factors using the methods described above. We must note that
our inclusion of $\gamma$-Z box corrections for the total weak charge of the nuclei while also fixing $q_p=0.0713$ from the Qweak experiment~\cite{Qweak:NIM} necessitates different values of $q_n$ for $^{48}{\rm Ca}$ and $^{208}{\rm Pb}$. We show the weak charge properties used for both nuclei in Table~\ref{tab:qweak}.

The analysis as described above assumes plane waves for the electrons, however for heavy nuclei, particularly Pb, this is a poor approximation due to the effect of Coulomb distortions. The analysis of the Coulomb distortions must be done for scattering states that are solutions to the Dirac Equation \begin{equation}[\alpha\cdot {\bf p} +\beta m_e+\hat V(r)]\psi=E\psi\end{equation} valid when $Z\alpha$ is relatively large. These have been accurately calculated~\cite{Horowitz:1998vv}. The Coulomb correction calculations are precise enough to support the experiment and pass several stringent tests. The numerical accuracy has been verified by multiple independent codes~\cite{PhysRevC.61.064307,Horowitz:1998vv,PhysRevLett.106.252501,ref:CooperRUNT}.

\subsection{Experimental Challenges}
\label{sec:exptchallenges}

\begin{figure}
	\centering
	\includegraphics[width=0.45\textwidth]{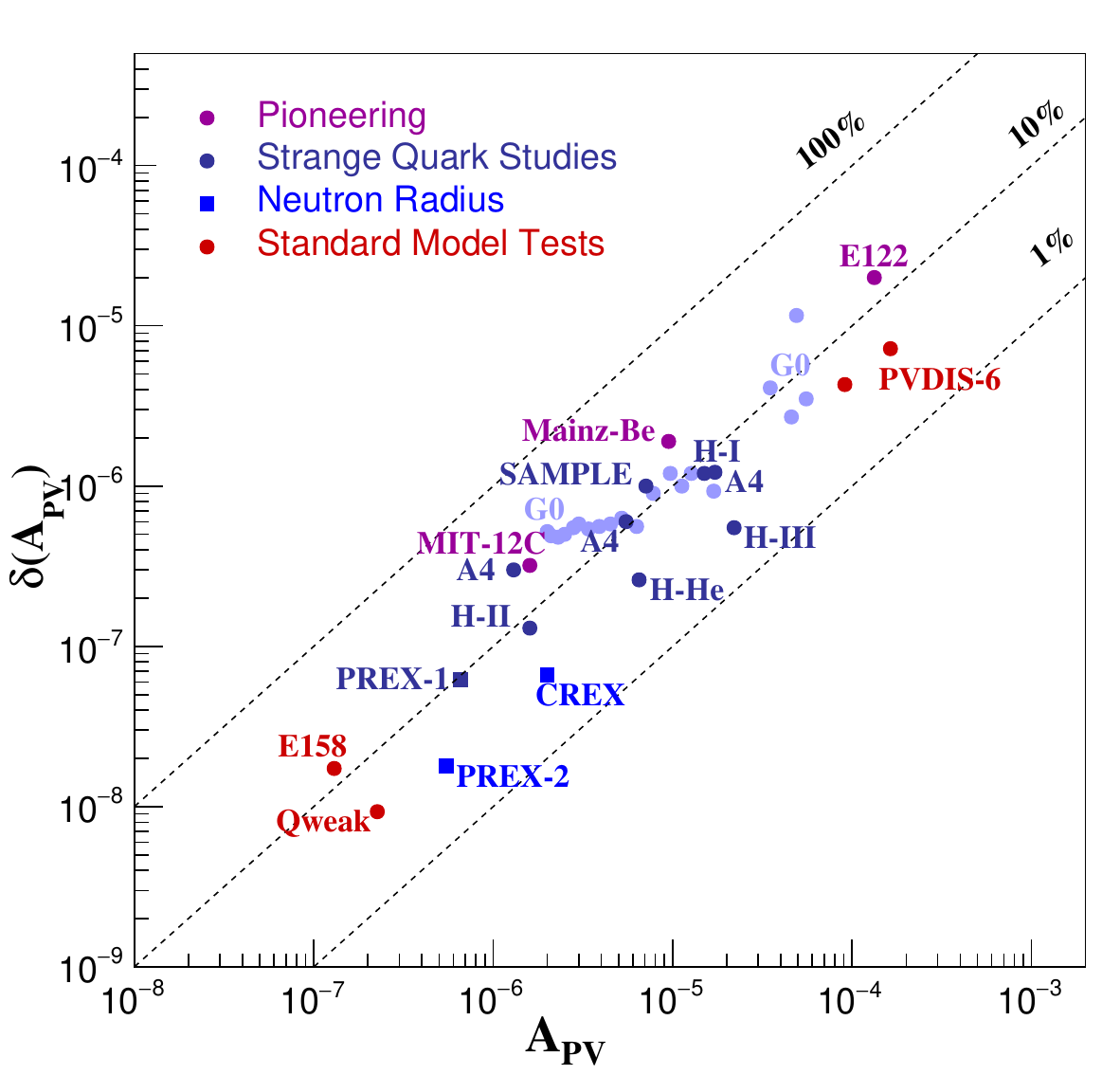}
	\caption{Total experimental error in the parity-violating asymmetry versus the asymmetry for a set of
    published PVES measurements. The diagonal lines indicate the fractional
error in the asymmetry measurements. Adapted from Fig. 3 in~\cite{FrontPhys111301}}
	\label{fig:APVvsdAPV}
\end{figure}

In order to put the difficulty of performing the PREX and CREX experiments into perspective, we show in Fig.~\ref{fig:APVvsdAPV} the size of the parity-violating asymmetry versus the quoted errors for all the published and future PVES experiments~~\cite{FrontPhys111301}.  PREX and CREX represent the newest generation of PVES experiments; only the Qweak~\cite{Qweak:2018tjf} experiment has published a smaller absolute precision for an asymmetry.
Another key challenge for these experiments is the rapid falloff of the scattering cross section with $Q^2$~\cite{PhysRevLett.38.152}, leading to a high sensitivity to the electron beam energy. Furthermore for PREX, operating a lead target, with its low melting point, is extremely challenging in an intense beam (see Sec.~\ref{sec:pbtarget}). 

One of the most challenging aspect of these experiments was the control of systematic beam effects. In PVES experiments one must prepare a clean parity transformation by changing the helicity state of the incoming electrons without changing any other property of the beam. In spite of the high-quality beam provided by the Continuous Electron Beam Accelerating Facility (CEBAF) the remaining helicity-correlated effects produced false asymmetries which had to be fully understood, kept small compared to the size of the effect being measured, and subtracted. The systematic studies and analysis procedure are described in full in Sec.~\ref{sec:analysis}.

%% file: z2_expApparatus.tex
\section{EXPERIMENTAL APPARATUS}\label{part:exp_app}
\label{sec:Apparatus}

\subsection{Experimental Overview}
\label{sec:exptmethod}

\begin{figure}[htb!]
	\includegraphics[width=0.45\textwidth]{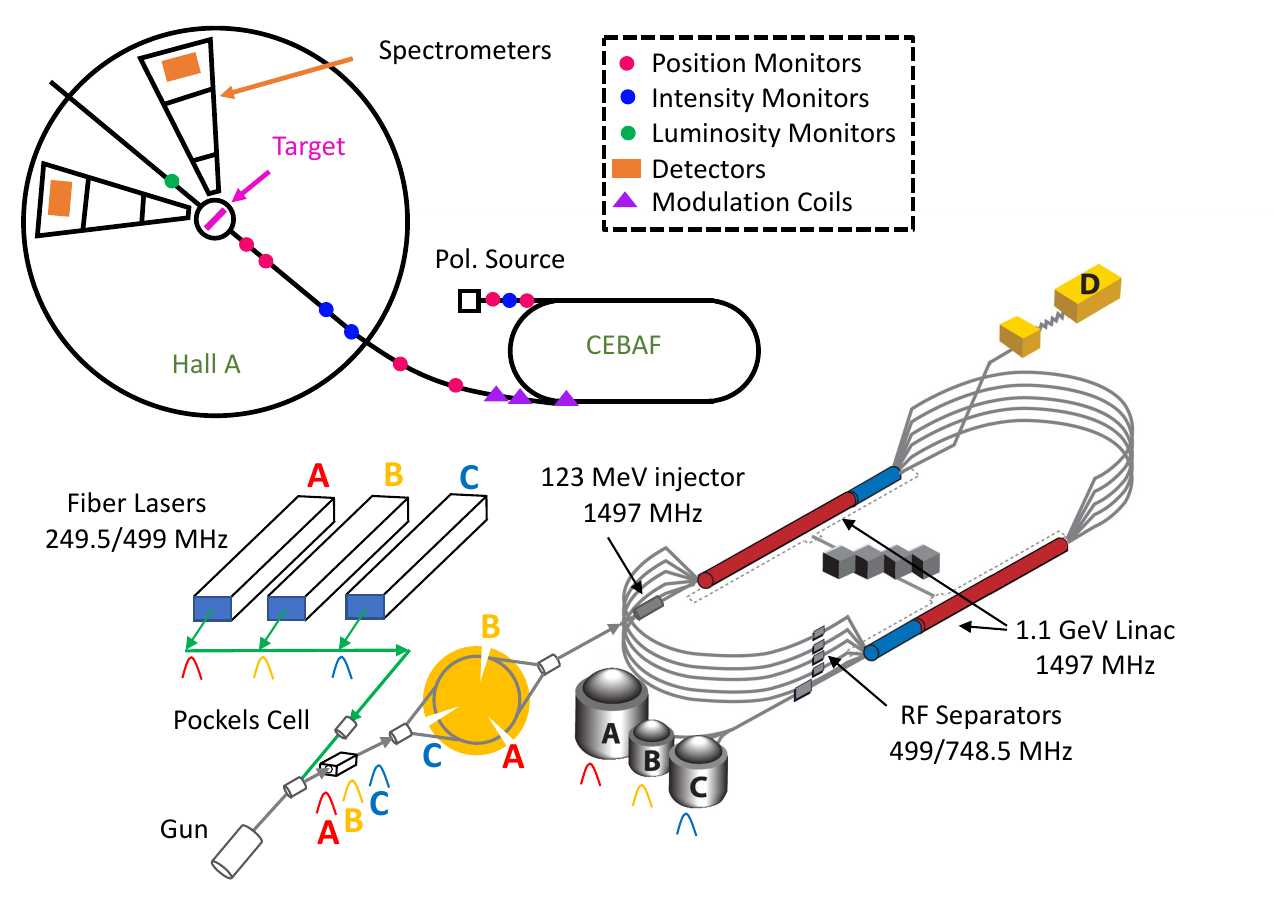}
	\caption{Schematic of the accelerator (bottom picture) and Hall A (top).  For PVES experiments at Jefferson Lab, the entire accelerator as well as the hall are a part of the experiment.  The figure starts at the injector and shows the linacs (linear accelerators) and recirculating arcs.  In the top figure, the target sits upstream of the spectrometers, and several critical beamline measurement elements are shown.}
	\label{fig:JLAB_HALLA}
\end{figure}

The experiments were performed in Hall A at the Thomas Jefferson National Accelerator Facility (JLab)
from June 2019 through August 2020
using a high-quality, highly (longitudinally) polarized electron beam incident on isotopically enriched, unpolarized, solid targets (see Sec.~\ref{sec:target}).  
Figure~\ref{fig:JLAB_HALLA} provides a schematic of the accelerator, including the injector.  PVES experiments are unique in that they involve all aspects of the accelerator, starting from the source, in order to tightly control the initial properties of the electron beam.

Circularly-polarized laser light is incident on a strained GaAs photocathode, producing longitudinally polarized electrons.  The direction of the polarization of the laser was adjusted by changing the voltage on a Pockels cell (PC) with frequencies up to 240~Hz, inducing a flip of the electron beam helicity at the same frequency. In addition, the direction of the electron spin was flipped slowly in two other ways in order to test systematics: through the insertion of a half-wave plate on the laser table (procedure takes approximately 1~min) and through the use of a beam spin-flipper (procedure takes several hours). The latter consists of a system of solenoids and Wien filters (Sec.~\ref{sec:WienFlip}). The spin in the injector is properly oriented using this system to account for the spin precession in the horizontal plane of the accelerator, and the electrons are injected into the CEBAF accelerator, a racetrack design with two linear accelerators connected by recirculating arcs~\cite{Adderley:2024czm}

\begin{figure}[htb!]
	\includegraphics[width=0.45\textwidth]{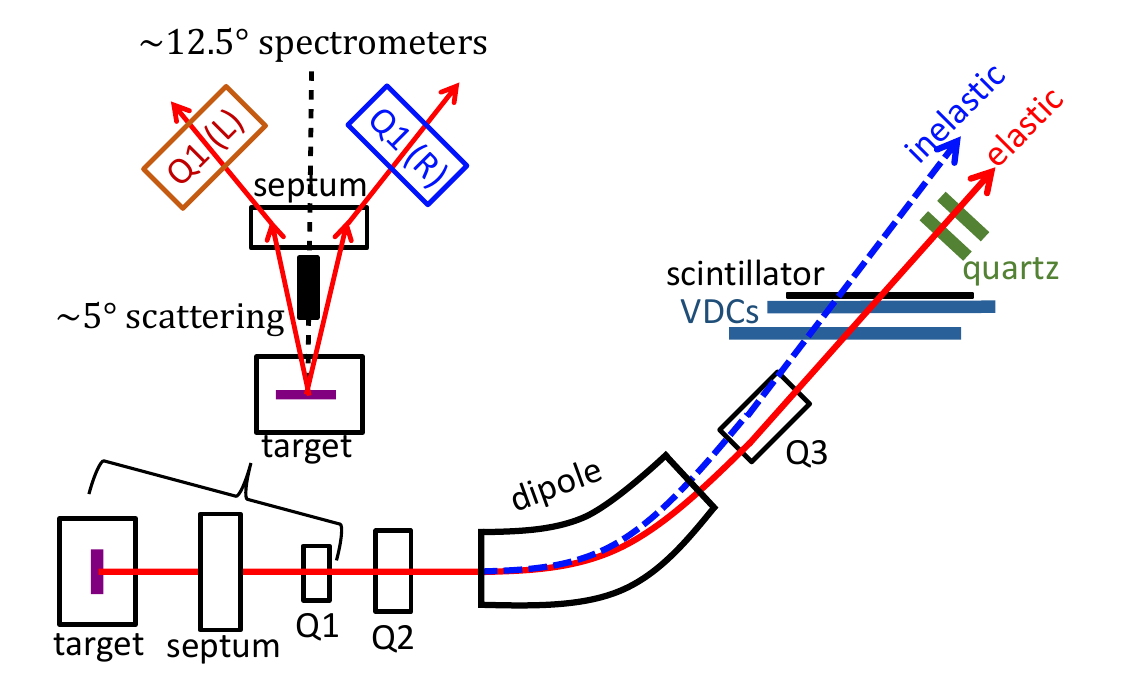}
	\caption{Schematics of the two HRSs in Hall A. Above - top view; below, side view of one of the HRSs.  The HRS magnet system (QQDQ) augmented by a septum magnet upstream of the acceptance-defining first quadrupole Q1 focuses elastically scattered electrons onto quartz integrating detectors, while causing non-elastic events to miss.  The septum magnets allow the experiment to reach $\approx5^\circ$ scattering angle, which increases the figure-of-merit.}
	\label{fig:HRS}
\end{figure}

Figure \ref{fig:HRS} shows top and side views of the High Resolution Spectrometers (HRSs) in Hall A~\cite{HallA_NIM}, which select and focus the scattered electrons onto the detectors at the focal plane. A septum magnet between the target scattering chamber and the entrance to each HRS provides pre-bending to allow the experiments to reach the further forward scattering angles than the HRSs alone can achieve. The electrons that scatter from the solid targets are collimated by precision collimators at the entrance to each HRS.
The detectors in each spectrometer arm consisted of a set of two thin fused-silica bars (henceforth also referred to as quartz), each coupled to photomultipler tubes (PMTs) which collect Cherenkov light created by the passage of electrons.  

The asymmetry measured in the experiment is the difference in the integrated yield of the quartz detectors in the two helicity states, where the yields are  normalized
to the incident charge.
The measured asymmetry must be corrected for background contributions, false asymmetries due to helicity-correlated beam properties, and the overall polarization of the beam, as described in Sec.~\ref{sec:analysis}. 

The asymmetry $A_{\rm PV}$ generally increases with $Q^2$ while the cross section decreases, which leads to an optimum choice of kinematics~\cite{Horowitz:1999fk}. For parity-violating neutron density experiments, the optimum kinematics is the point which effectively minimizes the error in the neutron radius, $R_n$. This is equivalent to maximizing  the following product, which is the figure-of-merit (FOM):
\begin{equation}
{\rm FOM} =   R \times A^2 \times {\epsilon}^2,
\label{eq:FOM}
\end{equation}
where $R$ is the scattering rate, $A$ is the asymmetry, $\epsilon = \frac{dA/A}{dR_n/R_n}$ is the sensitivity of the asymmetry to a small change in $R_n$ (the neutron radius), $dR_n/R_n$ is a fractional change in $R_n$ and $dA/A$ is a corresponding fractional change in $A$.  Note that the FOM defined for many types of parity-violation experiments is $R \times A^2$, but the neutron-density measurements must also fold in the sensitivity $\epsilon$.

Given practical constraints on the solid angle of the HRS, the FOM is maximized at smaller scattering angles. Using septum magnets we can reach down to $\approx 5^{\circ}$ scattering angle.
With a fixed scattering angle and the constraints of CEBAF, the optimum energy for these experiments was determined to be 953~MeV for PREX-2 and 2182~MeV for CREX. 

\subsection{Polarized Electron Source}
\label{sec:polsource}
In this section we provide a description of the polarized source and laser optics, including a new type of PC.  Recent references on the source can be found in \cite{Adderley:2022uql,RTP:2021}.  For older references see \cite{Adderley:2010zz,Sinclair:2007ez,Humensky:2002uv}.  Here we discuss the most important features for understanding the work presented in this paper.

\begin{figure}[tbp]
    \centering
    \includegraphics[width=.45\textwidth]{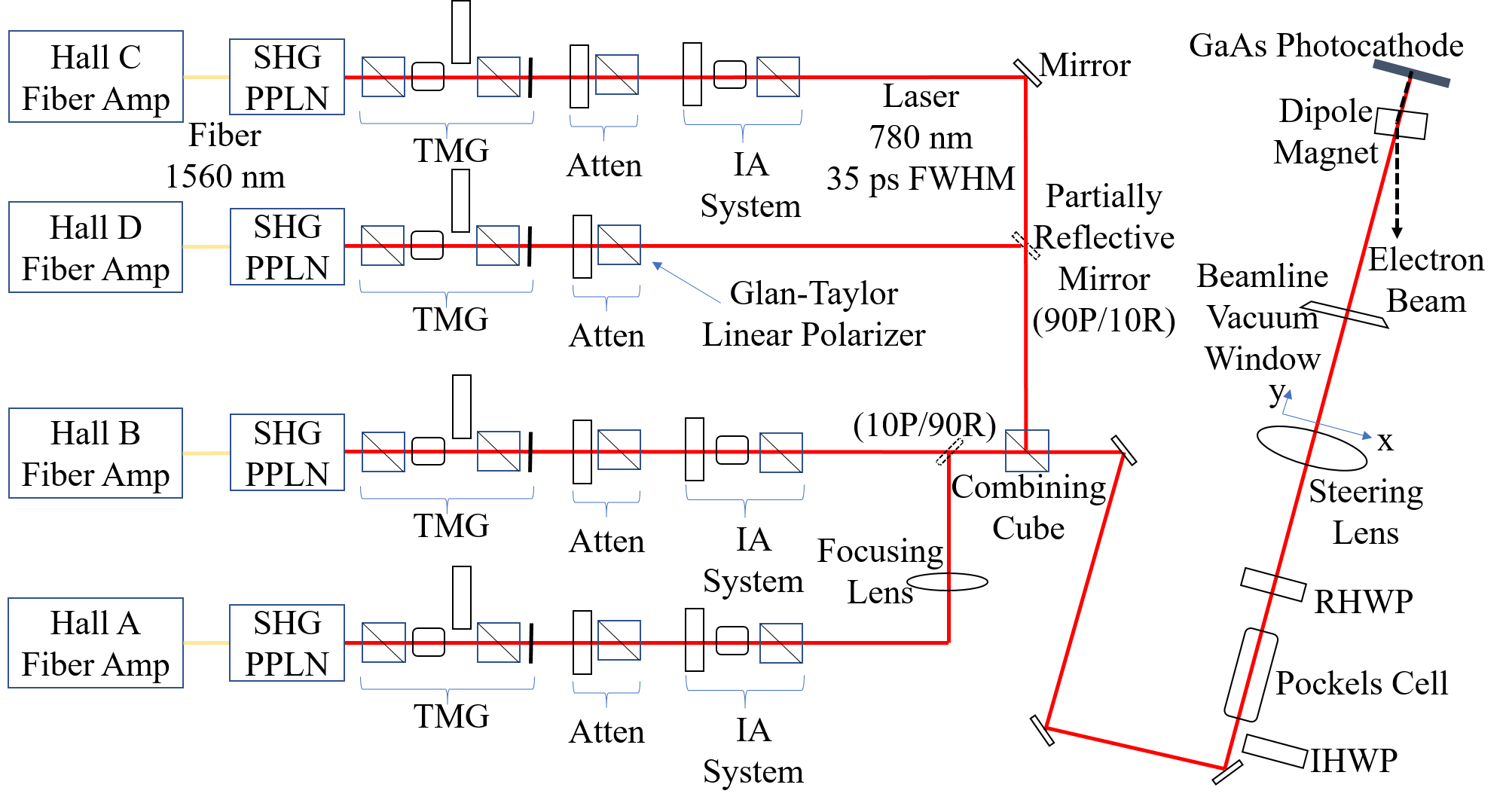}
    \caption{Laser table layout at the injector. For each experimental hall, an RF-pulsed fiber-amplifier-based laser system operating at 499 MHz or 249.5 MHz is used to generate a few Watts of laser power at 1560 nm. This light is then converted to 780 nm using second-harmonic generation (SHG) inside a Periodically Poled Lithium Niobate (PPLN) crystal. The Tune-Mode-Generator (TMG), which consists of a fixed linear polarizer, PC, insertable half-wave plate (IHWP), fixed linear polarizer, and a mechanical laser shutter, is used to generate machine-safe beam for accelerator tune-up. The laser power attenuator (Atten) consists of a remotely rotatable HWP and fixed linear polarizer and controls the power and thus the electron beam current delivered to the experiment. The Intensity-Attenuator (IA) system consists of a HWP, PC, and a fixed linear polarizer, and is used for intensity feedback for the other halls. Partially reflective mirrors and a beam combining cube are used to combine the four independent laser beams. The fixed linear polarizers (Glan-Taylor) serve to ``clean-up'' laser polarization. A final rotatable half-wave plate (RWHP) is used to minimize helicity-correlated position differences.}
    \label{fig:LaserTableLayout}
\end{figure}

The critical components of the CEBAF injector laser-table setup for the polarized electron source are shown in Fig.~\ref{fig:LaserTableLayout}. The components include clean-up Glan-Taylor polarizers, an insertable zero-order half-wave-plate for 780~nm (IHWP) to flip the helicity of the electron beam, the PC common to all laser beams used to make circularly polarized light, a rotatable half-wave-plate (RHWP) for 780~nm, a steering lens to focus the beam onto the photocathode and move the laser beam to different locations at the photocathode, the window at the vacuum chamber containing the photocathode, and finally the strained-superlattice 
GaAs photocathode which produces a highly polarized (80-90\%) electron beam . A second focusing lens was added just for the Hall A laser to better control the laser spot size and divergence at the PC.

The PC is an electro-optical switch for rapidly reversing the helicity of the laser at the polarized source. We have developed a new PC based on RTP (rubidium titanyl phosphate), a material that provides faster switching and higher repetition rate capabilities than the commercial KD*P (potassium dideuterium phosphate) PC~\cite{RTP:2021}.  RTP suffers less from piezo-electrical ringing.  However, RTP's uniformity is not as good as that of KD*P, making for poorer extinction ratios, and it is highly birefringent.  The new RTP cell was designed using electric field gradients to counteract this birefringence, leading to improved extinction ratios (in $\lambda$/2-wave configuration) and minimization of voltage-dependent beam steering (in $\lambda$/4-wave configuration). Acceptable RTP crystals require precision face-cut angles. The RTP cell provided fast flipping and adequate (nm-level) control of polarized electron beam steering.

\subsubsection{The PITA effect}
{\label{section:PITAeffect}}

\begin{figure}
	\centering
	\includegraphics[width=0.350\textwidth]{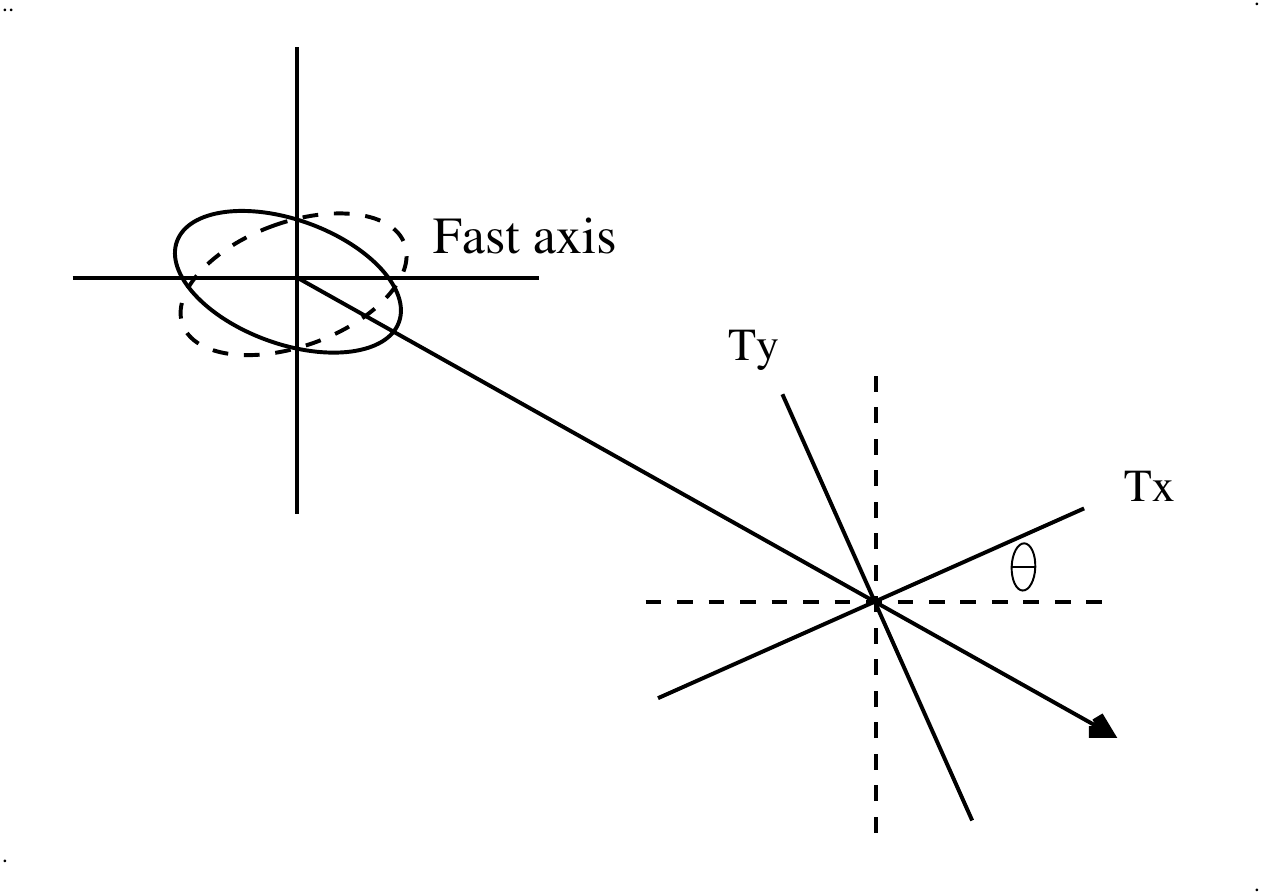}
	\caption{Origin of the PITA (polarization-induced transport asymmetry) effect.  At the PC, shown on the left, two imperfectly circularly-polarized laser helicity states are produced (left handed, dashed, and right handed, solid, lines).  The light is transmitted by a laser transport system, shown on the right, to the photocathode.  The intensity asymmetry arises from a product of three factors: an imperfect phase of the light between $x$ and $y$ axis, a non-zero asymmetry in the transport asymmetry, and the misalignment of the transport system with the fast axis of the PC.} 
	\label{fig:PITAeffect}
\end{figure}

An asymmetry in the intensity of the laser on the GaAs crystal of the polarized source arises from the ``PITA'' effect (Polarization-induced transport asymmetry), see Fig.~\ref{fig:PITAeffect}. This asymmetry is
\begin{equation}
\label{eq:intensityAsy}
A_I = \frac{I_+ - I_-}{I_+ + I_-} ,
\end{equation}
where $I$ is the light intensity and the subscripts $\pm$ refer to the sign of the HV on the PC, which gives rise to the two helicity states.

The polarization of a propagating electromagnetic wave is determined by the magnitudes and the phase relationship between two orthogonal components of the electric field vector. For perfectly circularly  polarized light, the phase is $\frac{\pi}{2}$. In practice, a PC usually produces elliptically polarized light. The absolute value of the phases are very close to $\frac{\pi}{2}$ for both helicity states. Since they are prepared independently, one can introduce two independent phase errors $\alpha$ and $\Delta$ to the two phases $\delta_{\pm}$:
\begin{equation}
\delta_{\pm} = \pm(\frac{\pi}{2} + \alpha) + \Delta \;  .
\end{equation}

The optical transport system downstream of the PC (system of mirrors and vacuum window) acts like a polarizer with the two axes ($x$ and $y$) having slightly different transport efficiencies, $T_x$ and $T_y$.  The fractional difference between these transport efficiencies $\epsilon$ is typically of order $\epsilon \approx10^{-2}$. Designating the angle between the transport system axes with respect to the fast axis of the PC as $\theta$, the PITA asymmetry is 
\begin{equation}
\label{eq:PITA}
        A_{\rm PITA} = \epsilon \Delta {\rm sin}(2 \theta) \; .
\end{equation}

It can be seen that, while a non-zero $\alpha$ is relatively benign, a non-zero $\Delta$ induces a helicity-dependent phase shift, effectively a residual linear polarization that changes sign with helicity. The PC provides control over the PITA effect through changes to the applied voltages. Any beamline component (such as the vacuum window and the PC itself) with a static birefringence-induced $\Delta$ phase shift which is independent of the helicity state will also generate residual linear polarization, independently of the primary helicity-flipping voltage applied. It is possible to tune the PC to mitigate $\Delta$ phase shifts within the various beamline components, intentionally setting the PC voltages in a helicity-dependent way so as to cancel all other sources of $\Delta$ phase shifts, and therefore also minimize the PITA effect.

\subsubsection{Insertable (IHWP) and Rotatable (RHWP) Half-Wave Plates}
\label{sec:IHWP}

Inserting and retracting the IHWP in the laser path is a type of slow reversal of the sign of the circularly polarized light.  
During PREX-2 the polarization differed by about 1\% between runs with IHWP in and IHWP out, due to an unexpectedly large birefringence of the vacuum window. This effect was found after PREX-2 and fixed by replacing the window before CREX, eliminating this difference.

\begin{figure}
	\centering
	\includegraphics[width=0.35\textwidth]{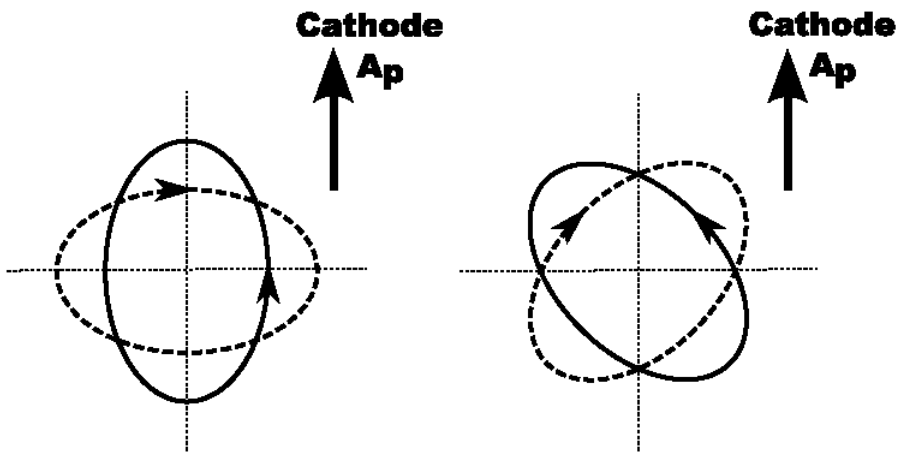}
	\caption{Maximal PITA effect sensitivity from the cathode's analyzing power ($\mathrm{A_p}$) due to residual linear polarization asymmetry between two imperfectly circularly-polarized laser helicity states (left handed, dashed, and right handed, solid, lines) is shown on the left, and the case with minimal PITA effect sensitivity by rotating the axes of the two states' residual linear polarization to $45^{\circ}$ with respect to the cathode's $A_p$ optical axis is shown on the right. Reproduced from \cite{Paschke_HCBA_2}.}
	\label{fig:simple_birefringent_axes}
\end{figure}

An RHWP is placed downstream of the PC (see Fig.~\ref{fig:LaserTableLayout})to assess PC alignment and to minimize helicity-correlated position differences. 
The RHWP changes the orientation of asymmetric linear polarizations that induce helicity-correlated beam asymmetries. Calibrations were performed to choose a rotation angle of the RHWP. For PREX, we chose to align its axis close to $45^{\circ}$ with respect to the GaAs cathode's linear polarization analyzing power axis to generally but not completely minimize $A_{I}$, as in Fig.~\ref{fig:simple_birefringent_axes}. For CREX, we chose an angle further away from $45^{\circ}$, for reasons explained in Sec.~\ref{sec:QasyFeedback}. To minimize steering effects during the RHWP scans, the voltages related to steering on the RTP Pockels cell were optimized with independent values for IHWP ``IN" versus IHWP ``OUT".  Position differences were reduced by adjusting the PC pitch and yaw angular alignments.

\subsubsection{Double Wien Flip Apparatus}
\label{sec:WienFlip}
\begin{figure}
	\centering
	\includegraphics[width=0.45\textwidth]{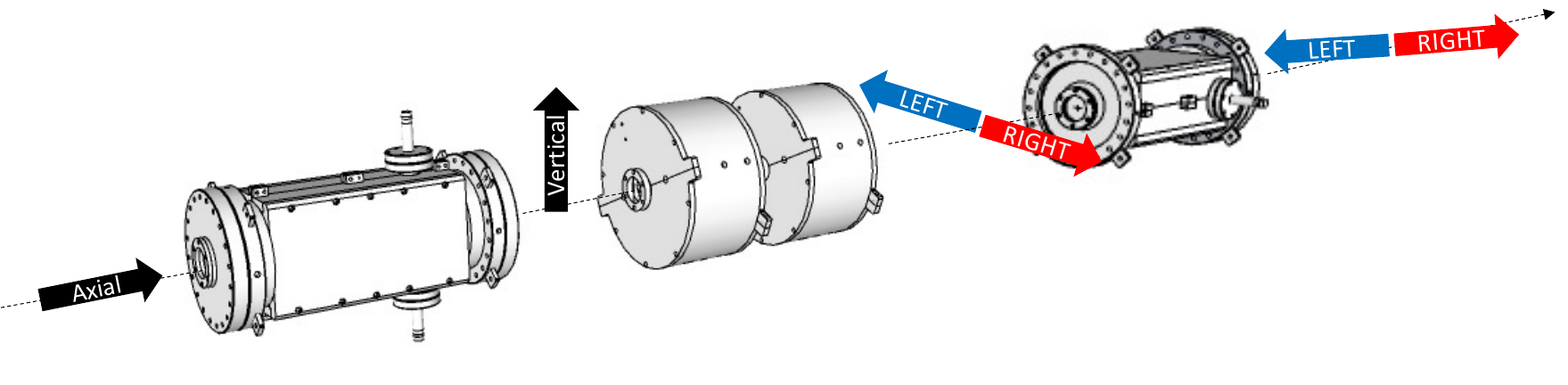}
	\caption{Schematic of the double Wien spin flipper used to optimize the spin direction and as a slow reversal of the electron beam polarization.  This is achieved by a Vertical Wien filter which rotates the spin from the axial to vertical direction, followed by a pair of solenoids which selectively rotate the spin from the vertical direction to either a Left or Right direction, providing an overall sign change of the longitudinal polarization. A Horizontal Wien rotates the spin by an amount required to ensure that left (right) is parallel (anti-parallel) to the momentum at the experiment after spin precession in the accelerator.}
	\label{fig:Double_Wien}
\end{figure}

To ensure longitudinal polarization for the beam when it arrives at the target, and to provide another method of slow helicity reversal, a double Wien spin flipper \cite{Adderley:2022uql} was deployed in the beamline, downstream of the polarized source and before the first accelerating cavity (see Fig.~\ref{fig:Double_Wien}).

The first ``vertical" Wien filter rotates the polarization vector of photo-produced electrons perpendicular to the plane defined by the two accelerating sections.
The solenoids then rotate the spin to the horizontal direction and the final ``horizontal" Wien filter adjusts the spin orientation to compensate for spin precession in the accelerator. Thus, the double Wien spin apparatus can flip the direction of the spin relative to the momentum, achieving an overall change of the sign of the polarization.  
While the solenoid settings impact accelerator optics, their effect on the focusing predominantly depends on $B^2$, so the sign flip's impact on the accelerator tune is small. 
Thus, taking data with both spin orientations helps to suppress possible higher-order systematics, such as a helicity-correlated beam spot size or shape caused at the polarized source (Sec.~\ref{sec:hcLaserSpotSize}).
The diagnostic tool for tuning the vertical Wien filter
and solenoids is the Mott polarimeter~\cite{Aulenbacher:2018weg} while the horizontal Wien filter setting is verified by the
M\o ller polarimeter in the Hall~\cite{NIMA2023167506}.

\subsubsection{Helicity Control Electronics}
\begin{figure}[!htbp]
	\centering
	\includegraphics[width=0.45\textwidth]{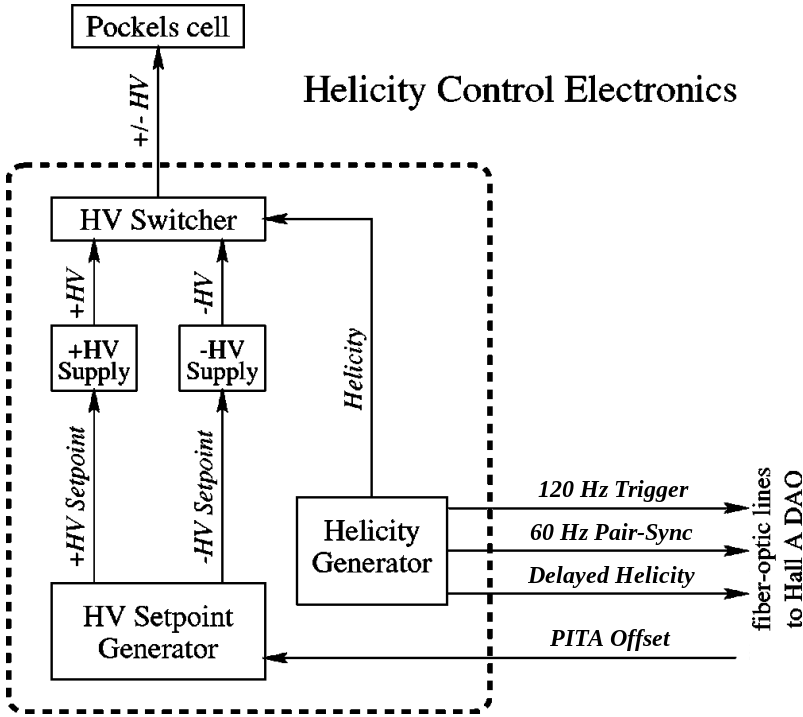}
	\caption{Schematic of the helicity-control electronics. The helicity-state logic signal is transmitted via fiber optic cables and determines the sign of the high voltage on the PC. The PC and helicity-signal generator are both electrically isolated from the beamline and data acquisition components elsewhere in the laboratory (dashed box). Adapted from \cite{HAPPEX_long_paper}.}
	\label{fig:helicity_control_electronics}
\end{figure}

\begin{figure}[!htbp]
	\centering	\includegraphics[width=0.45\textwidth]{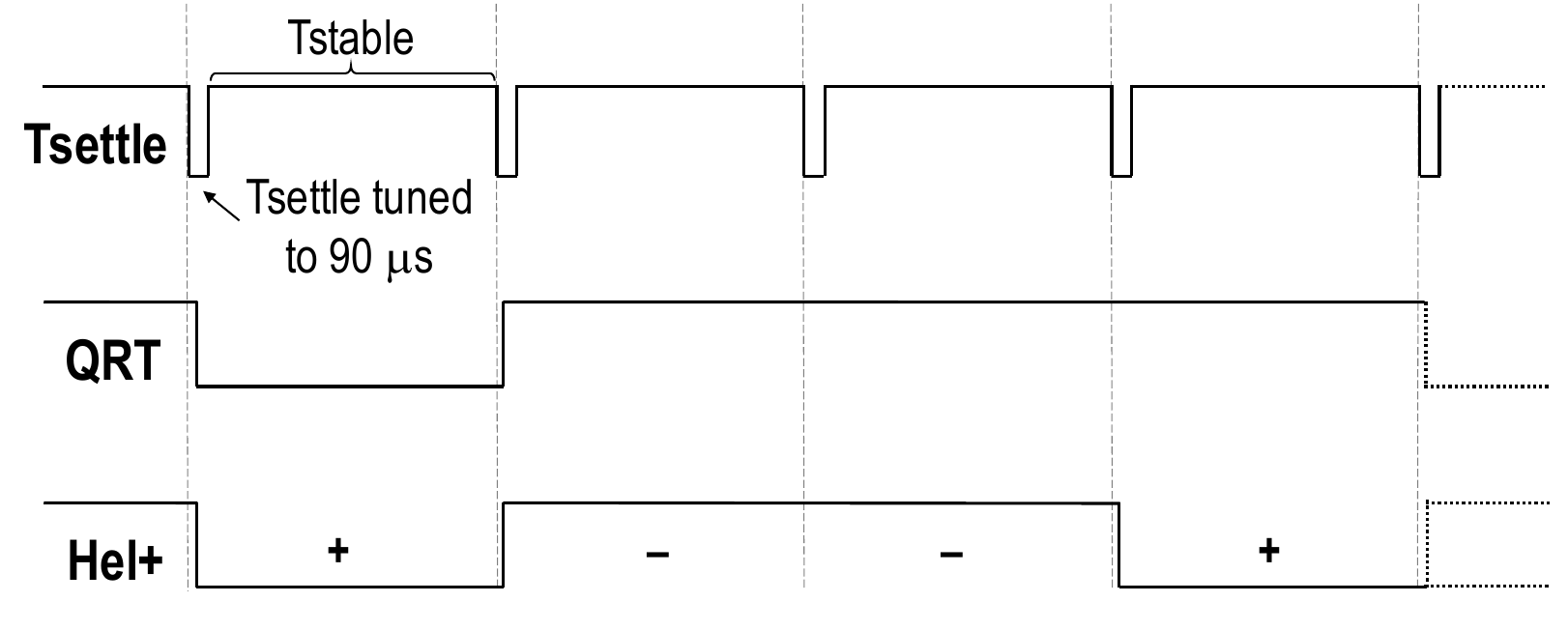}
	\caption{The helicity logic signals timing diagram for CREX, which ran in quartet mode at a helicity-flip rate of 120 Hz. PREX-2 ran in both quartet mode and octet mode, which is two opposite polarity quartets following each other at twice the helicity-flip frequency. The QRT (quartet) and Helicity signals change state one microsecond after the T$_{\rm settle}$ signal begins. The helicity multiplet polarity, the sign of the first window in a multiplet, is chosen by the Helicity Control Board according to the predictable pseudo-random pattern for each new QRT.}
	\label{fig:helicity_logic}
\end{figure}

The helicity control electronics, described in Ref.~\cite{Adderley:2022uql}, start with the helicity control board~\cite{HCB_UserGuide} in an electrically isolated VME crate in the Injector Service Building (Fig.~\ref{fig:helicity_control_electronics}).
The ``helicity window'' timing is defined by the $T_{\rm settle}$ and $T_{\rm stable}$ periods (Fig.~\ref{fig:helicity_logic}). 
The $T_{\rm settle}$ time 
when the helicity state is potentially changing or unstable serves as a veto signal for data collection.

Helicity events are grouped into multiplets. The helicity sign of the first event of multiplet is determined by the Helicity Board using a 30 bit pseudo-random number generator.  
The logical helicity signals themselves are transmitted via fiberoptic cables to allow complete ground-isolation of the helicity-generator circuit, reducing the possibility of electronic pickup
in other parts of the experiment. 

The randomly-alternating multiplet pattern largely cancels 60 Hz line or other noise and the pseudo-random selection of multiplet sign and the simultaneous production of both a ``Hel+'' and ``Hel-'' helicity signal cancels potential power load or electronic pickup signals. 
Further, the helicity signal sent to the experiment's DAQ is delayed by 8 multiplets, removing any correlation with the true helicity of the electron beam during that multiplet window. This was verified by beam-off electronics studies and DAQ studies where constant voltages were applied, which showed that the electronics noise was controlled at the few ppm level, with no noticeable accumulated asymmetry signal at the ppb scale.

For PREX running we utilized a helicity pattern first at 120 Hz with quartets and later at 240 Hz with octets, while CREX used 120 Hz quartets. For both experiments $T_{\rm settle}$ was chosen to be 90~$\mu s$, 
small compared to the 4.16~ms scale of the helicity windows (at 240~Hz helicity flip).

\subsubsection{Setup of the Laser and Electron Beam}
\label{Sec:SetupLaser}
The PC alignment procedure~\cite{RTP:2021} minimized the degree of linear polarization for both helicity states.   
Birefringent gradient effects were minimized by adjusting the pitch and yaw of the PC to minimize the position differences as measured with a quad photodiode.
The analyzing power of 
6\%~\cite{Adderley:2022uql} for the strained superlattice GaAs/GaAsP photocathode was obtained by measuring the photocathode quantum efficiency with linearly-polarized light rotated through $180^{\circ}$ using the rotating halfwave plate.  

\subsubsection{Helicity-Correlated Laser Spot Size}
\label{sec:hcLaserSpotSize}
Measurements on the laser table constrained the helicity-correlated variations in the spot size $\sigma$ (standard deviation of width) for the electron beam to $\Delta \sigma / ( 2 \sigma ) < 10^{-4}$.
Laser measurements were performed with a linear-array photodiode (LAPD) detector.  
The laser beam spot size $\sigma$ is measured as
\begin{equation}
\sigma = \sqrt{ \frac{\sum I(x_i)(x_i - \bar x)^2 }{\sum I(x_i)}} ,
\end{equation}
where $x_i$ and $I(x_i)$ are the beam position and the density on the $i^{\rm th}$ element of the array.  
To infer a spot-size asymmetry in the electron beam, a model is used which takes into account the transport of the laser to the photocathode and the measured analyzing power of the cathode (see Sec.~\ref{Sec:SetupLaser}).  
The electron beam spot size was also analyzed independently with beam position monitor (BPM) data, which led to a comparable upper bound, see Sec.~\ref{sec:hcSpotSize}.

\subsection{Control of Electron Beam Properties} 
\label{sec:syscon}

\subsubsection{Beam Properties}
The experiments ran at beam currents of 70~$\mu$A (150~ $\mu$A) for PREX-2 (CREX).  PREX-2 ran briefly at higher currents, up to 85 $\mu$A. The beam energy was $953 \pm 1$~MeV (PREX-2) and $2182 \pm 1$~MeV (CREX)~\cite{santiesteban2021precise}. The intrinsic spot size of the beam, measured with a wire scan, 
was adjusted by quadrupole magnets on the beamline 
to have a width of $80\leq\sigma \leq 150$ $\mu m$ at the target, which reduced the heat load on the target foils.  

The beam was rastered (Sec.~\ref{subsection:raster}) with a size of 6~mm$\times4$~mm to spread the beam over the face of the target and thereby reduce the instantaneous heating. The beamline instrumentation for controlling the beam trajectory and determining the position and current was the same as in previous parity-violation experiments~\cite{HAPPEX_long_paper,Qweak:NIM}.

\subsubsection{Helicity-Correlated Beam Asymmetries}
\label{sec:HCBA}
Changes in the electron beam parameters correlated with helicity reversal can give rise to a systematic shift in the experimental asymmetry, as explained below.  Control of these systematics started with a
careful setup of the laser optics, specifically the PC (Sec.~\ref{Sec:SetupLaser}) and RHWP (Sec.~\ref{sec:IHWP}) to 
reduce
the asymmetries. They were further reduced
online with feedback (Sec.~\ref{sec:QasyFeedback}) and their effects partly canceled by the slow reversals (Sec.~\ref{sec:IHWP},~\ref{sec:WienFlip}).

\subsubsection{Intensity Asymmetry and Feedback }
\label{sec:QasyFeedback}

The intensity feedback system manipulates the PC voltages in the injector to correct for helicity-correlated intensity asymmetries ($A_{I}$, see Eq.~\ref{eq:intensityAsy}) caused by the PITA effect (see Sec.~\ref{section:PITAeffect}).
The ``PITA slope", which is how the induced intensity asymmetry 
depends on 
 the applied external voltage to the PC, was calibrated, and used in a feedback correction loop every 7.5 seconds~\cite{Paschke_HCBA_2}. 

By not fully rotating the RHWP to $45^{\circ}$ with respect to the cathode, the remaining sensitivity of the cathode's analyzing power axis with respect to the RHWP-rotated incident linear polarization is used to minimize the impact of static imperfections in the PC's birefringence. The lever arm was sufficient both to generate an opposite linear polarization to cancel other birefringence sources such as the vacuum window, as well as for intensity feedback. Additionally, the birefingence was minimized by a rotation of the cathode itself to align its axis with respect to that of the vacuum window. The feedback minimized the net $A_{I}$ below 100~$\rm{ppb}$ such that its systematic error  on $A_{\rm PV}$ was below 1~$\rm{ppb}$.

Alongside the PITA intensity asymmetry feedback on the Hall A beam current, we also performed intensity feedback on the Hall C beam current using the IA system, similarly to the PITA feedback. This was done to minimize the impact of large beam intensity asymmetries in Hall C on the beam properties in Hall A. Feedback was not needed on the beam for Halls B and D, as their intensities were much lower, so their intensity asymmetries did not affect the beam properties in Hall A. The ``intensity asymmetry'' (IA) cells are KD*P Pockels cells placed between two polarizers such that they do not affect the output laser polarization, and only affect the beam intensity and intensity asymmetry $A_I$, thus allowing control over $A_I$. A small helicity-correlated voltage, typically in the range of $\pm$5V, was applied making intensity asymmetry corrections on the order of a few hundred ppm. 

\subsubsection{Position Differences}
\label{sec:Position Differences}

Helicity-correlated position-dependent differences are produced by spatial gradients in the $\Delta$ phase shift found in birefringent optical components in the injector beamline optical, or in the analyzing power of the residual linear polarization. Additionally, position differences may be produced by electron beam clipping on an obstruction, which we avoided through a careful tuning of the electron beam to clear all apertures~\cite{Paschke_HCBA}. 
These asymmetries are potential sources of false asymmetries and are discussed in Sec.~\ref{sec:BeamCorrectionsTheory}.  Position-dependent differences from optical elements with $\Delta$ phase shift gradients in the injector, as well as spot-size asymmetries, are suppressed by the rotation of the RHWP and by tuning of the PC voltages to values that minimize the analyzing power; these values are determined in dedicated calibration runs. Additionally, the PC was used to feed back on 
helicity-correlated differences in beam positions, as described next.

\subsubsection{Adjustments to PC During the Run}
\label{sec:AdjustPC}
The measured position differences during the first part of PREX-2 were $\approx$50~nm for 1 day of data taking after the PC alignment and RHWP optimization. 
Based on the expected sensitivities to beam position differences, and our precision goal for the experiment, our requirement was to maintain these differences below 10~nm. Therefore 
a procedure for a slow feedback (on the time scale of 1 day) was employed for the rest of the run. 
The slope that relates the PC voltage to the position difference in the experimental hall were measured.
The position differences were measured over 1 day and manual adjustments were done to the RTP PC using the calibrated slopes. The initial 50~nm differences were canceled with a setup that had similar amplitude but opposite sign.
Subsequent feedback steps were smaller as the position differences converged to zero.  A similar procedure was utilized during CREX on the $x$ and $y$ positions as well as the energy.

\subsubsection{Electron Beam Spot Size Variation}
\label{sec:hcSpotSize}

A helicity-correlated beam spot size, such as a breathing mode where the radius has a dependence on beam helicity, can arise from spot-size variations of the laser beam at the polarized source, see Sec.~\ref{sec:hcLaserSpotSize}.  This is a  ``second order'' beam correction that is not accounted for in our model for beam corrections (see Sec.~\ref{sec:BeamCorrectionsTheory}) and was therefore important to keep small. One of the important purposes of the Wien Flip (Sec.~\ref{sec:WienFlip}) was to further suppress this effect via cancellation. Here we discuss upper bounds on this effect as determined using BPMs in the injector.  

The standard BPMs measure the beam position using the induced signals on 4 BPM wires, located $90^{\circ}$ apart in azimuth. Three linear combinations of the four BPM wire channels are used to determine beam current $I$, and positions $x$, $y$.  A fourth linear combination can be computed which is proportional to the elliptical component of the spot size asymmetry, called $\epsilon_{\rm BPM} $:
 \begin{equation}
    \epsilon_{\rm BPM} = \frac{a^2 (xp + xm - yp - ym)}{8 \sigma^2 (xp + xm + yp + ym)} ,
\end{equation}
here a $\approx7$ mm is the radial position of the signal detection wires in the BPM vacuum chamber, $\sigma$ is the electron beam spot size at the BPM, which is typically 200~$\mu$m, and $xp, xm, yp, ym$ are the wire signals at $\pm 45^\circ$ to the beam. The wires with index $p$ are radially opposite in the chamber to those with $m$ for orthogonal positions $x$ and $y$.

To use this formula, one must correct for the position sensitivity in 
$\epsilon_{\rm BPM}$.
Once this is done the electron beam data at the injector, measured parasitically prior to the experiments, yielded 
$\epsilon_{\rm BPM} < 2 \times 10^{-4}$,
consistent with the laser table measurements. These measurements were not sensitive, however, to any circularly-symmetric spot-size variation.

\subsubsection{Beam Polarization Components}

The electron beam spin direction at the injector was manipulated using a set of horizontal and vertical Wien filters  (see Fig.~\ref{fig:Double_Wien}) such that it was fully longitudinal at the target.  A transverse polarization component creates a parity-conserving asymmetry. To first order, the effect of any 
vertical polarization will cancel in the experimental apparatus due to the left-right symmetry of the spectrometers and detectors. Symmetry-breaking differences between left and right due to beam position offsets and slight differences in position and energy sensitivity and acceptance between the two arms  were quantified and the systematic error was computed (Sec.~\ref{sec:Atmeasurement}).

The horizontal polarization of the beam precesses through each of the arcs in the accelerator and in the beamline  entering the hall~\cite{Higinbotham2009}. 
At a beam energy of 953~MeV for PREX-2, this translates into 286$^\circ$ of precession, whereas at the energy used for CREX of 2182~MeV the precession 
was 655$^\circ$. 
The setup of the injector horizontal launch-angles at the Wien filter are accurate to 
within 1$^{\circ}$ at the beam energies used in PREX-2 and CREX, 
mainly due to uncertainty in the Wien angle. 
At PREX-2 and CREX energies even relatively large energy changes of 500~ppm (
above the observed fluctuations of a few 100~ppm) would have a negligible effect on the total precession ($\ll 1^\circ$). 

The PREX-2 and CREX experiments were sensitive to any horizontal component of the transverse polarization as there was no cancellation in the up-down direction. Special detectors called ``${\rm A}_{\rm T}$ detectors'' (see Sec.~\ref{sec:detectors}) were placed close to the focal plane of the spectrometer in a region with increased sensitivity to transverse polarization and were used to measure and correct for contamination from the parity-conserving asymmetry (Sec.~\ref{sec:Atmeasurement}). 

For the purpose of measuring the analyzing powers for the transverse asymmetries~\cite{AnResults}, the double Wien spin flipper was set to produce vertical polarization 
by using the vertical Wien (see Fig.~\ref{fig:Double_Wien}) to set the spin direction to vertical, and  turning off the solenoids; the horizontal Wien then does nothing,  leaving the  spin vertical.

\subsubsection{Adiabatic Damping}

\label{sec:adiabatic_damp}
The CEBAF accelerator may be thought of as a folded transfer line,
and the different segments of the accelerator have specific optics purposes~\cite{CEBAF}.  Following each linac is a splitter and matching region which separates the different energy beams and defines the beta function for entry into the following arc or the extraction line to the experimental halls.  Beta functions and emittances (Courant-Snyder or Twiss parameters) are measured in the matching regions following each splitter and quadrupole fields are modified to match the beta functions intended for entry into the following region.  If beam envelopes are within 50\% of design values following the adjustment the machine is deemed to be ``well matched''. 

There are skew quadrupoles before the first cryomodule and after each successive cryomodule in each linac to null skew quadrupole terms in the accelerating cavities, thereby minimizing $xy$ coupling. If a region of the accelerator beamline is well matched and free of $xy$ coupling, the helicity-correlated position differences can be reduced, or damped, proportionally to $\sqrt{(\alpha/P)}$ where $\alpha$ is a constant and $P$ is the momentum~\cite{edwards_syphers}.
The position differences become, ideally, reduced by a factor of $\sqrt{(1.0 \ {\rm GeV} / 5 \ {\rm MeV)}} \approx 14$, for PREX-2 beam energy, between the 5 MeV region and the target. 
Deviations from this ideal reduction factor can occur due to two effects, namely $xy$ coupling and mismatched beam-lines.
The Courant-Snyder parameters calculated at different sections of the accelerator for various orbits are a measure of the quality of matching, with a constant value at all sections for all orbits indicating perfect 
matching.

\subsubsection{Beam Modulation System} 
\label{sec:BeamModulationSystem}

\begin{figure}[!ht]
    \centering
    \includegraphics[width=0.45\textwidth]{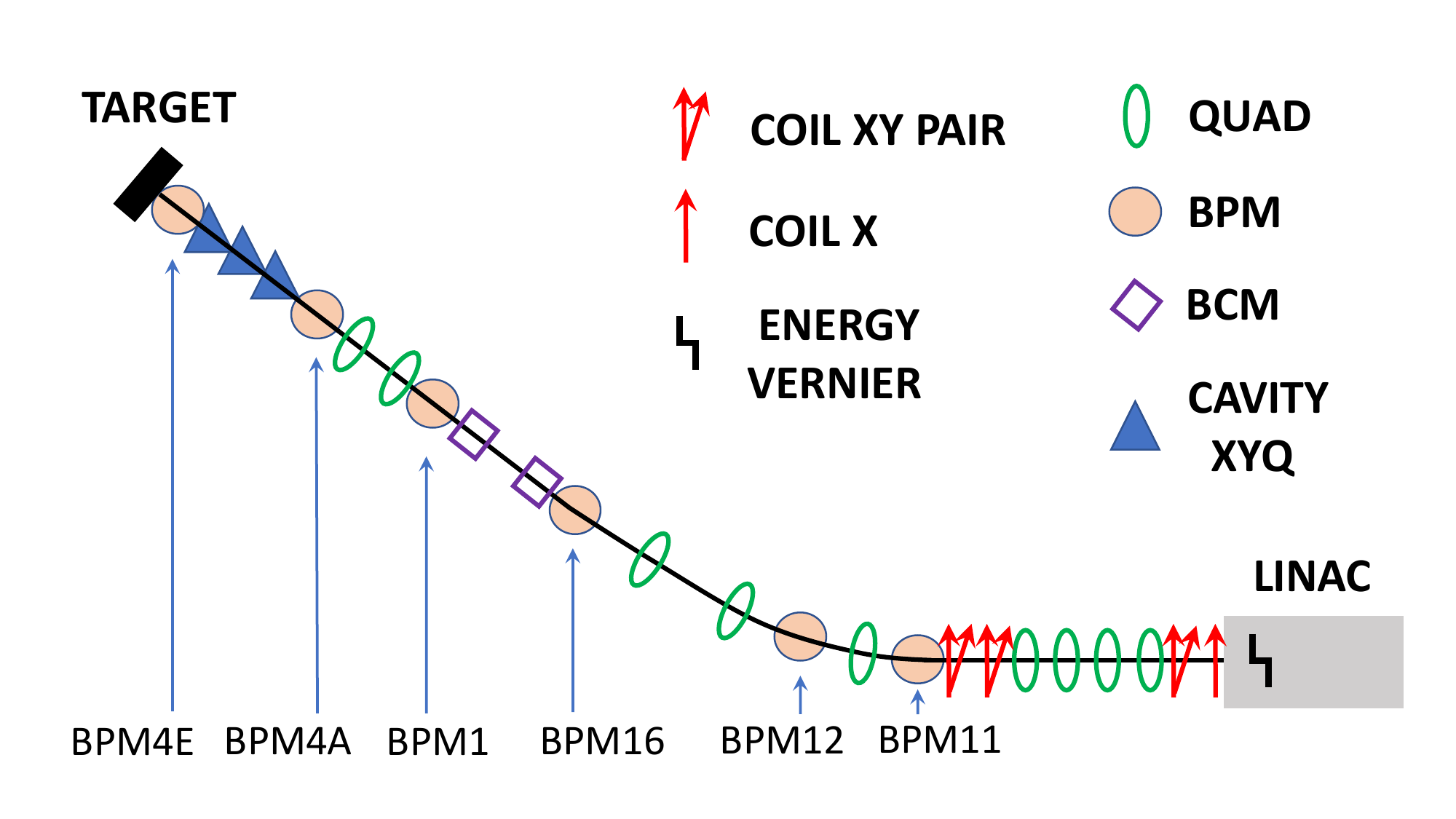}
    \caption{Schematic of the Hall A beam modulation system and other beam instrumentation from the last part of the South LINAC to the target.  Shown are the seven steering coils that were used, an energy vernier in the 
    linac, the quadrupoles for beam optics (QUAD), the beam position monitors (BPM), the beam current monitors (BCM), and the microwave cavity triplets (XYQ).  The BPMs used in the analysis (Sec.~\ref{sec:BeamModulation}) are shown in the bottom of the graphic. The XYQ consist of three separate cavities which provide beam position in two orthogonal directions as well as charge~\cite{Ursic:1997zz}. Only the utilized elements are shown.}
    \label{fig:bmw}
\end{figure}

The sensitivity and corrections to the integrated scattered flux due to fluctuations in the beam parameters are extracted from measurements of the detector's response to controlled modulations of the five main beam parameters (see Sec.~\ref{sec:BeamCorrectionsTheory}), namely the coordinates $x$ and $y$ in the horizontal and vertical planes perpendicular to the beam direction, the corresponding angles $\theta_x$ and $\theta_y$ in those planes, and the energy $E$. These modulations were occasionally made concurrently with production data-taking, representing a few percent of the total production data. 

Beam positions were measured at two points in the field-free region near the target using wire-antenna BPMs~\cite{HallA_NIM} (BPM4A and BPM4E), and in several other positions along the Hall A beamline.  A BPM in the middle of the arc of magnets leading into Hall A is particularly sensitive to the energy; it is called BPM12. The beam modulation system consists of seven magnetic beam-steering coils located several meters upstream of the main bend into Hall A and one energy vernier along the accelerator’s south linac.  The energy vernier varies the phase of an accelerator cavity to modulate the energy. Each steering coil was driven by the output of a 16 bit VMIVME-4145 Waveform Generator~\cite{VMIVME}, which is controlled by the data acquisition system. The system layout on the beamline is shown in Fig.~\ref{fig:bmw}.    

\begin{figure}[!ht]
    \centering
    \includegraphics[width=0.5\textwidth]{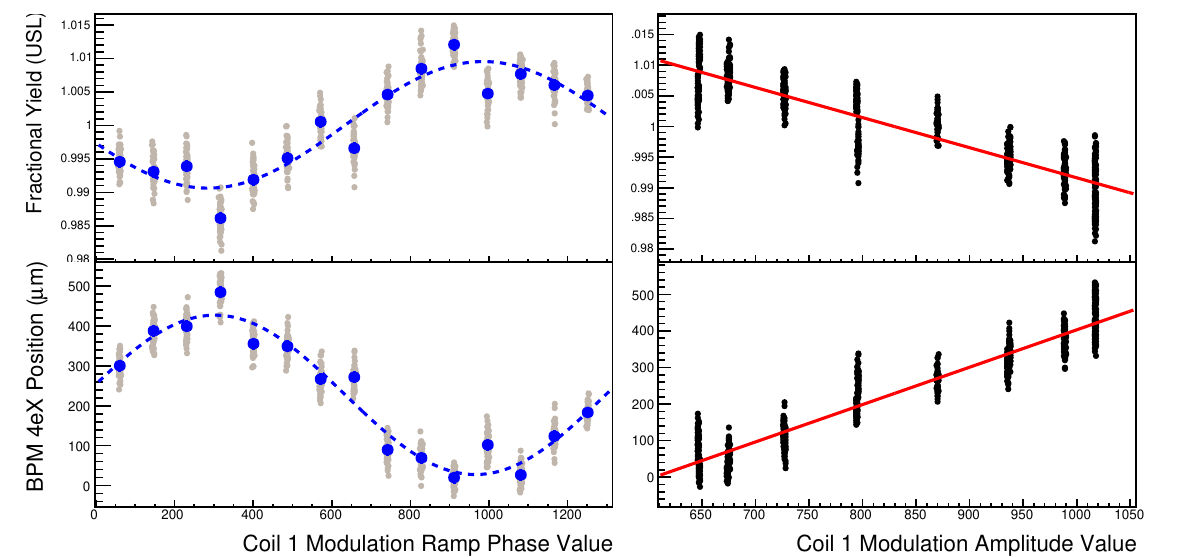}
    \caption{Typical response of the detectors (top left) and a beam position monitor (bottom left) to the beam modulation system. Measured responses (slopes) are shown on the right panels.  The modulation (sine curve) is larger than the natural variations (vertical point distributions at each reported phase value).}
    \label{fig:typicalCoilResponse}
\end{figure}

The modulation coils are chosen to fully span the beam parameter's phase space. Three of the coils provide redundancy to cross-check the orthogonality and completeness of a four-coil measurement. A 16 step 
coil cycle forms a sine wave at the 240 Hz sampling rate and 50 such sine waves form a complete coil cycle lasting 3.33 seconds. A ``supercycle'' is a group of modulation cycles of all the coils and the energy vernier, with a 16.4 s between cycles. Each supercycle lasts for $\approx 3$ minutes and then idles for the next 10 minutes of production data taking. The typical peak-to-peak beam deflection at the BPMs is $\approx \pm 300 \; \mu$m and clearly apparent over the BPM resolution ($\approx 0.5 \; \mu$m) and intrinsic beam fluctuations ($\approx 100 \; \mu$m), see Fig.~\ref{fig:typicalCoilResponse}.  The figure also shows typical slopes which measure the response of the BPMs to the coils that are used in Sec.~\ref{sec:BeamCorrectionsTheory}. For example, $\frac{\partial M_{4ex}}{\partial C_{1}}$ denotes the response of the monitor labeled ``4ex'' to the coil no. 1.
 
\subsubsection{Synchronized Beam Raster}
\label{subsection:raster}

To mitigate localized target heating from the small ($\approx 100 \mu{\rm m}$) beam spot we utilized a fast magnetic raster system~\cite{CYan2005} to spread the beam across the face of the target.
Moreover, we synchronized the two raster frequencies (generating deflections in the $x$ and $y$ directions) of $\approx$ 25 kHz such that each integrated helicity window spans the same pattern on the face of the target.

\begin{figure}
    \centering
    \includegraphics[width=0.45\textwidth]{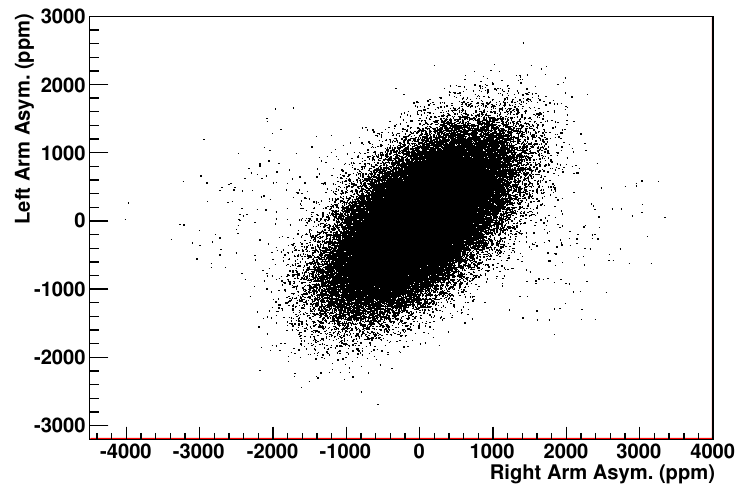}
    \includegraphics[width=0.45\textwidth]{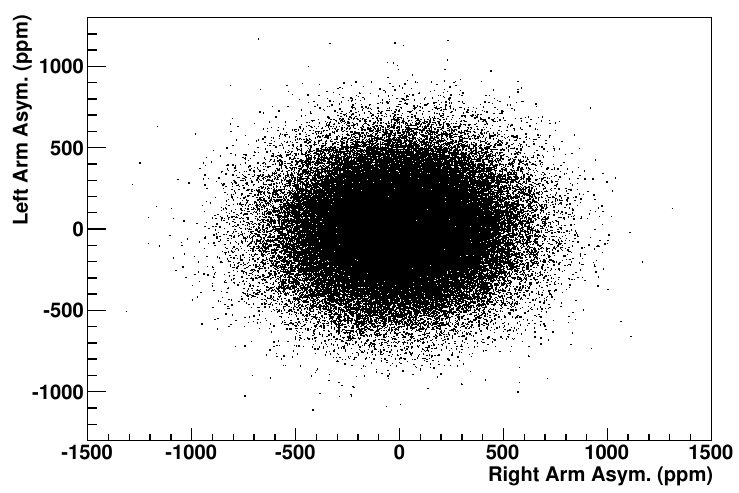}
    \caption{PREX-1 data, showing the left- and right-arm 
    asymmetry distribution correlations.
    The effects of beam position and energy fluctuations have been removed by regression  
    (see Sec.~\ref{sec:beamparsens}) in both plots. 
    When the raster was not synchronized (top) there was residual correlation between the two detector asymmetries. This correlation goes away (bottom) when the raster was properly synchronized to the helicity-flip frequency.}
\label{fig:LR_asym}
\end{figure}

The synchronization ensures that the raster executes the same orbit between two helicity cycles, canceling luminosity fluctuations when one takes the difference. Figure~\ref{fig:LR_asym} shows the correlation of pulse-pair asymmetries measured on the two HRS detector systems during PREX-1 before and after the synchronization
was implemented.  Since the two HRS see different electrons, these asymmetries should be uncorrelated. The extra correlation is a sign of a common-mode noise, which would increase the running time required for the experiment.  The successful reduction of common mode noise with raster synchronization is demonstrated in the figure.

The raster frequencies were adjusted to integer multiples of the $f_{\rm hel} = 120$ Hz helicity-flip frequency, 
\begin{equation} 
f_A = A \times f_{\rm hel} \hskip 0.08in ; \hskip 0.08in  f_B = B \times f_{\rm hel} \hskip 0.08in ; \hskip 0.08in C = A - B .
\end{equation}

If the integer difference $C$ is large and neither $A$ nor $B$ are integer multiples of $C$, this pattern repeats after a longer time and covers the face of the target more uniformly, reducing the heat density. The reduction in heat load was confirmed with computational fluid dynamics (CFD) simulations. 
We used raster frequencies of $A$ = 213 $f_{\rm hel}$ and $B$ = 205 $f_{\rm hel}$ for the 120 Hz running, such that $C$ is 8 and divides neither $A$ nor $B$. With $f_{\rm hel} = 119.99976$ Hz, 
the two frequencies were tuned to $f_A = 25.55994888$ kHz and $f_B = 24.5999508$ kHz for the 120 Hz running of both PREX-2 and CREX. The 240 Hz running for PREX-2 used $f_A = 25.68010272$ kHz and $f_B = 24.72009888$ kHz, setting $A$ = 107 and $B$ = 103. The raster frequencies were controlled by two independent Agilent frequency generator channels which were synchronized to the helicity frequency by hand until they were locked to the helicity signal on an oscilloscope. The frequencies were monitored on an oscilloscope over the course of both experiments and were observed to remain stable to the precision indicated by the reported significant figures shown above.

\subsection{Targets}\label{sec:target} 

The two main production targets were isotopically enriched ${}^{208}$Pb and ${}^{48}$Ca. In addition, a water target was used to determine the central scattering angle $\theta$ and several other targets were deployed for calibrations.  The production targets as well as a $^{40}$Ca target were installed on a horizontal arm with 16 target stations cooled with gaseous helium with an inlet temperature of 14~K and pressure of 12 atm.  The calibration target ladder moved at a $45^{\circ}$ angle to the horizontal, containing 5 target stations including the water cell, and was cooled by water circulating at 1 gal/min within the frame of the ladder.

\subsubsection{Lead Targets}
\label{sec:pbtarget}
The trade-off between the gain in rate with thickness and the loss of rate from the elastic peak due to radiative effects led to an optimum thickness of about 10\% of a radiation length. Isotopically enriched, chemically pure 0.55 mm thick lead targets were used. The lead targets, as well as the calcium targets (see Sec.~\ref{sec:calcium_target})  were purchased from the Oak Ridge Isotope Program. The lead was 99.1\%  ${}^{208}$Pb with 0.7\% of ${}^{207}$Pb, 0.2\% of ${}^{206}$Pb, and negligible amounts ($< 10^{-4}$) of other elements. 

In order to achieve the required scattered count rate, 
most of the data for PREX-2 was collected with a 70~$\mu$A beam current, while for the last 12\% of the data we increased the current to 85 $\mu$A. 
A number of techniques were developed to mitigate the challenge of the low lead melting 
point (327 ${}^{\circ}$C).  The target was operated at cryogenic temperatures and sandwiched with diamond foils to help conduct away heat.  The beam was rastered to evenly distribute the heat deposition over a sufficiently large area.  With these techniques we found that the lead targets had a lifetime of about a week of production data collection. 

During PREX-1, three lead targets were sandwiched between pairs of diamond foils with different thicknesses. The thinnest pair was 0.113 \& 0.116~mm, a second pair was 0.129 \& 0.150~mm, while the thickest was 0.206 \& 0.206~mm.
A direct correlation was observed between the thickness of the diamond backing and the amount of charge a target would take before melting.
Based on this experience PREX-2 employed 0.255-mm thick diamond foils. The contents of the production ladder are listed in Table~\ref{tab:ColdProductionTargets}. 
The two 255 $\mu$m thick chemical vapor deposition (CVD) diamond foils sandwiched the lead foil to improve thermal conductance to a 0.75-inch thick copper frame which was cooled to 14~K with cryogenic helium gas. 

\begin{figure}[tbhp]
\centering
\includegraphics[scale=0.26]{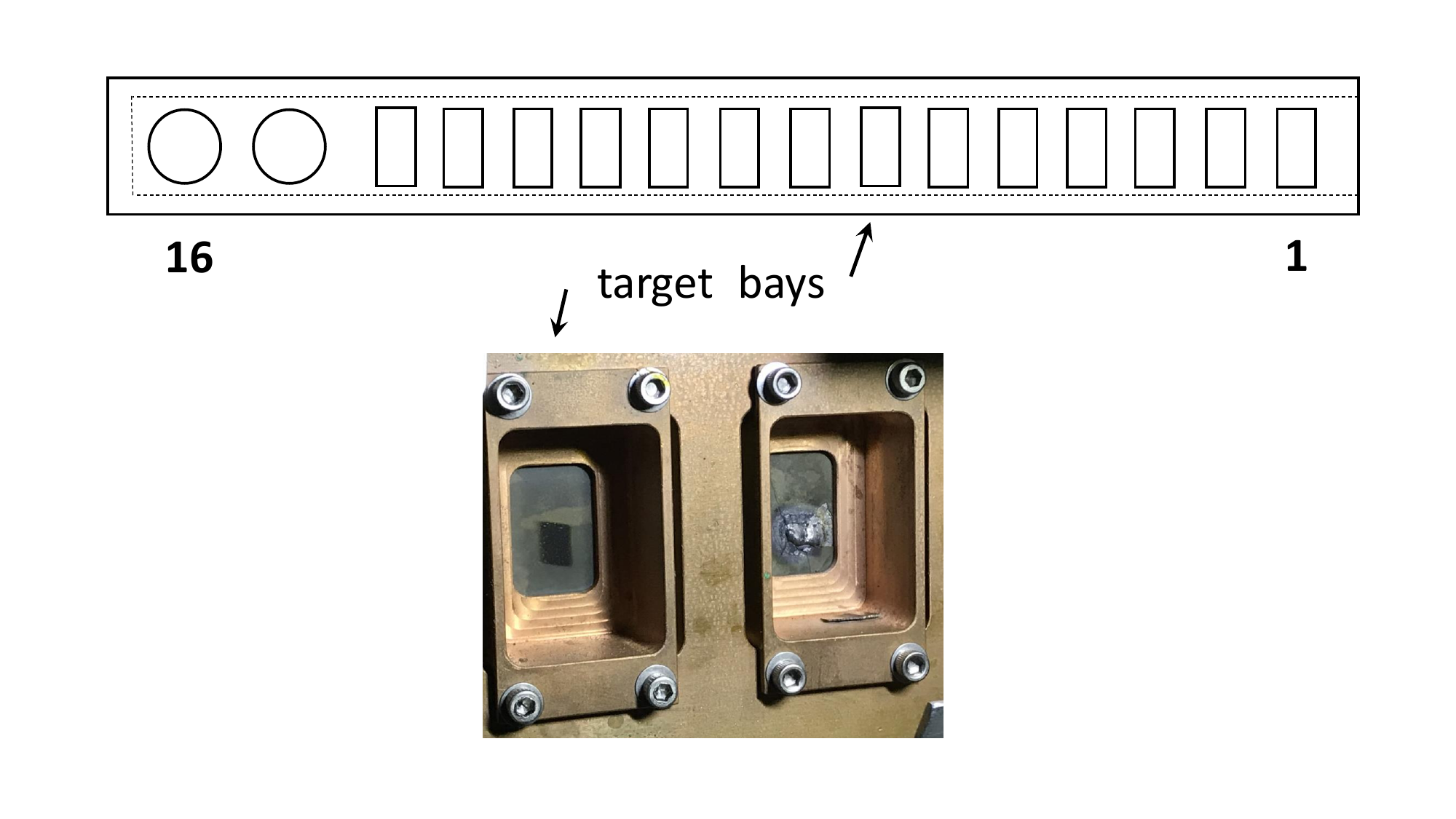}
\caption{Schematic of the Production Target Ladder (top). Shown are the 16 target bays in a 0.75 inch thick copper frame, cooled by 14~K gaseous Helium. The coolant circulates in a channel around the perimeter. Position 1, referenced in Table~\ref{tab:ColdProductionTargets}, is on the right in the schematic. Photograph of two of the lead/diamond target bays (bottom).  The target on the left remained intact and shows the burn area of the rastered beam.  The target on the right was one of the targets that failed after about 1 week of running.}
\label{fig:targetLadder}
\end{figure}

Thermal calculations using the ANSYS-CFD engine Fluent showed that the bare target could withstand 10~$\mu$A of beam current. The diamond backing increased it to 100~$\mu$A as long as there was good contact between the lead and diamond.
To improve the thermal contact a thin ($\le 20~\mu$m) layer of Apiezon L vacuum grease, a pure hydrocarbon with high thermal conductivity, was applied to the lead/diamond interface. It added negligibly to the carbon background subtraction error. A silver-based paste compound used for heat-sinking in the semi -conductor industry was applied between the diamond and the copper around the perimeter of the foil, out of the central area where the beam 
impacted the target. Good contact was maintained by pressing the target stack together using size \#4-40 Belleville washers torqued by 0.16 turns or 0.10 mm
of compression to apply 50 N
of force on each of 4 screws (see photo in Fig.~\ref{fig:targetLadder}).  

During PREX-2 we observed that after about 1 week of running the detected rates dropped quickly indicating that the lead/diamond sandwich could no longer conduct away the beam heating.
This material failure is likely due to a degradation of the diamond crystal structure caused by the electron beam and the consequent reduction in its thermal conductivity. At a critical conductivity value the lead melts, resulting in catastrophic loss of target thickness. Different structures of carbon, ranging from diamond to graphite, have  orders-of-magnitude different thermal conductivities. 

We deployed ten isotopically enriched lead targets on our target ladder (see Fig.~\ref{fig:targetLadder} and Table~\ref{tab:ColdProductionTargets}). One of the $^{208}$Pb targets was prepared with a graphite backing because graphite is known to be very robust at high beam currents, though it has a lower thermal conductivity than diamond, allowing only 60~$\mu$A. Furthermore, two natural lead targets, one with diamond backing and one with graphite backing were deployed as backups. None of the backup targets were used since only five $^{208}$Pb targets failed. 
Throughout their lifetime the targets developed density non-uniformities, and so the precision raster synchronization was critical (see Sec.~\ref{subsection:raster}). 
Low-current calibration data, triggered on individual scattered electrons, were regularly collected to evaluate the thickness of the lead relative to the diamond. 

To correct for the background from scattering from the diamond foils, one must know both its asymmetry and the relative fraction of events that scattered from the diamond. $^{12}$C is an isoscalar spin-0 nucleus whose parity-violating asymmetry is measured \cite{Souder:1990ia}. The relative scattering rates from the $^{208}$Pb and the diamond foils depend on their thicknesses. 
The $^{208}$Pb foil was $0.550 \pm 0.028$~mm thick. This was inferred from its measured mass ($m$) and area ($A$), and known density $\rho$ using $\rho t = \frac{m}{A}$. 
The conservative 5\% uncertainty on this was estimated from the variation in the measured thicknesses of multiple ``identical'' foils obtained from the same source (the largest variation was 3.8\%) and the variation of thickness in a single foil observed in spot checks with a microscope. The diamond foil thickness of $0.255 \pm 0.013$ mm was measured directly using calipers, corresponding to a 5\% uncertainty.
Foil to foil variations in measured thickness were within this uncertainty (the largest being 3.6\%). With these thicknesses, the $^{12}$C background fraction was estimated to be $6.3 \pm 0.5\%$. The correction to the asymmetry is detailed in Sec.~\ref{sec:DiamondFoilBackground}. 

\begin{table}[htbp]
\caption{Listing of the production targets (see Fig.~\ref{fig:targetLadder}). Shown are the material, the thickness of the material, the type of backing, if any, and the total thickness of the two foils in the backing. ``Natural Pb'' means naturally-occurring 99.9\% chemically pure lead.  The thicknesses are nominal. }
\label{tab:ColdProductionTargets}
\begin{ruledtabular}
\begin{tabular}{c|cccc}
Position
& Material & Thickness & Backing & Tot. Back.  \\ 
  &  & mg/cm$^2$ & & mg/cm$^2$ \\
\hline
1    & Natural Pb    & 556   & Graphite  & 176    \\ 
2    & Natural Pb     & 556   & Diamond  & 180    \\ 
3    & $^{208}$Pb  & 630   & Graphite  & 176    \\ 
4    & Graphite  & 445   & None  & N/A    \\ 
5 -- 13 & $^{208}$Pb  & 630   & Diamond  & 180    \\ 
14    & Carbon Hole & N/A   & N/A  & N/A   \\ 
15 ($1^{\rm st}$)  & $^{48}$Ca   & 1016   & None  & N/A    \\ 
15 ($2^{\rm nd}$)   & $^{48}$Ca   & 992   & None  & N/A    \\ 
16 & $^{40}$Ca & 1004   & None  & N/A    \\ 
\end{tabular}
\end{ruledtabular}
\end{table}

\subsubsection{Calcium Targets}
\label{sec:calcium_target}
Two $\approx$ 1 g/${\rm cm}^2$ ($\approx$ 6 mm thick) isotopically enriched $^{48}$Ca targets were used for the CREX experiment. They occupied position 15 in the production ladder. A thick $^{40}$Ca target was also deployed for background and transverse asymmetry studies. The targets were each mounted in a hole in the cold copper frame (see Fig.~\ref{fig:targetLadder}). Since the frame was cooled with gaseous helium to 14~K and calcium is a good thermal conductor, the target did not require a backing and was projected to withstand beam heating at up to 150 $\mu$A beam current with the raster on. The calcium targets were in the form of a puck and installed on cold heat sinks in cylindrical sockets, secured with top-hat style flanges with a 1 cm bore. Ten \#4-40 screws torqued to 0.28 N-m for a total clamping pressure of 79 MPa secured the flange to the heat sink.

The $^{48}$Ca targets were 99.9\% chemically pure and the only important isotopic contamination was several percent of $^{40}$Ca. The first (second) target was 96.1\% (91.7\%) $^{48}$Ca and 3.84\% (7.96\%) $^{40}$Ca. We obtained 7.8\% of our data with the first puck and the other 92.2\% with the second, stacked calcium target. The first target was lost due to a beam mis-steer event that resulted in the melting of the $^{48}$Ca puck. The second target was assembled from three calcium disks in a stack with a total thickness that was 2.4\% thinner than the first puck.

\subsubsection{Calibration Targets Ladder}
\label{sec:OpticsLadder}

An additional set of targets was mounted on a separate target ladder, the Calibration Targets Ladder, for both experiments (see Table~\ref{tab:OpticsTargets}). This target ladder included a thin lead target for studies of the inelastic levels, a thin tungsten target, and a graphite carbon target for spectrometer optics calibrations.  As was done on the production target ladder, we deployed a ``carbon-hole target'' consisting of a 0.1 g/${\rm cm}^2$ carbon foil with a 2 mm hole in the middle for positioning the ladder and checking the size of the rastered beam. Finally, we had a water target (5~mm thickness of water held in a container with 0.05~mm thick stainless steel walls) for angle calibration for the HRSs (see Sec.~\ref{sec:PointingMethod}).

\begin{table}[htbp]
\caption{Calibration Target Ladder.
Shown are the material and the thickness of the material. ``Natural Pb'' means naturally occurring 99.9\% chemically pure lead.}
\label{tab:OpticsTargets}
\begin{ruledtabular}
\begin{tabular}{@{\hspace{3em}} cc @{\hspace{3em}}}
Material & Thickness  \\ 
  &  mg/cm$^2$  \\
\hline
Natural Pb   & $61.2 \pm 0.5 $  \\ 
Tungsten  & $17.5 \pm 0.2 $    \\ 
Graphite  & $83.3 \pm 0.3 $    \\ 
Carbon Hole  & N/A    \\ 
Water   & $1080 \pm 20 $    \\ 
\end{tabular}
\end{ruledtabular}
\end{table}

\subsection{Spectrometers}
\label{sec:spect}

The High Resolution Spectrometers (HRSs)~\cite{HallA_NIM}, each a pair of quadrupoles followed by a 6.6 m long dipole and a third quadrupole, were used
to spatially isolate elastically scattered electrons. Additionally, septum magnets were used between the target and the HRS arms in order to reach a smaller scattering angle ($\approx 5^\circ$). The ``hardware momentum resolution'' (the width of the distribution for mono-energetic electrons with no event-by-event corrections) of the spectrometer system is better than $10^{-3}$.

The spectrometer systems served the following purposes: 
(1) select scattered electrons from the target with a narrow momentum bite 
and implicitly remove electrons that scattered from other sources;
(2) separate the elastically scattered electrons from inelastic background, hence enabling the flux integration method;  
(3) focus the elastically scattered electrons into a small spot ($2 \times 2$ cm$^2$) where we placed the quartz detectors (see Sec.~\ref{sec:detectors}); 
(4) perform measurements of the momentum transfer distribution.

\begin{figure}[t]
    \centering
    \includegraphics[scale=0.20]{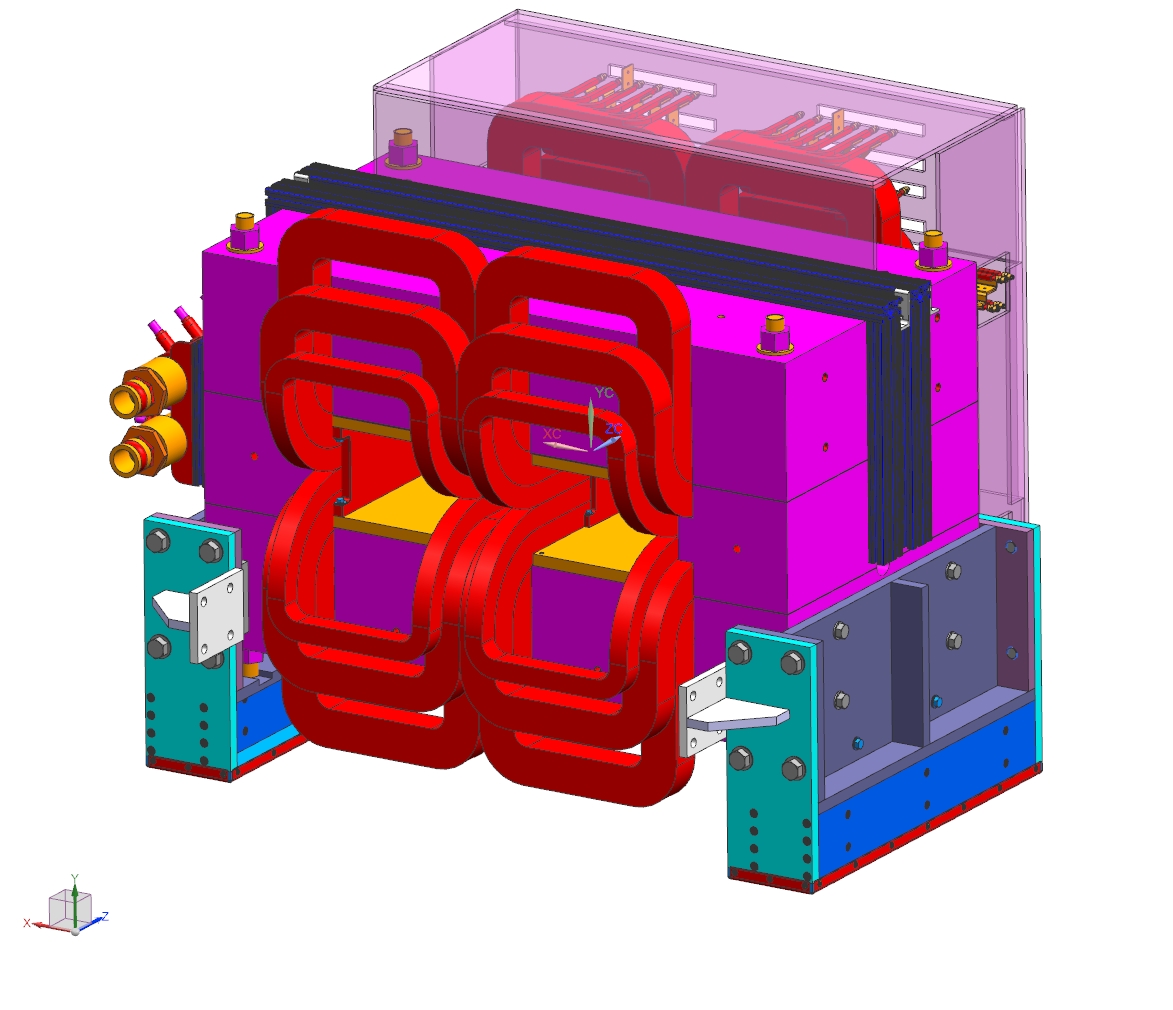}
    \caption{CAD drawing of the septum magnet system, including magnet coils (red), shielding blocks (magenta) and low-conductivity water cooling system pipes.}
    \label{fig:SeptumMagnet}
\end{figure}

The lowest scattering angle of each HRS is 12.5$^\circ$, thus requiring a pre-bend angle of about 8$^\circ$ to achieve the optimized figure-of-merit for both PREX-2 and CREX via the two septum magnets (see Fig.~\ref{fig:SeptumMagnet}) that are custom-made horizontally-bending room-temperature dipoles.  Magnetic field calculations were performed to ensure the septum could operate without any modifications between PREX-2 and CREX. Moreover, the design minimized  the amount of stray field on the beamline so as to suppress backgrounds and ensure clean transport of the unscattered beam to the beam dump. Design considerations also ensured no momentum resolution degradation of the HRSs while providing the necessary pre-bending.

The scattering chamber was connected to the HRSs via vacuum attachments which ran through the bore of the septum magnets on either side of the beamline  (left-L and right-R, looking downstream). At the entrance to the HRSs, precision acceptance-defining collimators ensured that the acceptance in the LHRS and RHRS was identical. 
Each collimator was made of 4.1 cm thick lead and was placed in the aperture of Q1. Each had a 18~cm vertical aperture and accepted electrons with scattering angles of between 3$^{\circ}$ and 8$^{\circ}$. The collimators in both spectrometer arms were surveyed and installed within $\pm 1$~mm to preserve left/right and up/down symmetry of the apparatus.
The scattered particles that passed through the acceptance-defining collimators were directed vertically upward to a well-shielded region where the detector systems were placed. This design achieved a solid angle acceptance of 3.7~msr and momentum acceptance of $\pm 4.5$\%, together with good position and angular resolution in the scattering plane, and an extended target acceptance. 

\subsection{Detectors}
\label{sec:detectors}
\begin{figure}[!ht]
\centering
\includegraphics[scale=0.25]{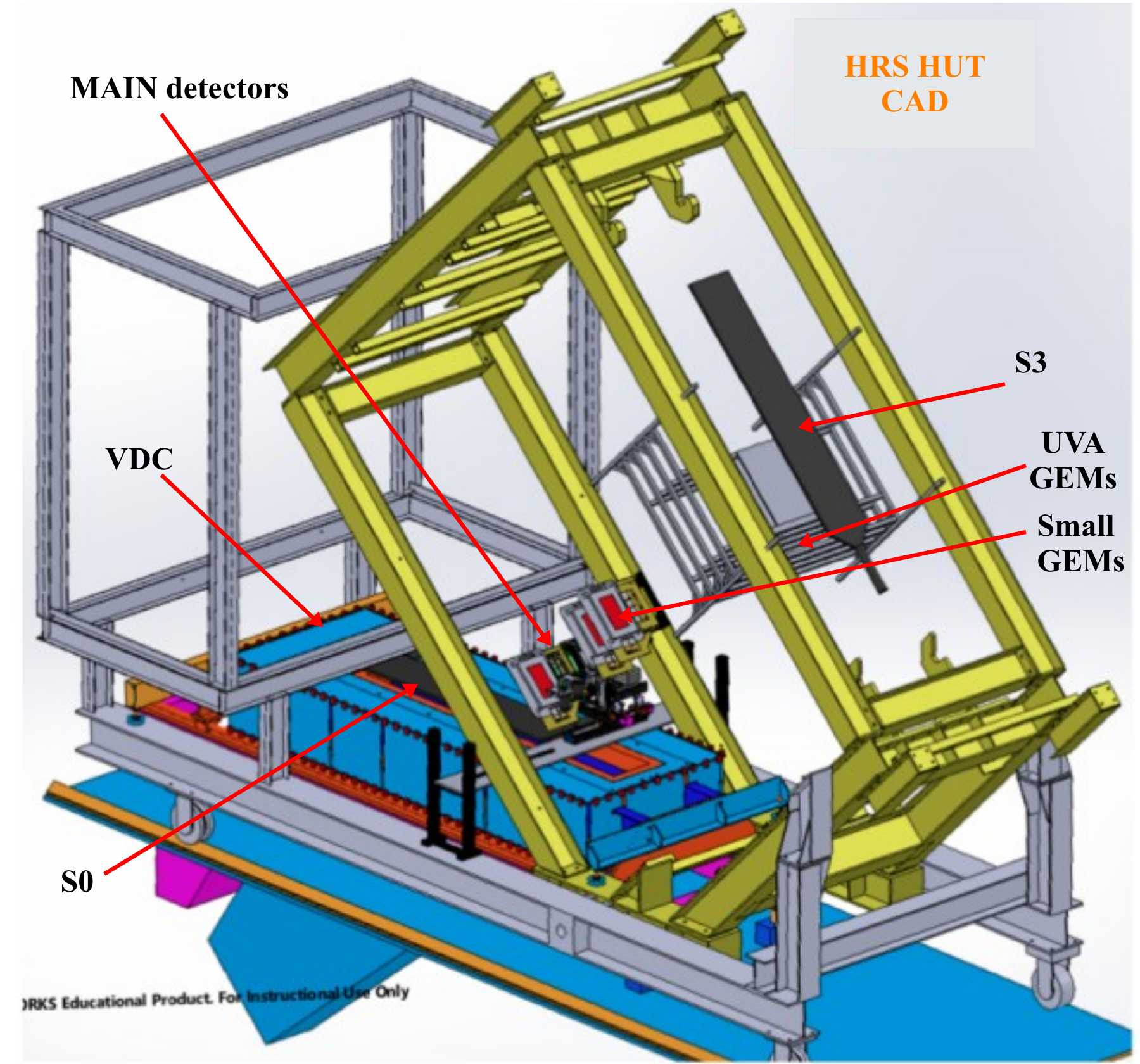}
\caption{CAD rendering of the PREX-2 and CREX detectors shown within the I-beam framing of the HRS detector hut.  Scattered electrons enter the hut from the bottom left of the figure.  The relative positions of the various detectors are shown.  The pair of $A_T$ detectors, not shown, were mounted in the vicinity of the GEM package, but on the outer edges of the beam envelope. }
\label{fig:PREXII_CREX_detectors_CAD}
\end{figure}

PREX-2  and CREX used the same detector system. In each HRS, the detector package was composed of: (1) Cherenkov detectors (both the main and the auxiliary A$_T$ detectors), (2) a Gas Electron Multiplier (GEM) chamber tracking system, and (3) a standard Hall A Vertical Drift Chamber (VDC) tracking system with scintillator paddles for triggering~\cite{HallA_NIM}. Each Cherenkov detector was made out of a fused-silica bar.
A CAD view of the major components of the detectors mounted in the HRS detector hut frame is shown in Fig.~\ref{fig:PREXII_CREX_detectors_CAD}. 

\begin{figure}[!ht]
\centering
\includegraphics[scale=0.25]{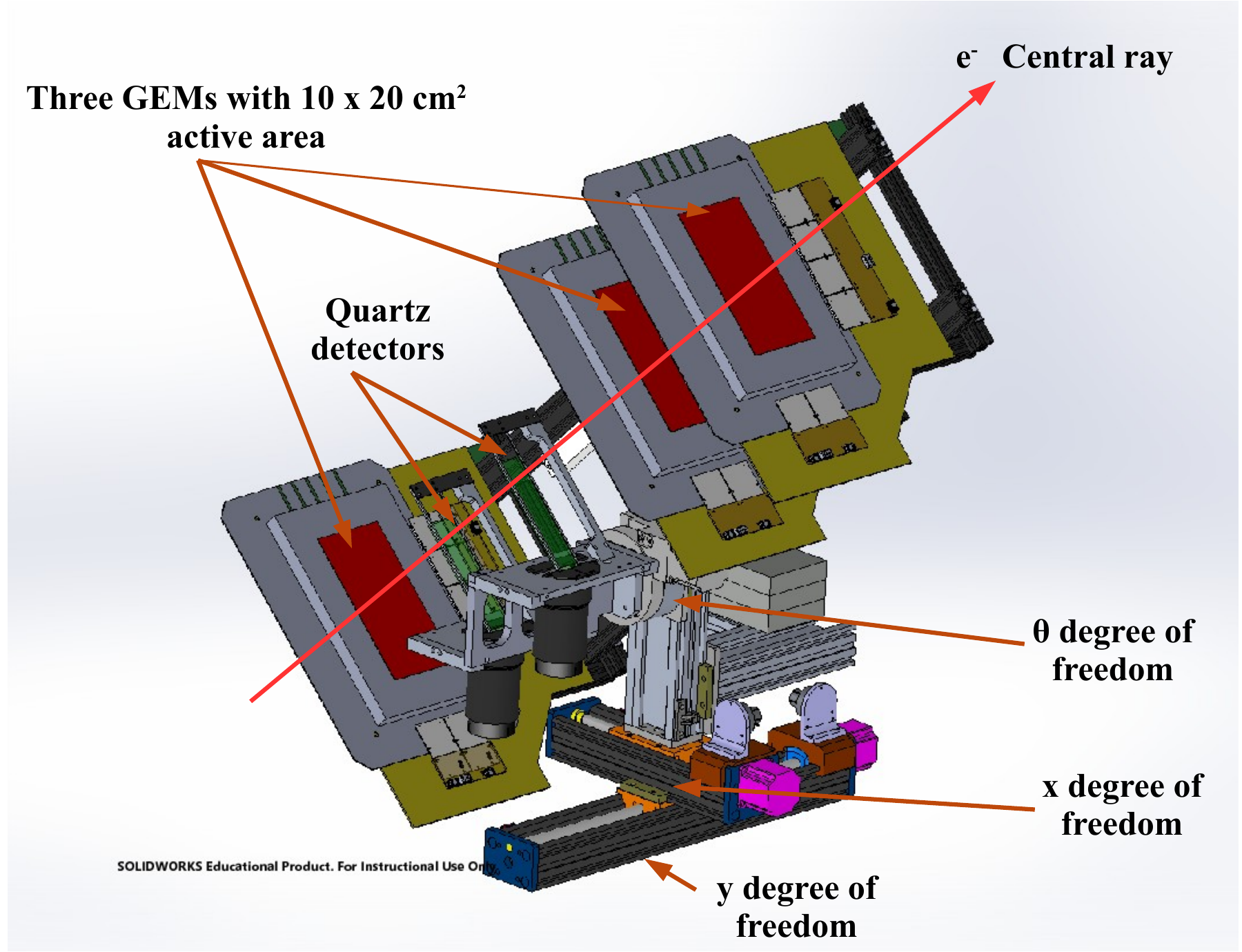}
\caption{CAD rendering of the right HRS PREX-2 and CREX focal plane detector package. The light-tight quartz covers have been removed for viewing the quartz tiles inside; the quartz is colored green for viewing clarity.}
\label{fig:PREXII_CREX_Focal_Plane_detectors_CAD}
\end{figure}

The main Cherenkov detectors were used both for the integration mode physics asymmetry measurements, as well as in counting mode together with the standard Hall A detector system (VDC and trigger scintillators) for HRS optics calibration, elastic peak-quartz alignment checks, and $Q^2$ measurements--these latter applications required precision particle tracking.  The right HRS focal plane detector system, with GEMs and motion system ($\hat{x}$, $\hat{y}$, and $\hat{\theta}$ degrees of freedom) is shown in Fig.~\ref{fig:PREXII_CREX_Focal_Plane_detectors_CAD}. The left HRS detector package was a mirror image of the right HRS package.

\subsubsection{Quartz focal plane detectors}
Each HRS used a tandem-mount main detector system and a pair of auxiliary A$_T$ detectors. They were located in the vicinity of the focal plane such that the upstream quartz detector was 1.3 m downstream of the upstream VDC plane. The detector orientation ensured approximately normal incidence of scattered
electrons on the quartz face.  The quartz size ($16~\mathrm{cm} \times 3.5~\mathrm{cm} \times 0.5~\mathrm{cm}$) was chosen just large enough to intercept all the elastically-scattered electrons and small enough to minimize the background from inelastic electrons. 

The quartz thickness was chosen to minimize the ratio of the root-mean-square to the mean $\frac{\rm RMS}{\rm Mean}$ in the photo-electron (PE)
distribution, as modeled by a Landau distribution. The distribution's tail causes excess noise beyond pure counting statistics in the measured asymmetry. The effective statistical width $\sigma_{\rm meas}$ of the asymmetry is inflated by this excess noise according to
\begin{equation}
    \sigma_{{\rm meas}} = \frac{1}{\sqrt{N}}\sqrt{1 + \left(\frac{\rm RMS}{\rm Mean}\right)^2},
\end{equation}
where $N$ is the number of electrons traversing the detector (``counting statistics'').

The upstream main detector in each HRS was used for
the main asymmetry analysis. PREX-2 used the downstream main detectors for redundancy checks while CREX used them only in counting mode for periodic checks of the alignment of the elastic on the detector. 
Each quartz tile was made of radiation-hard, optically polished, high-purity Spectrosil 2000 artificial fused-silica dry-butted directly to a 2-inch diameter window of a Hamamatsu Photomultiplier Tube, PMT, model R7723Q~\cite{HamamatsuPMT}, and covered with a 3D-printed acrylonitrile butadiene styrene (ABS) plastic case, except for the electron interception region, which was covered with 
0.0762 mm thick black polyimide Kapton film.  The PMT was also kept inside a 3D-printed plastic case which was fitted with a mu-metal shield.  The auxiliary A$_T$ detectors were used for monitoring any parity-conserving asymmetry background from residual transverse polarization of the electron beam. The  A$_T$ detectors were identical to the main detectors and were installed $\approx$~2 m downstream of the upstream VDC plane.

The main detector motion control system was primarily composed of Velmex BiSlides and rotary stages. Position transducers and optical encoders for position read-back validation, a LabJack USB ADC for the encoder readouts and a Raspberry Pi connected to ethernet allowed remote control.  The Pi used a custom Qt-based control GUI and cameras with live-streaming views of the main and auxiliary A$_T$ detectors.  The main detectors and GEMs were installed in a single extruded aluminum frame, so they could be moved together, while each A$_T$ detector was controlled independently. The position transducer signals allowed independent monitoring of all four detector positions.

Each main detector saw integrated rates of $\approx$~2.2~GHz during PREX-2 and $\approx$~28~MHz during CREX. The scattered flux rates in each A$_T$ detector was roughly one-fifth of the main detector rate. 
Prior to the experiments, several simulations and beam tests were performed at MAMI and SLAC to study the photo-electron yields from the detectors. Measurements of the PMT gain and precision non-linearity characterizations were done before and after the experiments in a bench test.

\subsubsection{Gas Electron Multipliers}
\label{sec:GEMS}
A set of Gas Electron Multiplier (GEM) detectors~\cite{GEM:1997}, capable of high-precision particle tracking at high particle fluxes, was used in  each HRS to cross-check the focal-plane track distributions obtained with the VDCs.
The pair of GEM detector sets consisted of  three large  area GEM detectors, each measuring 50 cm $\times$ 60 cm, and three smaller GEM detectors measuring 10 cm $\times$ 20 cm. All GEMs had two dimensional $xy$-style strip readout with  a pitch of 400 $\mu$m in both directions. As shown in Figs.~\ref{fig:PREXII_CREX_detectors_CAD} and \ref{fig:PREXII_CREX_Focal_Plane_detectors_CAD}, the GEM detectors were arranged parallel to each other and placed downstream of  the VDCs, first the 10 cm $\times$ 20 cm GEM detectors and then  the larger  GEM detectors. The two main detectors were placed between the first and second small GEM detectors, while the two A$_T$ detectors were situated between the small and large  GEM detector sets. 

\subsubsection{Small Angle Monitors}
A system of eight Small Angle Monitors (SAMs) were installed about 7 m downstream of the target scattering chamber. These were arranged symmetrically about the beam pipe every 45$^{\circ}$ in azimuth. The SAMs were used to monitor the large flux ($\approx 100$~GHz) of particles that were scattered at extremely small angles ($<1^{\circ}$).  Each SAM consisted of a quartz Cherenkov radiator connected via an air light guide to a 2-inch diameter Hamamatsu R375 PMT as the photosensitive device.  The SAMs were used during the experiment to monitor target-density fluctuations,
to measure the minimum noise level of the integrating-mode detector electronics, and to monitor possible beam-related false asymmetries.

\subsection{Data acquisition }
\label{sec:daq}
Two data acquisition (DAQ) systems were used for PREX-2 and CREX, one running in an integration mode and the other running in counting mode. 
Both used the CODA (CEBAF Online Data Acquisition) framework, a Jefferson Lab toolkit for building DAQ systems~\cite{CODAref}. 
Along with a custom-made trigger supervisor, different VME and FastBus DAQ crates running at different physical locations were controlled and synchronized. A local controller placed in each crate collected data from all front-end modules and buffered it before sending it via the network to the Event Builder (EB) running on a Linux workstation. The EB built events from data sent by various controllers and passed them to the Event Recorder (ER) to write to disk.  A Mass Storage tape Silo (MSS) was used for long-term storage.  
Using the Event Transfer (ET) component of CODA, data were read online for real-time analysis and intensity asymmetry feedback (Sec.~\ref{sec:QasyFeedback}). Additional user-defined data, such as slow controls information, were integrated into the events.  

\subsubsection{Integrating Mode DAQ}
\label{IntegratingDAQ}

The detector and beam monitor signals were integrated during the helicity period with a small blank-off period ($T_{\rm settle}$, see Fig.~\ref{fig:helicity_logic}) of about $90$~$\mu$s, and digitized by 18-bit sampling ADCs originally built for the Qweak experiment~\cite{Qweak:NIM}. 
The readout devices included the 18-bit ADCs, scalers, and logical input registers with the helicity information and other status bits. The integrating ADCs received the detector PMT output signals, pre-amplified using an I-to-V converter~\cite{Qweak:NIM}.  
Hardware synchronization was checked to ensure that the event fragments from the four VME crates belonged to the same helicity window. The helicity information itself
was used to check synchronization. In addition, a triangle-waveform signal injected into voltage-to-frequency converters, whose output went into a scaler in each crate, allowed a synchronization check.

\subsubsection{Counting Mode DAQ}
\label{sec:CountingDAQ} 

The counting DAQ, deploying the standard HRS DAQ used by previous experiments in Hall A~\cite{HallA_NIM}, used  
signals from two scintillators (named S0 and S3) for generating the trigger (see Fig.~\ref{fig:PREXII_CREX_detectors_CAD}).  The S0, installed just above the VDC in 
each HRS, was a $185\mathrm{~cm} \times 25\mathrm{~cm} \times 1\mathrm{~cm}$ 
plastic scintillator paddle with two PMTs, one on each 25 cm side.
Installed  $\approx$ 3 m downstream of the upstream VDC plane, S3 was a single plastic paddle ($71\mathrm{~cm} \times 9\mathrm{~cm} \times 1\mathrm{~cm}$) with a single PMT on one side along its length for PREX-2. For CREX, S3 was a logical (OR) of PMT signals from three such paddles with $\approx 1$ cm overlap along the length between adjacent paddles. 
A high-rate random pulser trigger was used for electronics tests and pedestal determination. 

The counting-mode event consisted of data from the VDCs (for particle tracking), the scintillators, the upstream (UPQ) and downstream (DNQ) main quartz detectors, the A$_T$ detectors (ATIN and ATOUT), and the GEM detectors. 

\begin{figure}[htbp]
    \centering
    \includegraphics[width=0.45\textwidth]{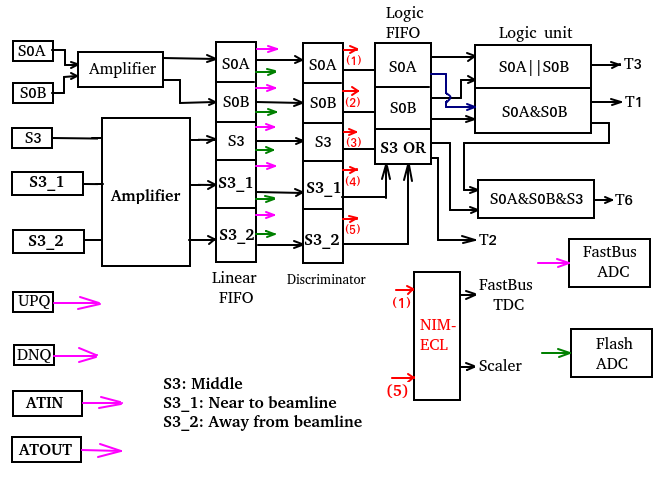}
    \caption{Circuit for CREX Counting mode DAQ for one HRS. T1, T2, T3, and T6 refer to different trigger conditions. See Sec.~\ref{sec:CountingDAQ} for details.}
    \label{crexcircuit}
\end{figure}

These data were read out in a Fastbus system with LeCroy 1881M ADCs and 1877 TDCs for the analog and digital signals, respectively. Schematics of the electronics setup for the counting mode DAQ are shown in Fig.~\ref{crexcircuit}. For both PREX and CREX, we ran two independent counting mode DAQ systems, one for each of the spectrometer arms. 
These DAQs also recorded the beam position monitor signals integrated over 4 $\mu$s, concurrent with the scintillator trigger through a Flash ADC (JLab fADC250).  We also measured the beam spot size on the target by recording the current in the raster coils with the event trigger provided by scintillators.  
Since the scattering rate is proportional to the target thickness, these data provided a snapshot of the target density and uniformity, which was important during data-collection to verify that the target was intact and to measure the ratio of diamond backing to lead material for the $^{208}$Pb target. The data were analyzed using the standard Hall A analyzer~\cite{PoddSoftwareGit} augmented by software packages for GEM-based track reconstruction. A dedicated online analysis package was developed to check detector response and the quality of the data. 

\subsubsection{Slow Controls}
\label{sec:slow controls} 
Besides the two DAQ systems, we used a slow-controls system called EPICS~\cite{EPICS} to record different experimental parameters such as collimator temperature, beam position, current and energy, detector gas flows, magnet currents, vacuum levels, high voltage values, {\it etc.} EPICS sampled data from different Input/Output Controllers (IOCs) and recorded their average values over an adjustable time window. This provided the real-time status of the system and helped us to diagnose or avoid problems.

\subsection{Simulation}
\label{sec:simulation}

Monte Carlo simulations were used in the design of the collimation and shielding before the experiments took data, as well as in the data analysis for tasks such as determining the acceptance function of the spectrometers, the size of radiative corrections to the measured asymmetries, and to determine certain backgrounds.
The optimization of the collimator design and radiation shielding design were primarily done  using {\sc geant4}~\cite{PREXG4_git}. The {\sc fluka} simulation framework~\cite{PREX_FLUKA_git} was used to estimate activation of materials inside the hall. Irradiation dose estimations were used for planning personnel access near the target and downstream areas, as well as planning for decommissioning the target chamber and collimator. 

The PREX-1 experiment faced two main radiation dose challenges. First, the high-$Z$ target and downstream beam-intercepting collimator distributed significant power into the hall. 
Secondary showers from the collimator inside the beam pipe downstream of the target caused significant damage to electronics. 
Second, a significant dose was produced at the accelerator site boundary of Jefferson Lab, measured using a set of radiation monitors placed around the accelerator complex boundary \cite{Pavel_sitedose_1996}. 
The site-boundary dose is mainly caused by upward-going, high-energy neutrons which can penetrate the hall walls and dirt overburden and shower in the air above the hall. The allowed total site-boundary dose 
is 10 mrem per calendar year (JLab internal limit) and 100 mrem/year (Dept.\ of Energy limit). 
The Jefferson Lab Radiation Control group measured a 1.34~mrem cumulative dose at the site boundary during the PREX-1 experimental period. 

Minimization of the site-boundary dose was one of the experimental design considerations, relying on detailed simulations to minimize the radiation produced 
both outside of the hall and at the locations of sensitive electronics inside the hall. The primary radiation sources for these experiments were neutron and electromagnetic 
radiation from the scattering of the $\approx$1 GeV electrons from the lead target and the $\approx$2 GeV electrons from the calcium target. Secondary sources downstream of the target were also taken into account in the simulations.

\subsubsection{Radiation Shielding in the Hall}
The primary radiation from the target was intercepted using a beamline-centered collimator downstream of the target which allowed only particles that will eventually reach the beam dump to proceed downstream. This minimized the number of secondary sources between the collimator and the beam dump.
The PREX-1 collimator only absorbed about 10 W$/\mu$A while radiating about 18 W$/\mu$A  to the hall, contributing to the high radiation field observed during data taking. The goal for the updated collimator design was to absorb more power and radiate less energy to the hall. 
The collimator was mainly made of tungsten. This provided a high-$Z$ material to absorb particles scattered from the target. Sufficient tungsten was employed to slow down neutrons and absorb most of the electromagnetic power incident on the collimator. The inner core was a tungsten-copper alloy to better conduct heat into the chilling water, while the outer layer was pure tungsten to increase the stopping power. 

\begin{figure}[htbp]
	\centering
	\includegraphics[width=0.15\textwidth]{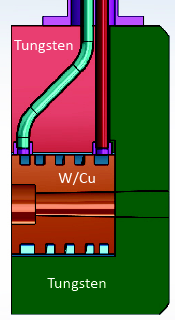}
	\caption{Side view of PREX-2 beam-intercepting primary collimator.  Beam is incident from the left. The collimator box is 16 cm long and 29 cm tall.}
	\label{fig:simulation_beam_intercepting_collimator}
\end{figure}

 The PREX-2 collimator was optimized to absorb about 27 W$/\mu$A  but to  only radiate about 3.4 W$/\mu$A . The collimator design is shown in Fig.~\ref{fig:simulation_beam_intercepting_collimator}. 

\begin{figure}[h]
    \centering
    \includegraphics[width=0.45\textwidth]{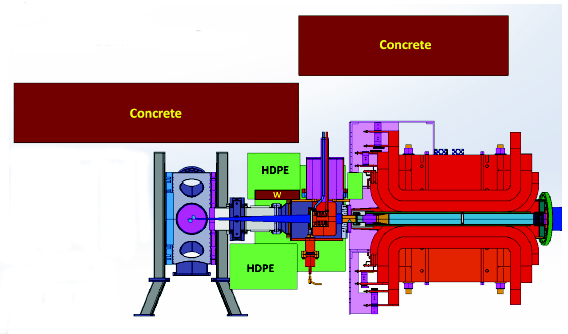}

    \includegraphics[width=0.45\textwidth]{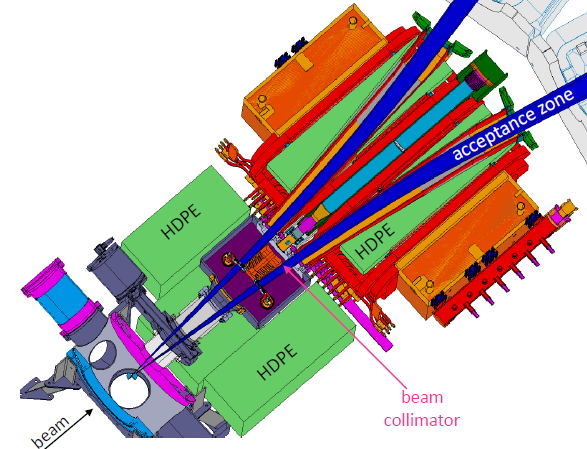}
    \caption{Shielding downstream of the target scattering chamber. The target and beam collimator were the primary radiation sources. Concrete, tungsten, and high density polyethylene (HDPE) were used to minimize the hall and site boundary radiation dose. Side and top views of the shielding are shown. To set the scale, note that the septum magnet, which is located just after the beam collimator, is about 1 meter long along the beamline.}
    \label{fig:simulation_shielding_volumes}
\end{figure}

Shielding was installed around the collimator to minimize the radiation field inside the hall. The shielding optimization was done above the target and around the collimator area. High-$Z$ materials were used to moderate high energy neutrons, high-density polyethylene (HDPE) was used to further moderate neutrons down towards thermal energies, and finally concrete was used to stop electromagnetic radiation and to locally shield electronics. The shielding design is shown in Fig.~\ref{fig:simulation_shielding_volumes}. 

The improvement made for PREX-2 shielding using simulations is apparent when the neutron-equivalent dose at a given location is compared for CREX, PREX-2, PREX-1, and a previous high luminosity, high neutron dose experiment in Hall A (PVDIS~\cite{PVDIS:2014cmd}) as shown in Table~\ref{tab:simulation_equivalent_dose}. 

\begin{table}[htbp]
\caption{MeV neutron-equivalent dose in $10^9/{\rm cm}^2$ at the location of the power supplies for the HRSs, for different Hall A experiments, estimated using {\sc Geant4} simulation.}
\label{tab:simulation_equivalent_dose}
\begin{ruledtabular}
\begin{tabular}{ccccc}
HRS Power Supplies & PREX-2 & CREX & PVDIS & PREX-1 \\ \hline
Electron           & 14      & 21   & 16  & 125    \\ 
Neutron            & 8       & 15   & 10  & 105    \\ 
Total              & 21      & 36   & 26  & 230    \\ 
\end{tabular}
\end{ruledtabular}
\end{table}

\subsubsection{Site Boundary Dose}
The dose at the site boundary was minimized using concrete and HDPE shielding above the target chamber and collimator. 
More than 15 different possible configurations were studied to optimize the shielding design. The final design settled on placing two concrete blocks, each 40 cm thick, covering a 1.2 m $\times$ 3.3 m area above the target and the collimator, and placing HDPE shielding around the beam-line between the target chamber and the collimator as shown in Fig.~\ref{fig:simulation_shielding_volumes}.

Site boundary dose estimates were obtained from a combination of simulations and actual measurements for both PREX-2 and CREX experiments. During PREX-1 the site boundary dose of 1.34 mrem arose from 82 Coulombs (C) of electrons incident on lead targets. PREX-2 required about 170 C on the Pb target and CREX about 470 C on the calcium target. Simulations of previous experimental setups with significant boundary dose measurements, such as PREX-1,  provided estimates of the high-energy neutron rates that reached the roof of the hall. Scaling the total flux of high-energy neutrons during PREX-2 and CREX to these previous experiments (taking into account the different integrated luminosities) allowed a prediction of the boundary doses during these new experiments. This approach estimated a 0.9 mrem site boundary dose for PREX-2 and 1.1 mrem for CREX. The JLab Radiation Control group measured the boundary dose during PREX-2(CREX) and obtained a total dose of 1.24(1.30) mrem, only moderately larger than the estimates from simulation.

\subsubsection{Reliability of {\sc geant4} Simulation}
A few important studies were performed to verify the reliability of the {\sc geant4} simulation. The PREX-2 simulation was compared with the standard simulation from the JLab Radiation Control group for all boundary dose calculations at JLab, which is based on {\sc geant3}. A comparison of flux at different scattering angles for the lead target showed that {\sc geant4} and {\sc geant3} simulations agreed and produce more high-energy particles as compared to {\sc fluka}, but all the results agree to within a factor of two.  A comparison of the predicted neutron flux from a tungsten target using {\sc mcnpx} (a well-established neutron simulation code) \cite{mcnpx} showed that {\sc geant4} may slightly overestimate the production of high-energy neutrons.  Finally, the site boundary dose estimation from {\sc geant4} simulation was compared with predictions of {\sc fluka}, and with radiation monitor readings obtained during PREX-1. These benchmarks and comparison studies proved the reliability of the {\sc geant4} simulation utilized for PREX-2.

\subsection{Polarimetry } \label{sec:polarimetry}

Our goal for beam polarimetry was to determine the electron beam polarization to a relative precision of 1\% or better. Two independent methods were used to measure the polarization at the target, one based on M{\o}ller scattering and the other on Compton scattering, thus providing a redundant check on the precision. We also made use of a Mott polarimeter to measure the electron beam polarization at the source, and to adjust the amount of transverse polarization.

\subsubsection{M{\o}ller Polarimeter}

The Hall A M{\o}ller polarimeter makes use of the elastic scattering of polarized electrons from a polarized electron target, that is $$\vec{e}^{\,-}_{\rm beam}+\vec{e}^{\,-}_{\rm target}\to e^- + e^- .$$ In practice, we determine the beam polarization $P_{\rm beam}$ by measuring the scattering asymmetry $A_{\rm moller}$ for left- and right-handed polarized electrons which is given by $$A_{\rm moller}=P_{\rm beam}P_{\rm target}\langle A_{zz}\rangle  \; .$$ Here, $P_{\rm target}$ is the target polarization and $\langle A_{zz}\rangle$ is the analyzing power, averaged over the polarimeter acceptance.

For scattering through center-of-mass angles near $\theta_{\rm cm}=90^\circ$ the analyzing power is both precisely calculable and large. Therefore, knowing $P_{\rm target}$ lets us determine the incident beam polarization by detecting the scattered electrons in a spectrometer. Of course, in order to reach 1\% precision, the systematic uncertainties associated with the target polarization and spectrometer analyzing power must be aggressively minimized, in addition to needing to reach sufficient statistical precision.

\begin{figure}[tbp]
	\centering
	\includegraphics[width=0.45\textwidth]{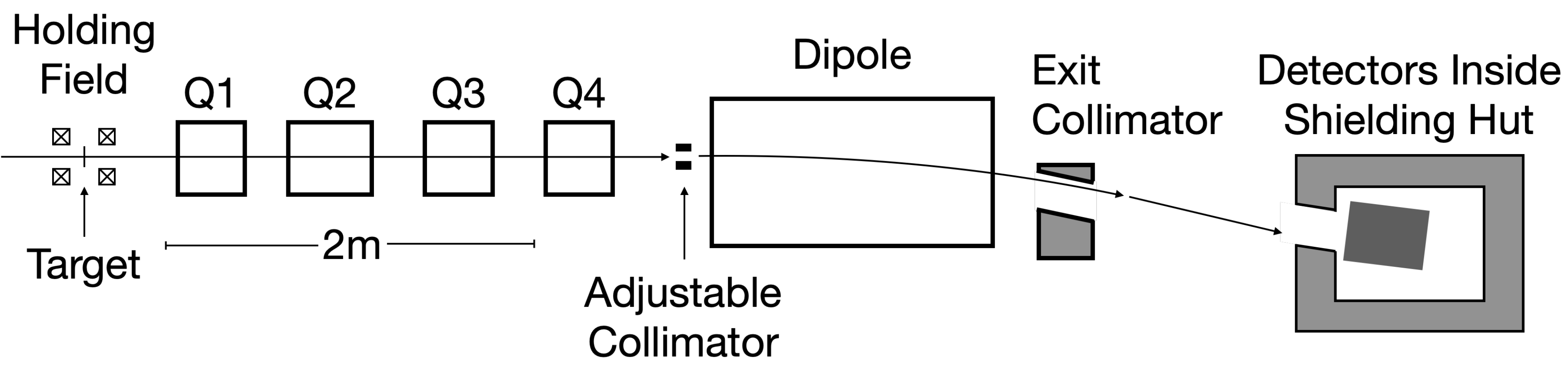}
	\caption{A schematic in elevation view of the M{\o}ller polarimeter. The scattered electron pair near $90^\circ$ in the center of mass are focused into a 
    momentum-analyzing dipole magnet and then detected in coincidence by shower counters in a shielded hut.}
	\label{fig:MollerCartoon}
\end{figure}

See~\cite{NIMA2023167506} for a complete description of the M{\o}ller polarimeter. Figure~\ref{fig:MollerCartoon}
shows a schematic of the spectrometer and detector system. A set of quadrupole magnets focused the scattered electron pair into parallel rays that entered the dipole magnet for momentum selection. The electrons then entered a shielded hut where they were detected by a set of shower counters. We required a coincidence between the two electrons. Scattering angles $\theta_{\rm cm}$ were restricted to within $\pm20^\circ$ of $90^\circ$ by collimators and the detector acceptance. The spectrometer acceptance and effective analyzing power were determined using a {\sc geant4} Monte Carlo simulation which was tested against data.

The polarized electron target was provided by a pure iron foil saturated by a 4~T magnetic field perpendicular to the plane of the foil and parallel to the electron beam~\cite{Debever:1997pdg}. We conducted a literature study~\cite{NIMA2022167444} to determine the spin polarization of saturated pure iron and find that we know the target polarization to $0.25\%$.

\subsubsection{Compton Polarimeter}

\begin{figure}[htbp]
	\centering
	\includegraphics[width=0.45\textwidth]{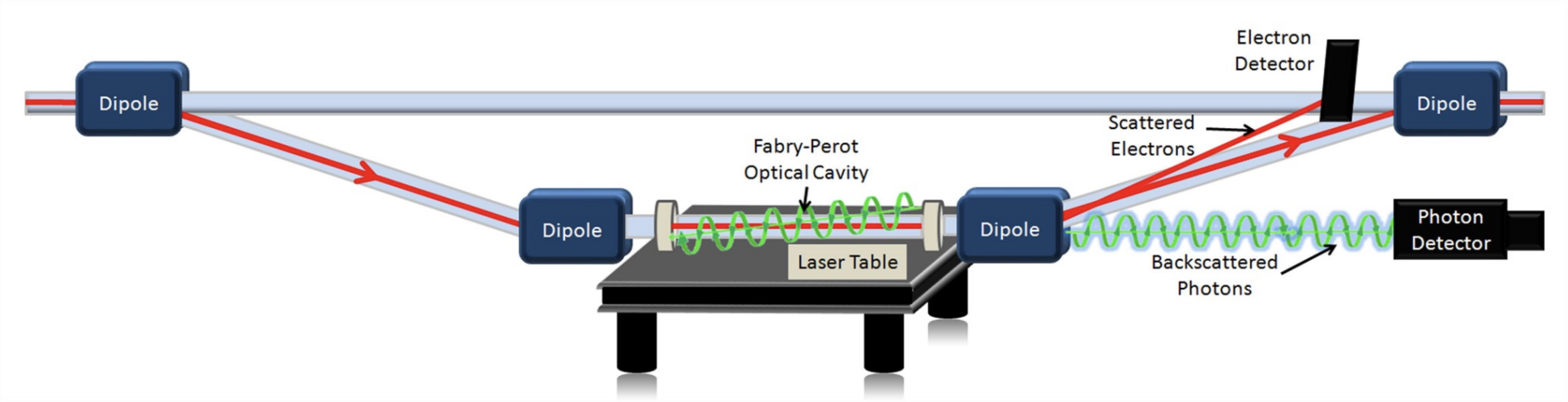}
	\caption{A cartoon view of the Compton Polarimeter. A dipole chicane passes the primary beam through a beam of laser light in an optical cavity and measures the back-scattered Compton photons to measure the electron beam polarization.}
	\label{fig:ComptonCartoon}
\end{figure}

Hall A employs a Compton-scattering polarimeter which allows continuous monitoring of the beam polarization and is able to achieve 0.5\% level uncertainties~\cite{Zec:2024iky}. The Compton polarimeter consists of a 4-dipole magnetic chicane which can divert the primary electron beam to scatter off the high-power green (532~nm) laser light stored in a Fabry-P\'erot optical cavity.  The back-scattered Compton photons were detected in a GSO scintillating photon detector, while the primary beam continued on towards the main experiment target.
The polarization measurement was done by measuring the helicity-dependent Compton scattering asymmetry and comparing it with the theoretical asymmetry for a completely polarized beam.  The acceptance of the detector was modeled using a {\sc geant4} simulation. 

\subsubsection{Mott Polarimeter}

The beam polarization was periodically measured in the accelerator's injector using the 5 MeV Mott polarimeter~\cite{Grames:2020asy}. 
The Mott polarimeter is sensitive only to transverse polarization and can simultaneously measure both the vertical and horizontal transverse components of beam polarization. The Mott polarimeter is used to set and measure the vertical polarization during transverse asymmetry measurements in the Hall. 
The beam transport from the injector to Hall A is symmetric and nearly flat, and thus preserves the vertical component of the beam polarization from the injector to Hall A. 

When longitudinal polarization is required for an experiment, the Mott polarimeter is then used to zero the beam's vertical polarization component. By adjusting the spin rotation from the two spin solenoids (in the double Wien spin flipper, see Fig.~\ref{fig:Double_Wien}) the vertical component is reduced to zero. Then, the initial horizontal angle of beam polarization is set by the Horizontal Wien. The angle of the Wien is determined from the accelerator linac energies and the total beam energy delivered to the hall. Finally, since the knowledge of the initial Wien angle is limited by the uncertainty on the beam energy, a set of Moller polarization measurements was performed by varying the Wien angle at the injector to determine the optimal horizontal Wien angle to deliver fully longitudinal beam at the Hall target.

%% file: z3_analysisV2.tex
\section{ANALYSIS}
\label{sec:analysis}

The raw asymmetry $A_{\rm raw}$ is computed from the detector flux $F = D/I$ where $D$ is the detector signal integrated over a fixed period of time and $I$ is the beam current integrated over the same period.  
For an individual asymmetry measurement between matched periods of right ($R$) and left ($L$) handed helicity, the asymmetry is defined as
\begin{equation}
    \label{eq:Araw}
A_{\rm raw} = \frac{F_R - F_L}{F_R + F_L}.
\end{equation}

The raw asymmetry is corrected for beam fluctuations ($A_{\rm beam}$) to form the corrected asymmetry $A_{C}$. 
\begin{equation}
\label{eq:asyCorr}
    A_{\rm C} = A_{\rm raw} - A_{\rm beam} 
\end{equation}
Additional corrections for detector non-linearity $A_{\rm NL}$, transverse asymmetry $A_{\rm T}$, electron beam polarization $P_e$, and signal fractions $f_i$ and asymmetries $A_i$ for several backgrounds, are applied as averages over the full data set to arrive at the 
measured parity-violating asymmetry
\begin{equation}  
    A^{\rm meas}_{\rm PV} = \frac{1}{P_e}\frac{A_{\rm C}-A_{NL} - A_T -A_{BL} - P_e\sum\limits_i A_i f_i}{1-\sum\limits_i f_i}    
\label{eq:asyCorrected}
\end{equation}
Here, the blinding term $A_{\rm BL}$ (see Sec.~\ref{sect:blinding}) is also removed in this calculation, as a final step in the analysis.  

In the remainder of this section we detail the various selection criteria and corrections that lead to our determination of $ A^{\rm meas}_{\rm PV}$ for $^{208}$Pb and $^{48}$Ca.

\subsection{Event Selection and Run Conditions}
\label{sec:event_selection}

At the initial stage of the asymmetry analysis, a set of 
event-selection cuts were made in order to 
select data with stable 
conditions, thereby reducing the effect of noise and instability in the various detector responses. To avoid introducing bias to the asymmetry measurement, cuts were only applied to the data at the helicity event level, i.e., the 120~Hz or 240~Hz samples. 
Requirements were set on the stability of the data 
based on the mean and the root-mean-square (RMS) of the data in a moving time window, and therefore did not consider the
helicity information. The length of the moving time window was typically $\approx 6$ seconds, much longer than that of a helicity pattern ($\approx 33$~ms). Therefore, data at the helicity pattern level are free from potential effects implicitly imposed by cuts at the event level. 

The event cuts in the asymmetry analysis software evaluated data quality based on the following aspects simultaneously:
\begin{itemize}
\item{Detection of failures in the ADCs, e.g., saturation, null, or frozen read-back.}
\item{Validation of the event timing sequence and the reported helicity information.}
\item{Rejection of detector signal sizes outside of customized lower and upper limits.}
\item{Requirement of stable beam parameters, e.g., beam current and position, within
a sliding time window, in order to reject beam trips and the time period shortly after a trip when the instrumentation is unstable.}
\item{Tagging of events taken during the beam modulation period.}
\end{itemize}

Data during PREX-2 were taken at 3 different beam currents (50~$\mu$A, 70~$\mu$A, 85~$\mu$A), 2 different frequencies (120~Hz, 240~Hz), while cycling through 7 different lead targets. Data during CREX was divided into 3 run periods: Wien Right Spring 2020, Wien Left Spring 2020, and Wien Right Summer 2020. During both experiments, a close watch was kept on the on-line data stream for healthy running conditions. The prompt analysis flagged problems during running. At the end of PREX-2 (CREX), we had collected 114 (383) Coulombs of beam on target which passed the event cuts.  Analysis averaging was performed over different subsets or timescales of the experiment.  A ``minirun'' is a collection of $\approx$~9000 multiplets, which contains enough statistics to be able to examine trends in the data. The minirun is the timescale over which regression analysis is performed (Sec.~\ref{sec:beamparsens}).  Finally, a ``slug'' is a collection of miniruns over which the IHWP condition (IN or OUT) was the same, corresponding to approximately 8 hours of continuous running.

\subsection{Calculation of Raw Asymmetries}
\label{sec:rawasym}
The data used to compute the asymmetry had to pass requirements on data quality (see Sec.~\ref{sec:event_selection}) but no helicity-dependent cuts were applied. 

Data was collected with quartet patterns at 120~Hz, designed to optimize cancellation of 60~Hz variations and slow drifts, with the pattern (RLLR) or (LRRL). In addition to quadruplets at 120 Hz, the other patterns used were octets at 240 Hz and pairs at 60 Hz. The octet helicity pattern was formed from two subsequent sets of opposite quartet patterns. In each case, the sums of flux measurements in right-handed windows ($F_R$) and left-handed windows ($F_L$) were used to calculate $A_{raw}$ (Eq.~\ref{eq:Araw}).

\subsection{Pedestals and Linearity}
\label{sec:pednonlin}
The PMTs for the main detectors, the $A_T$ detectors, and the SAMs have dark currents associated with them, as do the various beam monitors. 

These dark currents, together with the ADC design and pre-amplifier settings, define the pedestal (or zero signal baseline) of the detector or monitor. The pedestals were subtracted from the signals acquired during production data taking. The pedestal is 
independent of the electron beam helicity state.
If $F_R$ ($F_L$) is the flux detected in a detector for right (left) helicity states, and $S_{\rm ped}$ is the pedestal signal, the detector asymmetry is given by
\begin{equation}
A_{\rm det} =   \frac{F_R-F_L}{F_R+F_L+2S_{\rm ped}} .
\label{eq:Adet}
\end{equation}
Note that the $2S_{\rm ped}$ term affects the physics asymmetry resulting in a need to determine the detector and beam monitor pedestals regularly. A new pedestal calibration was performed whenever there was any change in detector configuration, for example, a high voltage change or detector alignment change. 

The first step in the pedestal calibration procedure was to calibrate the Unser~\cite{Unser:1981fh} monitor, which is a stable reference beam-current monitor. Once the Unser was calibrated, it was used to normalize one of the BCMs. The normalizing BCM, the one chosen to charge-normalize the detector signal, was then used to calibrate the other BCMs, BPMs, and detector pedestals. All pedestal calibrations were performed with dedicated runs where the beam current was increased incrementally in steps.

Detector non-linearity was an essential source of systematic uncertainty that needed to be understood and controlled. The main contribution to the detector non-linearity error was the PMT non-linearity, which is a variation in the PMT gain when the input varies in intensity. Due to the high flux in PREX-2 
the PMTs needed to be operated at a lower gain setting than for a typical PMT application. At these lower gains, the PMT signal output can behave non-linearly with respect to either the light received by the photocathode or the dynode amplification stages. For a PMT, the non-linearity is caused by a systematic imbalance between the divider supply current, which flows through the voltage divider circuit, and the amount of signal current flowing along the array of dynodes due to the continuous flow of a cascading shower of amplified photo-electrons. Other components in the electronics may also exhibit some degree of non-linearity.

The non-linearity was studied in bench tests (using two LEDs) before and after the experiments and in-situ during the experiments. From the LED bench tests, the non-linear response of each PMT was found to be less than 0.3\% at PREX-2 and CREX gain settings and cathode current levels. The in-situ non-linearity was monitored via PITA scans (see Sec.~\ref{section:PITAeffect}) and current-ramp calibration runs. The PITA scans were limited by beam corrections and could not give a precise measurement of non-linearity.  The current-ramp pedestals agreed with the bench test pedestals within 0.5~\%. 
Therefore, we did not make any non-linearity correction 
(i.e. $A_{NL}=0$ in Eq.~\ref{eq:asyCorrected}), but assigned a relative systematic uncertainty due to non-linearity equal to 0.5\% for PREX-2 and 0.3\% for CREX.

\subsection{Beam Asymmetries}
\label{sec:BeamAsymmetries}
\begin{figure}
     \centering
     \includegraphics[width=0.35\textwidth]{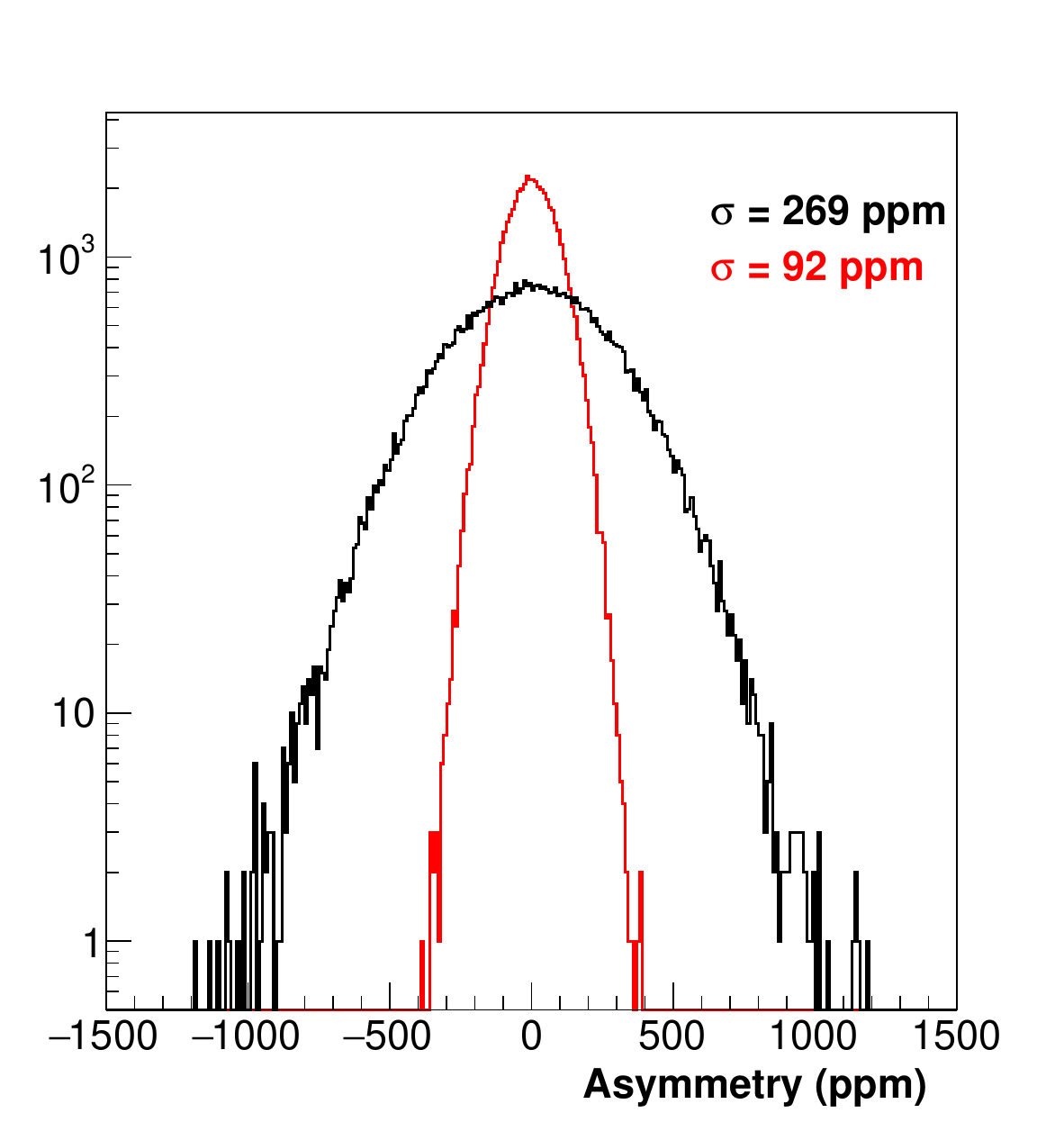}
     \caption{ Distribution of beam corrected asymmetries (red) sampled from the PREX-2 data compared with the raw detected asymmetries (black) from the exact same data set.  A similar reduction of width occurs for CREX.  The reduced $\sigma$ is close to $\frac{1}{\sqrt{N}}$ for $N$ electrons in a pair of helicities as expected from counting statistics.
     }
     \label{fig:prex_correction_histogram}
 \end{figure}

Helicity correlations in the beam positions, angles, and energies must be evaluated and subtracted from the raw asymmetries (Eq.~\ref{eq:asyCorr}).  The intensity asymmetry (Eq.~\ref{eq:intensityAsy}) was studied and minimized on the laser table (Sec.~\ref{sec:polsource}), and upper bounds on the helicity-correlated spot size variations were measured (Sec.~\ref{sec:hcSpotSize}). The beam asymmetry corrections, discussed below, accomplish three main tasks: (i) removing fluctuations in the asymmetry distribution which are caused by variations in the beam parameters, as shown in Fig.~\ref{fig:prex_correction_histogram}, (ii) removing the average beam asymmetry $A_{\rm beam}$ which appears in Eq.~\ref{eq:asyCorr} due to helicity correlations in the beam parameters, and (iii) evaluating the systematic errors due to these required corrections.

\subsubsection{Beam Corrections}
\label{sec:BeamCorrectionsTheory}

The required correction to the raw measured asymmetry depends on five beam parameters $X_{k}$, which are the two $x$ and two $y$ coordinates of the incoming beam upstream of the target and the beam energy, as well as the sensitivity $\alpha_k$ of the detected flux to each parameter:
\begin{equation}
\Delta A=\sum_{k=1}^5\alpha_k\Delta X_k
\label{eq:Ai_Xj}
\end{equation}

While conceptually useful, the beam parameters $\Delta X_k$ are not independently measured by beam monitors, so beam corrections are performed using sensitivities $\beta_j$ assigned to independent beam monitor measurements $M_j$:
\begin{equation}
\Delta A=\sum_{j=1}^N\beta_j\Delta M_j
\label{eq:MonCor}
\end{equation}

Consider a set of five beam position monitors, so that $ M_j( X_k)$ are linear functions of the five beam parameters.
In order for the beam monitors to span the full phase space of the beam envelope, the five monitors must be positioned so that the set of functions $M_j(X_k)$ are not degnerate.  This property is established by using a  pair of horizontal and vertical monitors in a long section of the beam line upstream of the target with no focusing elements, and a fifth monitor at a position where the beam has significant dispersion. 
A total of 12 monitors in the Hall A beam line were read out in these two experiments. 
The additional monitors, beyond the minimal set of 5 monitors described above, 
were included in the analysis to improve precision. 

The sensitivities $\beta_j$ required for correction are evaluated using both the intrinsic jitter in beam parameters and dedicated calibration data using purposeful modulation of the beam. While each calibration technique has been employed independently for corrections in previous PVES measurements, the analysis presented here incorporates these two techniques in a novel unified framework to optimize control of random noise without sacrificing systematic accuracy.

The importance of this analysis is highlighted in Fig.~\ref{fig:prex_correction_histogram}, comparing the distribution of raw measured asymmetry $A_{\rm raw}$ with that of the beam corrected asymmetries $A_{\rm C}$ for a typical sample of multiplets from the PREX-2 dataset.  The large reduction in random noise is crucial for achieving the precision goals of the experiment, while the final average correction requires accurate measurement to avoid introducing systematic error. 
The discussion below describes multiple analyses used to investigate the characteristics of the calibration system, beam monitors, the beam jitter, and the systematic beam asymmetries in order to demonstrate a thorough understanding of this sensitive correction.  

\subsubsection{Multivariate Regression}
\label{sec:beamparsens}
The technique of multivariate linear regression can be employed to remove the correlation between dependent variables (in this case, measured asymmetries) and independent variables (beam monitors).  In the ideal case, this might be described as a technique for using the intrinsic jitter in the beam parameters to calibrate the sensitivity of the flux measurement to the beam parameters. By construction, this technique will maximally reduce random noise in the asymmetry that is correlated with beam monitors. It also has an advantage of returning precise results with relatively small amounts of data, so it is useful for rapidly updating sensitivity measurements and tracking changes in the apparatus.  

The primary complication for reliable correction using regression arises from the limits of resolution in measurements of some components in the beam phase space.  In particular, angular variations in the trajectory are typically measured much less precisely than position variations, relative to their size.  In this case, 
sensitivities 
which maximize the suppression of correlations may fail to accurately apply the appropriate average correction, if doing so would introduce additional noise due to beam monitor resolution. For this reason, regression is used as a diagnostic tool, but is {\it a priori} considered less systematically robust than other approaches. 

Regression correction coefficients $\beta_j^r$ are obtained by minimizing the $\chi^2$
\begin{equation}
  \chi^2=\sum_i\left(A^i-\sum_j\beta^r_j\Delta M^i_j\right)^2/\sigma^2,
  \label{eq:regchisq}
\end{equation}
where $\sigma$ is approximated to be the same for each of the multiplets $i$, and $j$ are the individual beam monitors.
The superscript $r$ indicates that the coefficients are from regression analysis. A similar notation will be used to distinguish correction coefficients from other analyses. 
The $\beta^r_j$ are obtained by solving the simultaneous equations
\begin{equation}
    \frac{\partial \chi^2}{\partial \beta^r_j}=0.
\end{equation}
Therefore the beam asymmetry correction for the $i$-th multiplet is 
\begin{equation}
    \Delta A^i=\beta^r_j\Delta M^i_j,
\end{equation}
where in this and subsequent equations the repeated index $j$ implies a summation over the index. The method produces a set of $\beta^r_j$ that correct for beam-helicity correlated asymmetries and consequently minimize the width of the asymmetry. One drawback of the method is that if one of the beam monitors has significant instrumental noise, the corresponding $\beta^r_j$ magnitude will be reduced and introduce a systematic error while still minimizing the RMS width. The main contribution to the  $\Delta M^i_j$ must be the natural beam jitter, and the phase space spanned by this jitter must be the same phase space as the systematic beam differences.

A regression analysis is performed for each minirun ($\approx$ 9000 multiplets or about 5 minutes of uninterrupted data collection). Correlations are accumulated over the entire minirun, and corrections applied to each multiplet in order to enable diagnostic plots using the corrected data.  As shown in Fig.~\ref{fig:prex_correction_histogram}, a typical portion of PREX-2 showed a width of 92~ppm after removal of 253~ppm of beam noise from the $\sigma_{\rm raw}=269$~ppm raw asymmetry distribution. Since the quoted statistical error for the experiments is based on this distribution width, the beam corrections improve the statistical precision by nearly a factor of three. For CREX, the noise due to beam variations was larger, about 1500~ppm. The intrinsic statistical width was much larger, so the corrected width was 780~ppm compared to $\sigma_{\rm raw}\approx 1700$~ppm.

A total of 12 BPMs were available in the beam line leading up to the production target, which allowed 
an overdetermined analysis aimed at improving the precision of the regression correction.  The signal in a given monitor $M^i_j$ for event $i$ has contributions not only from the beam parameters $X^i_k$ but also instrumental noise $\delta^i_{jn}$, representing correlated noise with another monitor when $n \neq j$ and uncorrelated instrumental noise with $n=j$. This instrumental noise is in
either 
case not correlated to the beam parameters or the scattering asymmetry $A_i$.  Using more monitors than beam parameters in the multivariate regression reduces the effects of uncorrelated noise by averaging, and allows cancellation of correlated noise. 

\begin{figure}[!h]
    \centering
    \includegraphics[width=0.45\textwidth]{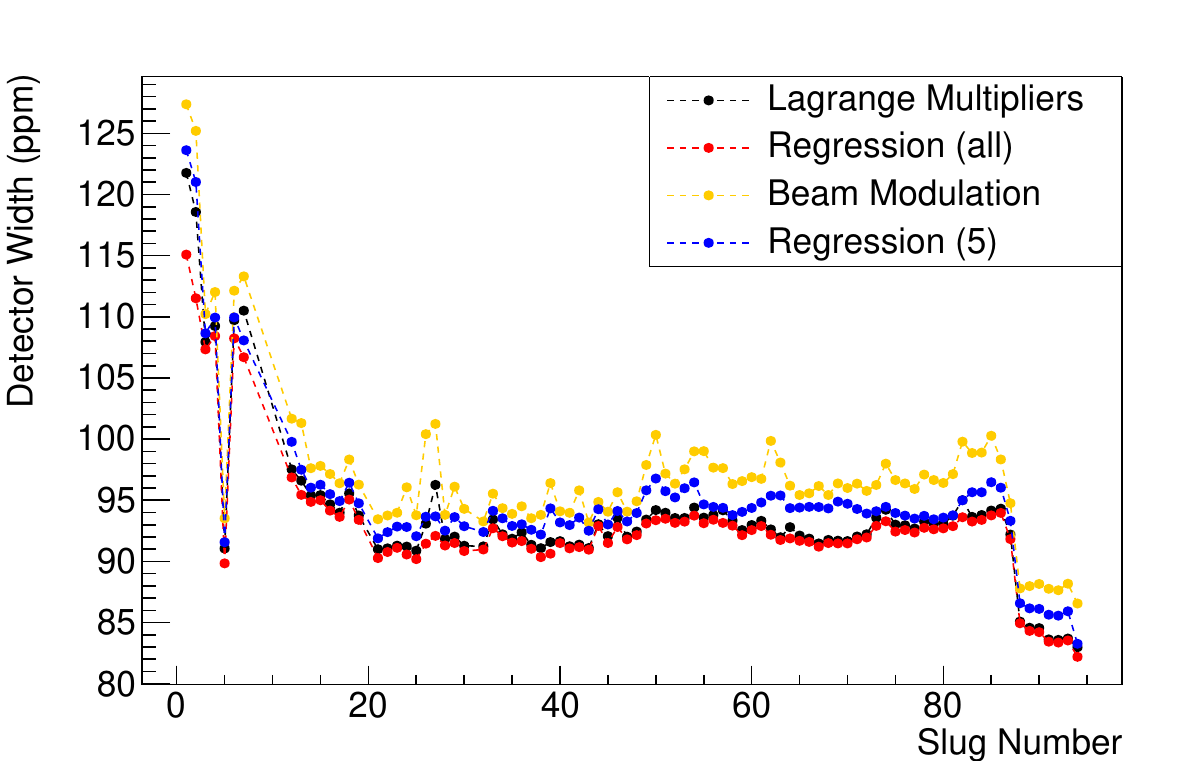}  
    \caption{History of the RMS width of corrected asymmetries for PREX-2 with different beam correction methods. 
    }
    \label{fig:RMS width of detector}
\end{figure}

As shown in Fig. \ref{fig:RMS width of detector} for PREX-2, regression using all 12 BPMs
achieved a width which was consistently a few ppm smaller than that obtained using a minimal set of five BPMs. This reduction in width suggests that beam monitor resolution was a small but non-negligible contribution of excess noise (i.e. above counting statistics) in the 5-BPM regression correction.

\subsubsection{Beam Modulation Analysis}

As described in Sec.~\ref{sec:BeamModulation}, a set of air-core coils and an RF acceleration cavity were used to modulate the beam as a technique to measure correction coefficients $\beta_j$.
The modulation coils are located within the beamline optics such that they collectively span the phase space of the beam trajectory. This requires that the responses of the five primary BPMs
to the five modulations $M_j(C_l)$ are not degenerate.  In this way, the $X_k,\ M_j,$ and $C_l$ each can serve as sets of beam parameters that span the phase space of the beam, where $M_j$ are measured, $C_l$ are known states of the beam modulation coils, and $X_k$ are the true beam kinematics.

As represented in Fig.~\ref{fig:typicalCoilResponse},  each individual coil or cavity was ramped during a beam modulation cycle, and the response of the integrating detectors $D$ and the beam position monitors $M$ were measured, yielding 
\begin{equation}
\frac{\partial \hat D}{\partial C_l},\ \ \frac{\partial M_j}{\partial C_l}.
\label{eq:ramping_data}
\end{equation}
Here $\hat D=D/\overline D$, where $D$ is any detector signal and $\overline D$ is the average signal. We use these data to compute the beam modulation coefficients $\beta^d_j$ to correct the asymmetries
\begin{equation}\Delta A^i=\beta^d_j\Delta M^i_j.
\label{eq:deltaAbeta}
\end{equation}
Here $\Delta A^i$ is the integrating detector asymmetry measured in multiplet $i$ and $\Delta M^i_j$ are the monitor differences for the same multiplet. Here the superscript ``d'' refers to the ``dithering'' analysis, which is another commonly used term for this technique. 

To determine $\beta^d_j$ we used the data from Eq.~\ref{eq:ramping_data} to determine how the detector signal depends on the beam monitors:
\begin{equation}
    \Delta \hat D=\frac{\partial \hat D}{\partial M_j}\Delta M_j=\frac{\partial \hat D}{\partial C_l} \frac{\partial C_l}{\partial M_j}\Delta M_j,
\label{eq:deltaDmatrix1}
\end{equation}
where we used matrix inversion
\begin{equation}
    \frac{\partial C_l}{\partial M_j}=\left(\frac{\partial M_j}{\partial C_l}\right)^{-1}. 
\label{eq:deltaDmatrix2}
\end{equation}
To use beam modulation data to correct asymmetries, we have
\begin{equation}
    \Delta A=\frac{D(C_l+\Delta C_l)-D(C_l-\Delta C_l)}{2\overline D}=\frac{1}{\overline D}\frac{\partial D}{\partial C_l}\Delta C_l,
\end{equation}
and therefore
\begin{equation}
    \frac{\partial A}{\partial C_l}=\frac{\partial \hat D}{\partial C_l},
\label{eq:AD_relation}
\end{equation}
with 
\begin{equation}
    \Delta A^i=\frac{\partial A}{\partial C_l}\frac{\partial C_l}{\partial M_j}\Delta M^i_j=\frac{\partial \hat D}{\partial C_l}\frac{\partial C_l}{\partial M_j}\Delta M^i_j
\end{equation}
Finally we have the beam modulation correction coefficients
\begin{equation}
    \beta^d_j=\frac{\partial \hat D}{\partial C_l}\frac{\partial C_l}{\partial M^i_j} 
\label{eq:betaDefinition}
\end{equation}
All of the required quantities 
were measured with the beam modulation data.
Note that it is not necessary to precisely normalize the modulation elements, as scale factors in the $C_l$ cancel.

\subsubsection{Beam Modulation Monitoring}
 \label{sec:BeamModulation}

 The modulation response $\partial M_j / \partial C_l$ was seen to be highly sensitive to the beamline optics configuration. Online monitoring of these responses during the experiment was necessary to maintain the necessary conditions for this beam asymmetry correction. This section describes how the system performance was quantified during the run, and summarizes this performance.

 The set of five primary BPMs comprises the horizontal and vertical measurements in the ``target BPMs'' (BPM4A and BPM4E), located on the quadrupole-free line near the target, and the horizontal measurement at BPM12 at a high-dispersion point in the arc leading into the experimental hall.  Three other BPMs were also read out: BPM11, which was also in the dispersive arc, and the non-dispersive BPM16 and BPM1. For PREX-2, it was useful to substitute a linear combination of the BPMs 11 and 12 as the energy sensitive monitor in the minimal set of 5 monitors, to enhance the sensitivity to energy.
The seven modulation elements $C_i$ are horizontal or vertical coils $1-6$, with $C_7$ referring to the energy modulation. The beamline configuration is illustrated in Fig.~\ref{fig:bmw}.

\begin{figure}
  \centering \includegraphics[width=0.5\textwidth]{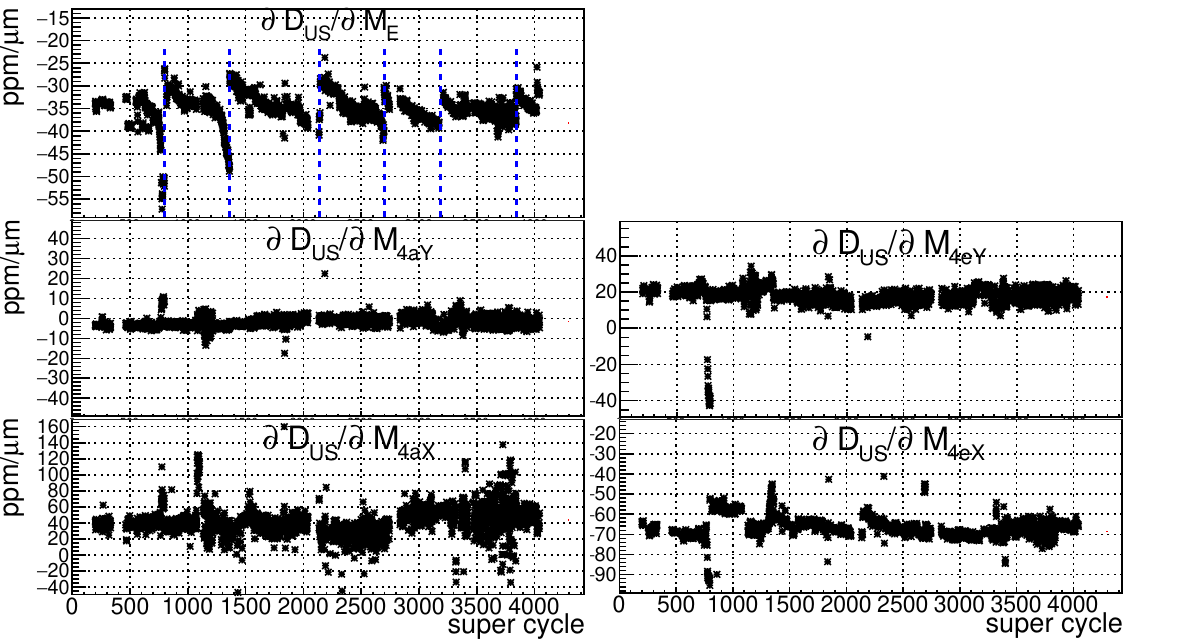}
  \caption{History plots of beam modulation slopes (Eq.~\ref{eq:betaDefinition}) in units of ppm/$\mu$m for PREX-2 data. The blue dashed lines indicate changes of the 
  lead target.  The detector $D$ used here is the upstream main quartz detector on the Left HRS; the results are similar for the other detectors. The plots, from left to right starting at the top, left are partial derivatives of $D$ with respect to monitor signals $M_{4ax}$, $M_{4ex}$, $M_{4ay}$, $M_{4ey}$, and $M_{E}$, where $E$ is a linear combination $E = 1.0 M_{11x} + 0.4 M_{12x}$ found to be a convenient representation that is very sensitive to energy.}
\label{fig:bmwslope}
\end{figure}

The key to the beam modulation analysis is to measure the beam modulation slopes $\beta^d_j$ to sufficient accuracy.  These slopes would become highly correlated and greatly increase the correction uncertainty as the beam modulation response matrix $\partial M_j / \partial C_l$ approached singularity.   A selection of four coils, along with the energy cavity, was made to best avoid degeneracy in the beam modulation matrices.

The history of $\beta^d_j$ over the PREX-2 data set is shown in Fig.~\ref{fig:bmwslope}.  Variations in the correction slopes were often the result of changes in the beam optics in the hall beamline.  These changes were required following changes in beam tune made upstream in the accelerator, and were necessary to maintain the intrinsic beam-spot size at the lead target or limit backgrounds in the Compton polarimeter.  There were also significant changes in slopes corresponding to the degradation of the lead target.  The vertical blue lines in the bottom plot shows the times when the lead target was changed. The downward deviation of the energy slope coincides with the onset of the lead target failure.  

Here we discuss some simple approximations made for an online analysis used to monitor for degeneracy in the modulation response matrix and to guide optics tuning to avoid it. As the $x$ and $y$ modulations were largely independent, it was sufficient to reduce the  5$\times$5 beam modulation model to two independent 2$\times$2 matrices, treating the horizontal position and angle in the target BPMs separately from the vertical. Taking just the horizontal responses, the detector responses to the steering coils may be written as

\begin{eqnarray}
\begin{pmatrix}
 \dfrac{\partial D_{k}}{\partial C_{1}}\\[0.4cm]
 \dfrac{\partial D_{k}}{\partial C_{3}}
 \end{pmatrix} 
 =
\begin{pmatrix}
 \dfrac{\partial M_{4ax}}{\partial C_{1}} & \dfrac{\partial M_{4ex}}{\partial C_{1}} \\[0.4cm]
 \dfrac{\partial M_{4ax}}{\partial C_{3}}& \dfrac{\partial M_{4ex}}{\partial C_{3}} 
\end{pmatrix}
\begin{pmatrix}
 \dfrac{\partial D_{k}}{\partial M_{4ax}}\\[0.2cm]
 \dfrac{\partial D_{k}}{\partial M_{4ex}}
 \end{pmatrix}
\label{eq:2x2bmwmatrix}
\end{eqnarray}
which can be rewritten as
\begin{eqnarray}
\kappa
\begin{pmatrix}
 1-\beta\\
 1
 \end{pmatrix} 
 =
\begin{pmatrix}
 1 & 1-\alpha \\
 1 & 1 
\end{pmatrix}
\begin{pmatrix}
 \dfrac{\partial D_{k}}{\partial M_{4ax}}\\[0.4cm]
 \Lambda\dfrac{\partial D_{k}}{\partial M_{4ex}}
 \end{pmatrix}
\label{eq:2x2bmwmatrixsimple}
\end{eqnarray}
 where $\alpha$, $\beta$ , $\kappa$ and $\Lambda$ are defined as

\begin{equation}
\alpha = 1- \cfrac{\cfrac{\partial M_{4ex}}{\partial C_{1}} ~ \cfrac{\partial M_{4ax}}{\partial C_{3}}}{\cfrac{\partial M_{4ax}}{\partial C_{1}} ~ \cfrac{\partial M_{4ex}}{\partial C_{3}}}
\label{eq:2x2bmwmatrixalpha}
\end{equation}

\begin{equation}
\beta = 1- \cfrac{\cfrac{\partial D_{k}}{\partial C_{1}} ~ \cfrac{\partial M_{4ax}}{\partial C_{3}}}{\cfrac{\partial M_{4ax}}{\partial C_{1}} ~ \cfrac{\partial D_{k}}{\partial C_{3}}}
\label{eq:2x2bmwmatrixbeta}
\end{equation}

\begin{equation}
  \kappa = \cfrac{\cfrac{\partial D_{k}}{\partial C_{3}}}{\cfrac{\partial M_{4ax}}{\partial C_{3}}} ~~~~~~~
\Lambda = \cfrac{\cfrac{\partial M_{4ex}}{\partial C_{3}}}{\cfrac{\partial M_{4ax}}{\partial C_{3}}} .  
\label{eq:2x2bmwmatrixkappalambda}
\end{equation}

The parameter $\alpha$ 
quantifies the closeness to singularity;
$|\alpha| < 0.1$ is problematic. The parameter $\beta$ is also a potential source of error. According to Eq.~\ref{eq:2x2bmwmatrixalpha} and Eq.~\ref{eq:2x2bmwmatrixbeta}, $\beta$ varies with $\alpha$, and the ratio $\beta$ / $\alpha$ should remain roughly constant with time. 
These parameters were monitored for each modulation cycle, throughout the data taking. 
These definitions allow us to find potential singularities in the beam modulation matrix by looking at the limiting behaviors, as follows
\begin{equation}
\frac{\delta \alpha}{\alpha}=\frac{1-\alpha}{\alpha}~\frac{\delta(1-\alpha)}{1-\alpha}
\label{eq:delta_alpha}
\end{equation}
and
\begin{equation}
\frac{\delta \beta}{\beta}=\frac{1-\beta}{\beta}~\frac{\delta(1-\beta)}{1-\beta}.
\label{eq:delta_beta}
\end{equation}

If $\alpha$ became too small, the beam tune was adjusted to make a non-singular beam modulation matrix with at least one set of beam modulation $x$ and $y$ driven coils. 

As a next approximation, we can add the energy response, including the modulation of energy and the horizontal measurement at the dispersive BPM12, along with the horizontal trajectory and measurements at the target BPMs, in a 3$\times$3 analysis,

\begin{eqnarray}
\begin{pmatrix}
 \dfrac{\partial D_{k}}{\partial C_{1}}\\[0.3cm]
 \dfrac{\partial D_{k}}{\partial C_{3}}\\[0.3cm]
 \dfrac{\partial D_{k}}{\partial C_{7}}
 \end{pmatrix} 
 =
\begin{pmatrix}
 \dfrac{\partial M_{4ax}}{\partial C_{1}} & \dfrac{\partial M_{4ex}}{\partial C_{1}} & \dfrac{\partial M_{12x}}{\partial C_{1}}\\[0.3cm]
 \dfrac{\partial M_{4ax}}{\partial C_{3}}& \dfrac{\partial M_{4ex}}{\partial C_{3}} & \dfrac{\partial M_{12x}}{\partial C_{3}}\\[0.3cm]
\dfrac{\partial M_{4ax}}{\partial C_{7}}& \dfrac{\partial M_{4ex}}{\partial C_{7}} & \dfrac{\partial M_{12x}}{\partial C_{7}}\\[0.3cm]
  \end{pmatrix}
\begin{pmatrix}
 \dfrac{\partial D_{k}}{\partial M_{4ax}}\\[0.3cm]
 \dfrac{\partial D_{k}}{\partial M_{4ex}}\\[0.3cm]
 \dfrac{\partial D_{k}}{\partial M_{12x}}
\end{pmatrix} ,
\label{eq:3x3bmwmatrix}
\end{eqnarray}
which can be rewritten as
\begin{eqnarray}
\kappa
\begin{pmatrix}
 1-\beta\\
 1\\
 \eta
 \end{pmatrix} 
 =
\begin{pmatrix}
 \Gamma & 1 & 1-\alpha \\[0.3cm]
 \delta & 0 & \alpha\\[0.3cm]
 1 & 0 & 0
\end{pmatrix}
\begin{pmatrix}
 \dfrac{\partial D_{k}}{\partial M_{12x}}\\[0.3cm]
 \dfrac{\partial D_{k}}{\partial M_{4ax}}\\[0.3cm]
 \Lambda\, \dfrac{\partial D_{k}}{\partial M_{4ex}}
 \end{pmatrix}, 
\label{eq:3x3bmwmatrixsimple}
\end{eqnarray}
where $\delta$, $\Gamma$, and $\eta$ are defined as

\begin{equation}
\delta = \cfrac{\cfrac{\partial M_{12x}}{\partial C_{3}}}{\cfrac{\partial M_{4ax}}{\partial C_{3}}} - \cfrac{\cfrac{\partial M_{12x}}{\partial C_{1}}}{\cfrac{\partial M_{4ax}}{\partial C_{1}}}
\label{eq:3x3bmwmatrixdelta}
\end{equation}

\begin{equation}
\Gamma = \cfrac{\cfrac{\partial M_{12x}}{\partial C_{1}}}{\cfrac{\partial M_{4ax}}{\partial C_{1}}}, ~~~~~~~
\eta = \cfrac{\cfrac{\partial D_{k}}{\partial C_{7}} ~ \cfrac{\partial M_{eax}}{\partial C_{3}}}{\cfrac{\partial D_{k}}{\partial C_{3}} ~ \cfrac{\partial M_{12x}}{\partial C_{3}}}.
\label{eq:3x3bmwmatrixgammaeta}
\end{equation}
Here the new key parameter is $\delta$, defined in Eq.~\ref{eq:3x3bmwmatrixdelta}. If $\delta$ is small ($<10$), the uncertainty in the correction due to the helicity-correlated energy difference will be large.

\subsubsection{Combined Modulation and Regression Analysis: Method of Lagrange Multipliers} 

As noted above, the regression analysis, by construction, removes all correlation between the measured asymmetry and beam monitors.  The required correction slopes are determined with high precision from the data being corrected, without using any separate calibration data.  However, the regression analysis is prone to overfitting the correction slopes, as correlations between the measured asymmetry and the beam monitors also contains the effect of instrumental noise in the monitors and are limited to the directions in phase space that are well populated by beam jitter. 

The modulation analysis, by contrast, requires averaging over a significant amount of calibration data which is accumulated only intermittently. It is intrinsically less precise and does not remove all correlation with beam monitor measurements. The modulation analysis has the advantage of calibration data that extends beyond the instrumental noise of the monitors, and therefore is unlikely to incorporate instrumental noise into the determination of the correction slopes.

The relative advantages of the two approaches can be combined in an analysis which develops corrections slopes from the correlations between the measured asymmetry and beam monitors while subjecting it to 
constraints implied by the modulation data.  We refer to this as the Lagrange multiplier analysis, as it employs the method of Lagrange multipliers to apply this constraint.

As before, we assume the asymmetry for the $i$-th multiplet depends on five beam parameters (Eq.~\ref{eq:Ai_Xj}), and corrections are applied using twelve beam monitors $\Delta M^i_{j}$ according to
\begin{equation}
\Delta A^i=\sum_{j=1}^{12}\beta^L_j\Delta M^i_{j},
\label{eq:Aicorrection}
\end{equation}
Here, the label ``$L$'' refers to the Lagrange multiplier analysis. The $\beta^L_j$ must be chosen so that the corrections from Eq.~\ref{eq:Aicorrection}  are the same as Eq. \ref{eq:Ai_Xj}.

To do this,
we define the linear function
\begin{equation}A(X_k)=A(M_j(X_k))=\sum_{j=1}^{12}\beta^L_jM_j(X_k).\end{equation}
Then
\begin{equation}\Delta A^i(X^i_k)=\sum_j\beta^L_j\Delta M^i_j(X_k), \quad
\Delta M^i_j=\frac{\partial M_j}{\partial X_k}\Delta X^i_k.\end{equation}

Eqs.~\ref{eq:Aicorrection} and \ref{eq:Ai_Xj} will be consistent if
\begin{equation}
\frac{\partial A}{\partial X_k}=\sum_{j=1}^{12}\beta^L_j\frac{\partial M_j}{\partial X_k}.\label{eq:alpha_beta}
\end{equation}
or, since the $X_k$ are independent variables,
\begin{equation}
\sum_{i=1}^5\frac{\partial A}{\partial X_k}\Delta X_k=
\sum_{k=1}^5\sum_{j=1}^{12}\beta^L_j\frac{\partial M_j}{\partial X_k}\Delta X_k.
\end{equation}
The above equations are equivalent since the $\Delta X_k$ are all independent variables.

Eq.~\ref{eq:alpha_beta}  will be satisfied if we require the following constraints on the  $\beta^L_j$ 
\begin{equation}
\frac{\partial \hat D}{\partial C_l}=\sum_{j=1}^{12}\beta^L_j\frac{\partial M_j}{\partial C_l}
\label{eq:L_constraint}
\end{equation}
which may be obtained from the beam modulation data. The proof is that if we multiply Eq.~\ref{eq:L_constraint} by the $5\times 5$ matrix $\partial C_l/\partial X_k$, we obtain
\begin{equation}
\frac{\partial \hat D}{\partial X_k}=\sum_{j=1}^{12}\beta^L_j\frac{\partial M_j}{\partial X_k}.
\end{equation}
Finally, the above equation together with the relation
\begin{equation}
\frac{\partial A}{\partial X_k}=\frac{\partial \hat D}{\partial X_k},
\end{equation}
which is derived like Eq.~\ref{eq:AD_relation}, gives Eq.~\ref{eq:alpha_beta}.
Any set of $\beta^L_j$ consistent with the beam modulation constraints from Eq.~\ref{eq:L_constraint} will correct the asymmetries without introducing systematic errors due to correlated noise in the beam monitors. However, the width of the asymmetry distribution does depend on the choice of the allowed $\beta^L_j$. Thus we are free to optimize the $\beta^L_j$ to minimize the width of the asymmetry distribution and thus minimize the statistical error.

The procedure is to minimize the $\chi^2$, which is calculated assuming the uncertainty $\sigma$ of each multiplet is the same,
\begin{equation}
\chi^2=\sum_i\left(A^i-\sum_j\beta^L_j\Delta M^i_j\right)^2/\sigma^2
\end{equation}
subject to the constraints of Eq.~\ref{eq:L_constraint}. This can be accomplished with the method of Lagrange multipliers. 

We define a function $\mathcal{L}$
\begin{equation}
  \mathcal{L}=\chi^2 + \sum_{k=1}^5\lambda_k\left(\frac{\partial \hat D}{\partial C_k}-\sum_{j=1}^{12}\beta^L_j\frac{\partial M_j}{\partial C_k}\right),
  \label{eq:LMchisq}
\end{equation}
and simultaneously solve the 12 linear equations
\begin{equation}
\frac{\partial \mathcal{L}}{\partial \beta^L_j}=0
\end{equation}
together with the five linear equations
\begin{equation}
\frac{\partial \mathcal{L}}{\partial \lambda_k}=0.
\end{equation}
The last five equations assure that Eq.~\ref{eq:L_constraint} is satisfied. The solution is a set of twelve $\beta^L_j$ that provide the narrowest width of the asymmetry distribution and still accurately correct for the beam parameters.  In application, the Lagrange multiplier method yields corrected asymmetry widths nearly as small as the regression analysis with twelve beam position monitors (see Fig.~\ref{fig:RMS width of detector}).  

\subsubsection{Beam Asymmetry Correction Results}
\label{sec:compareLagrangeRegress}

Ideally, the contributions from the electronic noise in the beam monitors are negligible, and the beam corrections from Lagrange multipliers and regression should agree.
Indeed, we observed that for the full data sets of PREX-2 and CREX, the two methods prescribe very consistent average beam corrections. 
However, the individual $\beta^r_j$ and $\beta^L_j$ are very different. The source of this difference is that the vast majority of the beam jitter can be described by fewer than five parameters, so even in the minimal set of five beam monitors, at least one beam monitor is essentially redundant. In this case, the minimum for the $\chi^2$ for regression is very shallow and different values for the $\beta^r_j$ will give approximately the same net correction.

We observed that when we chose linear combinations of the beam monitors that were uncorrelated, the $\beta^r_j$ agree much more closely with the $\beta^L_j$.  These linear combinations, denoted $\Delta M^{Ui}_j$, can be found by diagonalizing the beam monitors' symmetric correlation matrix 
\begin{equation}
S_{jk}=\frac{1}{N}\sum_i^N\Delta M^i_j\Delta M^i_k,
\end{equation}
where $N$ is the total number of multiplets in a minirun. The procedure for finding these uncorrelated monitor combinations is to solve the eigenvalue problem:
\begin{equation}
(S_{jk})\mathbf{Y}_l=\lambda_l \mathbf{Y}_l
\end{equation}
for the full set of twelve monitors. The secular equation for this eigenvalue problem has twelve solutions for $\lambda$, denoted $\lambda_l$, and there are twelve eigenvectors $\mathbf{Y}_l$ with $l$ indexed from 0 to 11.  These eigenvectors can be used to make the desired linear combination of the beam monitors
\begin{equation}
\Delta M^{Ui}_l=\sum_k\Delta M^i_k(Y_k)_l,
\end{equation}
where the eigenvectors have the properties of orthonormality:
\begin{equation}
\mathbf{Y}_l \cdot \mathbf{Y}_m = \delta_{lm},
\end{equation}
where $\delta_{lm}$ is the Kronecker delta. These orthonormal eigenvectors by definition diagonalize the symmetric correction matrix
\begin{equation}
\frac{1}{N}\left[\sum_i^n\Delta M^{Ui}_l\Delta M^{Ui}_k\right]=\left\{
\begin{aligned}
& \ 0  &\ (l \neq k) \\
& \ \sigma^2_l &\  (l = k) \\
\end{aligned}
\right. .
\end{equation}
The eigenvalue $\sigma^2_l$ is the square of the RMS for the monitor combination $\Delta M^{Ui}_l$.

Correction slopes can then be found using Eq.~\ref{eq:regchisq} or Eq.~\ref{eq:LMchisq} with $\Delta M^{Ui}_l$. As these are simply linear combinations of the beam monitors, applying beam corrections in the uncorrelated basis does not change either the central value or the width of the distribution of corrected asymmetry.
For the uncorrelated monitor combinations, the equations for the regression slope $\beta^{Ur}_j$ become uncoupled, and
\begin{equation}
\beta^{Ur}_j=\frac{\sum_i A^i\Delta M^{Ui}_j}{\sum_i(\Delta M^{Ui}_j)^2}.
\end{equation}
The denominator has the desired contributions from the beam variations and also a contribution from the instrumental noise. The degree to which a slope is affected by the monitor resolution will correspond to the ratio of variance of instrumental noise to that of beam jitter. 
For PREX-2, the comparison between the ratio of the smaller to larger eigenvalues is nearly a factor of 100, suggesting that the effect of resolution, and therefore the difference $\beta^{Ur}_j  - \beta^{UL}_j$, should be at most 1\% in the major eigenvectors, assuming that the various monitors have similar instrumental noise.

Table~\ref{tab:ev_bpm} gives the compositions of the uncorrelated monitor combinations for PREX-2, averaged over all miniruns in the data set. 
They are listed by the RMS of the helicity-correlated monitor differences (column 1 in Tab.~\ref{tab:UCM_Dx}), which is the square root of the eigenvalues, from largest to smallest. The first, which has large contributions from the $x$-direction monitors, describes primarily horizontal motion. The second, heavy in the $y$-direction monitors, describes vertical motion. The third, with large contributions from the horizontal monitors in the region of high dispersion, is sensitive to energy. The fourth and fifth, which are primarily vertical and horizontal respectively, have significantly smaller eigenvalues. While there is some variability in the compositions, the identities of these highest-ranked combinations were maintained throughout the run. 

Table~\ref{tab:UCM_Dx} summarizes the correction required for each of these uncorrelated monitors for PREX-2.  Each average is computed as an average over the $N$ miniruns, with the contribution from each minirun weighted by the inverse square of the statistical uncertainty of the main detector asymmetry, corrected using the Lagrange multiplier technique.   The first column, labeled Diff RMS, gives the average value of the square root of the eigenvalue, and provides a measure of the jitter in each uncorrelated monitor combination.  The second column is the average value of the uncorrelated monitor over the run, reflecting the average helicity-correlated beam difference.  Due to the care taken in configuring the polarized electron source, these average beam differences are not much larger than the 1$\sigma$ value of statistical convergence of beam jitter, shown in the third column.

 The slopes $\beta^{UL}_j$ determined with the Lagrange multiplier analysis varied over the experiment, due primarily to variations in the beam tune.  An approximate range which spans the majority of the observed 
 slopes
 is listed in the fourth column of Tab.~\ref{tab:UCM_Dx}.  The RMS width of the 
 distribution of these corrections
 (applied for each multiplet) and the average of these corrections, corresponding to  $\beta^{UL}_j * \Delta M^{U}_J$, is given for each of the $\Delta M^U_j$.  The largest contribution to beam noise and the largest correction slope were associated with the monitor combination (rank=2) most sensitive to energy.  This combination was also responsible for the largest average correction, $-70$~ppb, to the main detector asymmetry. Corrections from the combinations 5 through 12 were negligible. 
 We adopted this Lagrange multiplier method to determine the total beam correction
 for PREX-2,  $-60.4$~ppb.  The corrected asymmetry over the data set is $492.0 \pm 13.5$~ppb. 

The slopes and corrections for PREX-2 computed using the Lagrange multiplier analysis and regression are compared in Tab.~\ref{tab:UCM_compare}.  As expected, the two methods provided consistent slopes for the degrees-of-freedom that make the largest corrections, as reflected in averages of the difference in slopes between the methods and the RMS of the slope differences calculated for each minirun.  For the uncorrelated monitors with large jitter dominated by real beam motion (rank 0-4), the slopes and individual corrections agree at the 3\% level or better. The total correction calculated using regression is nearly indistinguishable from that calculated using the Lagrange multiplier analysis. 

These techniques were also employed for the analysis of the CREX data.  In CREX, monitor combinations and beam corrections were calculated independently in three stable periods of different beam optics. Due to the lower count rate in CREX compared to PREX-2, the statistical precision per minirun was lower and so the corrections slopes were calculated to a lower precision.  As with PREX-2, the largest uncorrelated monitor combinations could be associated with energy or position sensitivities, and energy was the largest contributor to the correction.  The corrections prescribed by regression were 
not significantly different from those calculated using the Lagrange multiplier method.  The total beam correction was $+53.5$ ppb for CREX, while the corrected asymmetry over the full data set was $A_C = 2080 \pm 84$~ppb.

\begin{table}[!h]
  \centering
  \caption{Table of the PREX-2 run's average composition of eigenvector monitor combinations $M^{U}_i$ in terms of individual monitors $M_j$.}
  \footnotesize
  \begin{ruledtabular}
  \begin{tabular}{c|r r r r r r r r r r r r}
    Rank&4aX&4eX&1X&11X&12X&16X&4aY&4eY&1Y&11Y&12Y&16Y\\
    \hline
    0&0.1&0.7&-0.6&-0.2&-0.1&0.3&-0.1&-0.1&0.0&0.0&-0.1&0.0\\
    1&0.1&0.2&0.0&-0.2&0.1&0.1&0.6&0.5&0.4&0.2&0.3&-0.1\\
    2&-0.1&-0.2&0.1&-0.7&-0.7&-0.1&0.0&0.0&-0.1&0.0&0.1&0.0\\
    3&0.0&-0.1&0.1&-0.1&-0.1&0.0&0.1&0.0&0.5&-0.2&-0.8&-0.2\\
    4&-0.4&-0.2&-0.5&0.5&-0.5&-0.2&0.1&0.1&0.0&0.1&0.1&0.1\\
    5&0.0&0.0&0.0&0.0&0.0&0.0&0.2&0.3&-0.5&-0.6&-0.1&-0.5\\
    6&-0.3&-0.5&-0.4&-0.4&0.5&0.3&0.1&-0.2&0.0&0.0&0.1&0.0\\
    7&0.1&0.0&0.0&0.0&-0.1&-0.1&-0.1&-0.4&0.3&-0.1&0.5&-0.7\\
    8&-0.1&0.0&0.0&-0.1&0.1&0.0&-0.7&0.5&0.4&-0.3&0.2&0.1\\
    9&-0.6&0.2&0.2&0.0&0.0&0.1&-0.2&0.2&-0.1&0.5&-0.1&-0.4\\
    10&0.5&-0.3&-0.3&0.0&0.1&-0.2&-0.2&0.3&-0.1&0.5&-0.2&-0.3\\
    11&0.2&-0.2&0.1&0.2&-0.3&0.9&-0.1&0.1&0.0&0.1&0.0&-0.1\\
  \end{tabular}
  \end{ruledtabular}
  \label{tab:ev_bpm}
\end{table}

\begin{table}[!h]
  \centering
  \caption{Beam jitter and helicity-correlated monitor differences, nominal correction slopes, and the implied corrections due to the eigenvector monitor combinations for PREX-2. See text for definitions.  }
  \footnotesize
  \begin{ruledtabular}
  \begin{tabular}{c|d{3.3} d{3.3} d{3.3} c d{3.3} d{3.3}}
    Rank & 
\headerwrap{0.1}{Monitor Diff RMS ($\mu$m)} &
\headerwrap{0.1}{Monitor Diff Mean (nm)} &
\headerwrap{0.1}{Diff RMS \mbox{/$\sqrt{N}$} (nm)} & 
 \headerwrap{0.16}{Nominal Slope Range (ppm/$\mu$m)}& 
 \headerwrap{0.16}{Correction RMS (ppm)}& 
 \headerwrap{0.16}{Correction Average (ppb)}\\
    \hline
    0   & 14.9 & -3.96 & 2.12 & [\,-10,5\,]    &     79  &  -22.3  \\
    1   &   9.6 &   2.31 & 1.38 & [\,-2,28\,]    &     57  &    22.5  \\
    2   &   7.1 & -1.83 & 1.01 & [\,27,47\,]    &   248  &  -70.4  \\
    3   &   3.3 & -1.61 & 0.46 & [\,-12,22\,]  &     25  &    -2.8  \\
    4   &   2.7 & -1.01 & 0.38 & [\,-25,-2\,]  &     35  &      9.7  \\
    5   &   1.4 &   0.16 & 0.20 & [\,-14,4\,]    &       5  &      1.3  \\
    6   &   1.0 &   0.15 & 0.12 & [\,-8,18\,]    &       6  &      0.0  \\
    7   &   0.8 &   0.02 & 0.11 & [\,2,20\,]      &       9  &      1.1  \\
    8   &   0.5 & -0.08 & 0.07 & [\,-22,14\,]  &       3  &      0.3  \\
    9   &   0.4 & -0.02 & 0.06 & [\,-10,20\,]  &       2  &      0.2  \\
    10 &   0.4 & -0.04 & 0.05 & [\,-5,15\,]    &       2  &      0.2  \\
    11 &   0.3 & -0.01 & 0.04 & [\,-15,5\,]    &       2  &      0.1  \\
    \hline
    Sum  & & & & & & -60.4 
  \end{tabular}
  \end{ruledtabular}
  \label{tab:UCM_Dx}
\end{table}

\begin{table}[!h]
  \centering
  \caption{Comparison of beam corrections calculated with the Lagrange multiplier method, $\beta^{UL}$ and regression $\beta^{Ur}$ for PREX-2.  }
  \footnotesize
  \begin{ruledtabular}
  \begin{tabular}{c|  d{3.3} d{3.3} d{3.3} }
   Rank & 
\headerwrap{0.16}{Slope Diff. Mean (ppm/$\mu$m)} & 
\headerwrap{0.16}{Slope Diff. RMS (ppm/$\mu$m)}&
\headerwrap{0.16}{Correction Difference (ppb)}\\
    \hline
    0   &   0.08 & 0.40 & -0.32  \\
    1   & -0.24 & 0.56 & -0.55 \\
    2   & -0.14 & 0.66 & 0.25    \\
    3   & -0.13 & 0.70 & 0.21   \\
    4   &  0.45 & 1.19  & 0.45    \\
    5   &  0.11 & 0.79  & 0.02  \\
    6   & -0.20 & 1.82 & -0.03 \\
    7    & -1.06 & 1.87 & -0.02  \\
    8   &  1.55 & 4.30 & -0.12  \\
    9   & -0.41 & 3.48  & 0.01   \\
    10 & -1.96 & 5.39 &  0.08    \\
    11 &  0.63 & 3.65  & -0.01  
  \end{tabular}
  \end{ruledtabular}
  \label{tab:UCM_compare}
\end{table}

\subsubsection{Additional Cross Checks}
\label{sec:beam_correction_cross_validations}

Additional cross-checks were performed in order to search for 
systematic failures in the beam correction techniques. Two such investigations are described here.

The first of these used the corrections slopes derived for correcting the measured asymmetry in order to predict the effect of the beam modulation cycles on the main detector signals. A residual sensitivity, $R_i$, is defined for a given beam modulation coil $i$ as 
\begin{equation}
\label{eqn:Residual sens}
R_i = \frac{\partial \hat D}{\partial C_i} - \sum \limits_j \beta_j \frac{\partial M_j}{\partial C_i},
\end{equation}
where $\hat D$ is the detector response, $C_i$ is the coil modulation impulse, $M_j$ is the beam position measured by monitor $j$, and $\beta_j$ is the correction slope. For modulation directions with significant responses, it is useful to characterize this residual sensitivity as a fraction of the effect of the modulation coil $\partial \hat{D}/\partial C_i$. 

In the case of the method of Lagrange multipliers, the beam modulation cycles are a constraint on the fit of correction slopes. In practice, the correction slope $\beta_j^L$ in each minirun was obtained with beam modulation sensitivities averaged over run periods in which the modulation 
sensitivities were seen to be approximately stable.  Modulations from all coils used in the analysis were successfully corrected with high accuracy (i.e. small $R_i$), demonstrating a negligible loss of fidelity due to this averaging.

As there are a total of 3 modulation coils in each direction (horizontal and vertical),  one pair of coils was not used in the analysis. The analysis selects the coils which provided the best resolution, so the ``redundant'' coils were not the same pair throughout the data set. The residual sensitivity of these redundant coils was a useful test, confronting the correction slopes with a beam motion which was not included in its calibration set.  It is also the case that this motion is excluded from calibration specifically because it is either small or seen to be highly redundant, so the value of this check is limited. The Lagrange multiplier analysis corrected the redundant coils with accuracy comparable to that of the coils used in the calibration set. No systematic offset in residual sensitivity was found.

This residual sensitivity test is more demanding for the regression analysis.  Table \ref{tab:residual_reg} summarizes the residual sensitivity based on regression slopes from PREX-2 data. Larger $R_i$ 
occur for vertical modulation coils (indicated as ``Y'' in the table). As seen in the first column of the table, sensitivity to the vertical direction is smaller than that for energy or the horizontal direction. The results also show a large RMS width in fractional residual sensitivity for Y, demonstrating the challenge for regression to fit
relatively insensitive degrees of freedom. For the more sensitive horizontal coils and the energy vernier, the central value and RMS width in the fractional residual sensitivity are comparable to the fidelity of the Lagrange multiplier analysis. 

\begin{table}
  \centering
  \footnotesize
    \caption{Beam modulation residual sensitivities in PREX-2 constraining coil data set with regression correction slope. unit: ppm/count}
\begin{ruledtabular}
  \begin{tabular}{c|c d{1.1} d{1.1} d{1.1} d{1}}
     Detector
    & \headerwrap{0.1}{Coil}
    & \headerwrap{0.16}{Nominal Sensitivity Mean } 
    & \headerwrap{0.16}{Residual Sensitivity Mean}
    & \headerwrap{0.16}{Residual Sensitivity RMS}
    & \headerwrap{0.16}{Fractional Residual Sensitivity RMS }  \\
    \hline
    \multirow{7}{*}{\parbox{0.1\linewidth}{\centering  Left Arm }}
&	1	(X)	&	-44.5	&	-0.4	&	1.3	&	3\ \%	\\
&	2	(Y)	&	-7.4	&	-3.9	&	0.8	&	11\ \%	\\
&	3	(X)	&	-14.0	&	0.0	&	0.5	&	4\ \%	\\
&	4	(Y)	&	-7.0	&	-1.0	&	1.6	&	23\ \%	\\
&	5	(X)	&	64.3	&	-1.0	&	2.0	&	3\ \%	\\
&	6	(Y)	&	8.3	&	1.4	&	2.7	&	32\ \%	\\
&	7	(E)	&	53.5	&	0.3	&	1.0	&	2\ \%	\\
    \hline 
    \multirow{7}{*}{\parbox{0.1\linewidth}{\centering Right Arm }}
&	1	(X)	&	49.4	&	0.2	&	0.9	&	2\ \%	\\
&	2	(Y)	&	-9.2	&	-5.1	&	0.5	&	6\ \%	\\
&	3	(X)	&	15.1	&	0.0	&	0.5	&	3\ \%	\\
&	4	(Y)	&	-5.8	&	-1.2	&	1.7	&	30\ \%	\\
&	5	(X)	&	-68.7	&	-0.3	&	1.6	&	2\ \%	\\
&	6	(Y)	&	8.5	&	1.9	&	2.6	&	31 \ \%	\\
&	7	(E)	&	32.9	&	0.4	&	0.8	&	2 \ \%	\\
  \end{tabular}
\end{ruledtabular}
  \label{tab:residual_reg}
\end{table}

A second test for systematic error in the beam corrections compares various correction techniques, aiming to isolate possible systematic differences even if those differences happen to cancel over the full data set.  
To this end, the difference in corrected asymmetry for any two correction techniques, $A^{I}$ and $A^{II}$, is calculated in each minirun as
\begin{equation}
 \Delta A_{i} = A^{I}_{i}-A^{II}_{i}, 
\end{equation}
This is averaged over all the miniruns in each slug, 
\begin{equation}
 \langle\Delta A\rangle_j = \frac{1}{N_j}\sum_{i}^{N_j}\Delta A_{i},
\end{equation}
where $\langle\Delta A\rangle_j$ is the average for the slug $j$ and $N_j$ is the number of miniruns in that slug.  This average was compared to the expected uncertainty given the fluctuations in this difference:
\begin{equation}
 \sigma(\langle\Delta A\rangle_j) = \frac{\sqrt{\sum_{i}^{N_j}(A^{I}_{i}-A^{II}_{i})^{2}}}{N_j}.
 \label {eq:sigma_DeltaA}
\end{equation}
In this way, the average offset between correction methods in each slug can be compared to the jitter between those methods within that slug.  The average, fit over all slugs, weighted by the inverse variance $\sigma^2(\langle\Delta A\rangle_j)$ , would highlight any systematic difference that is larger than random variations.

These averages, with corresponding measures of the quality of the fit to a constant value, are given in Tables~\ref{tab:delta_A_PREX} for PREX-2. 
The comparison of the 5-BPM regression (Reg-5) with other techniques demonstrates the effect of correlated instrumental noise, with deviations in the average asymmetry larger than the resolution, and a reduced quality of fit.  The beam modulation analysis restricted to 5 BPMS (BM-5) has less precision than the 12 monitor regression (Reg-12) and Lagrange multiplier (LM-12) techniques, but does not show a systematic deviation. 
The 12 monitor regression and Lagrange multiplier techniques demonstrate excellent consistency, avoiding the shortcomings that are expected in the 5 BPM analyses. The fit quality further suggests that the high degree of consistency between the corrections prescribed by the 12-monitor analyses, as shown in Tab.~\ref{tab:UCM_compare} is not an accident of cancellation over the PREX-2 dataset. 
For CREX the same analysis methods and cross-checks were performed showing similar results.

\begin{table}[!h]
  \centering
  \caption{Summary of $\Delta A$ between different beam correction techniques for PREX-2. See text for definition of the various methods.}
  \begin{ruledtabular}
  \begin{tabular}{l | l | d{3.3} d{3.3} d{3.3} c} 
    \headerwrap{0.15}{Method I} &
    \headerwrap{0.15}{Method II} &
    \headerwrap{0.13}{$\expt{\Delta A}$ (ppb)} & 
    \headerwrap{0.13}{$\sigma$($\Delta$A) (ppb)} & 
    \headerwrap{0.17}{$\chi^2$/ndf} & 
    Prob.\\
    \hline
     Reg-12 &LM-12  & 1.0 & 1.2 & 91.2/95 & 0.59 \\ 
     BM-5 & LM-12 &2.2 & 3.5 & 86.4/95 & 0.72 \\ 
     BM-5 & Reg-12 &-0.5 & 3.8 & 88.9/95 & 0.66 \\ 
     Reg-5 & Reg-12 &5.8 & 2.6 & 106.1/95 & 0.21 \\ 
     Reg-5 & BM-5 & 4.4 & 2.7 & 128.8/95 & 0.01 \\ 
  \end{tabular}
  \end{ruledtabular}
  \label{tab:delta_A_PREX}
\end{table}

\subsubsection{Systematic Uncertainty}
\label{sec:Certainty on the Beam Corrections}

Throughout the studies described above, the beam corrections appear robust.  The method of using an overdetermined set of BPMs in regression is expected to insulate the measurement from electronics noise in the monitors, and this protection is explicitly encoded in the Lagrange multiplier technique 
which utilizes purposeful calibration data as a constraint.  As shown in Tab.~\ref{tab:UCM_compare}, the slope differences between the these techniques in the top five eigenvectors have standard deviations near or below 1~ppm\slash$\mu$m and small average values. Although the lower-ranked eigenvectors have larger and noisier slope differences, the grand-averaged helicity-correlated differences of these eigenvectors are very small, such that these deviations do not reflect significant uncertainty in the beam correction. The two techniques prescribe beam corrections, averaged over the run, that are only negligibly different.

In order to set a scale on possible errors, the uncertainty of the corrections implied by the differences between the corrections found using the various different methods 
was used.  The slope difference RMS width in Tab.~\ref{tab:UCM_compare} and average monitor differences listed in Tab.~\ref{tab:UCM_Dx} would each represent a relative change in the correction as large as 2.7\%, and more commonly less than 1\%, whereas the actual average correction differences between the techniques are much smaller.  Taking a 3\% uncertainty on any given correction would seem to be a conservative upper estimate of the implied uncertainty.  The corrections are also dominated by a single eigenvector, and tests of these techniques demonstrate that they are particularly effective when dealing with prominent, independent, degrees of freedom.

A conservative estimate of 3\% on each monitor correction, summed incoherently, was used for PREX-2: 
\begin{equation} 
 \sqrt{\sum_i^{12} (\Delta A_i^U \times 0.03) ^2} = 2.3 \ \mathrm{ppb}. 
\label{eq:uncert_Abeam}
\end{equation}

With similar reasoning, the uncertainty for beam corrections for CREX was set to  5\% for each monitor contribution.  Each of the individual beam correction periods were independently summed. The result is
\begin{equation} 
 \sqrt{\sum_j^{3} \sum_i^{12} (f_j \Delta A_{j,i}^U \times 0.05) ^2} = 5.4 \ \mathrm{ppb},
\label{eq:uncert_Abeam_CREX}
\end{equation}
where $f_j$ is the fractional weight of each period $j$ used when averaging over the full data set. 

\begin{table}
\centering
\caption{Beam asymmetry correction and systematic uncertainty for PREX-2 and CREX.}
\begin{ruledtabular}
\begin{tabular}{@{\hspace{3em}} c c c @{\hspace{3em}}}
Experiment & $A_{beam}$ & $A_C$ \\
\hline
 PREX-2 & $-60.4 \pm 2.5$~ppb & 492.0 $\pm$ 13.7~ppb \\
 CREX & 53.5 $\pm$ 5.4~ppb & 2080.3 $\pm$ 83.8~ppb\\
\end{tabular}
\end{ruledtabular}
\label{tab:beam_correction_summary}
\end{table}

The grand-averaged beam-corrected asymmetry $A_C$,
for PREX-2 is 
\begin{equation}
 492.0 \pm 13.5 \ (\mathrm{stat.}) \pm 2.5 \ (\mathrm{syst.}) \ \mathrm{ppb}.
\end{equation}
And the same corrected asymmetry for CREX is
\begin{equation}
 2080.3 \pm 83.8  \ (\mathrm{stat.}) \pm 5.4 \ (\mathrm{syst.}) \mathrm{ppb}.
\end{equation}

\subsection{Backgrounds}
\label{sec:norm}

To calculate $A_{\rm PV}^{\rm meas}$, the
raw asymmetry $A_{\rm raw}$ needs to be further 
corrected for beam polarization, radiative effects, finite acceptance of the spectrometers, and contributions from various background processes as in Eqs.~\ref{eq:asyCorrected} and \ref{eq:asyCorr}. In this section, we describe the process used to make each of these corrections. The backgrounds are isotope contamination, diamond foil backing on the lead target, inelastic levels, rescattering in the spectrometer, including magnetic poletip scattering, and transverse asymmetries. These backgrounds are the terms $A_i$ and $f_i$ for asymmetry and fraction, respectively, for process $i$ in Eq.~\ref{eq:asyCorr}. The polarization factor $P_e$ is discussed in Secs. \ref{sec:mollerresults} and \ref{sec:comptonresults}.

\subsubsection{Isotope Contamination}
\label{sec:isotopeBgr}

To estimate the asymmetries for lead isotopes other than $^{208}$Pb, we use 
\begin{equation}
\label{eq:approxAsy}
  A_{\rm PV} \approx \frac{G_F Q^2}{\pi \alpha \sqrt2} \hskip 0.04in \left[ {\rm sin}^2 \theta_W  + \frac{1}{4} \left(\frac{N}{Z} - 1\right)\right]
\end{equation}for elastic scattering from $N \ne Z$ nuclei \cite{DONNELLY1990179}.  The calculated asymmetries for $^{207}$Pb and $^{206}$Pb are lower than the asymmetry of $^{208}$Pb by $\approx 0.84$\% and $\approx 1.7$\% respectively, and given their small assayed fractions (see~\ref{sec:pbtarget}), their effect on  $A_{\rm PV}$ is negligible ($<0.05$~ppb).

For the $^{40}$Ca correction, we use a theoretically calculated asymmetry \cite{ref:crex}  $A_{\rm PV} = 2430 \pm 30 $ ppb. The assayed target fraction of $^{40}$Ca, weighted by the amount of data taken from each calcium target, was 7.95\% which leads to a calculated correction to $A_{\rm PV}$ of $19 \pm 3$ ppb.

\subsubsection{Diamond foil Background}
\label{sec:DiamondFoilBackground}
The fraction of detected events in PREX-2 coming from the diamond ($^{12}$C) foils was determined from {\sc Geant4} simulation, using the measured foil and Pb target thicknesss and their uncertainties (see Sec.~\ref{sec:pbtarget}), and was found to be $f_C = 6.3 \pm 0.5\%$. The asymmetry from light isoscalar spin-0 nuclei is well understood ~\cite{Souder:1990ia, HAPPEX:2005qax} and was calculated at the present kinematics to be $A_C = 539 \pm 22$~ppb, rather similar to the measured $^{208}$Pb asymmetry. The net correction due to the $^{12}$C was therefore small, $0.7 \pm 1.4$~ppb.

\subsubsection{Inelastic Backgrounds}
\label{sec:inelasticBgr}

Being doubly magic nuclei, the first excited states of $^{208}$Pb and $^{48}$Ca have relatively high excitation energies (2.615 MeV and 3.831 MeV, respectively) compared to a typical heavy nucleus.  These relatively large first-excited state energies allow the experimental apparatus to geometrically accept nearly all elastically-scattered events while rejecting almost all inelastic events.  It is important to minimize the inelastic acceptance because $A_{\rm PV}$ from inelastic scattering is poorly known.

In order to eliminate the inelastic background, a large enough gap was maintained between the main detector acceptance and the first excited state throughout the experiment. To do this, the main detectors were aligned properly in the HRS focal plane using dedicated low-current counting-mode runs at the beginning of each experiment. Periodic alignment checks were performed to verify the main detector acceptance.

\begin{figure}
\centering
\includegraphics[scale=.30]{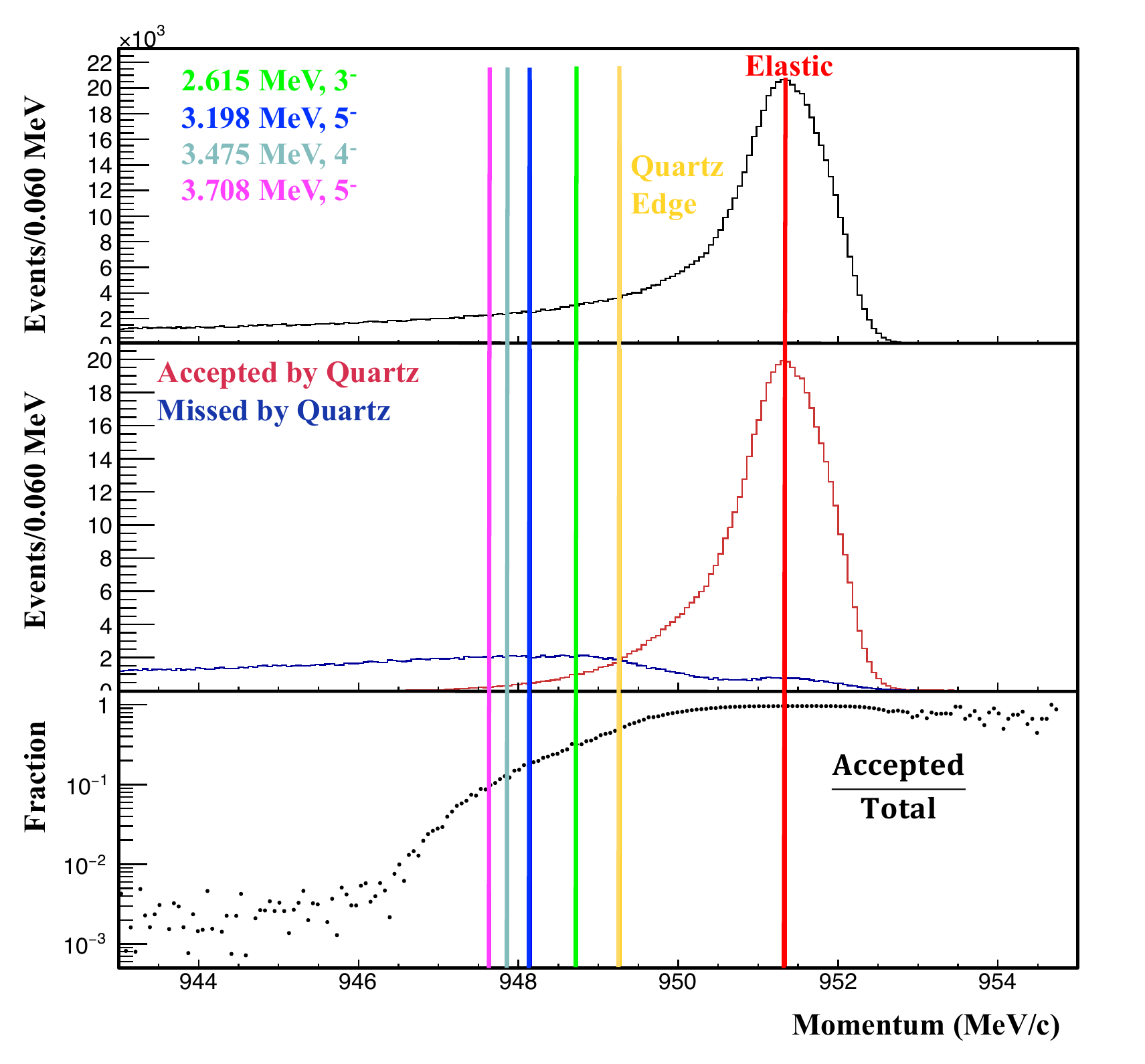}
\caption{The momentum distribution for $^{208}$Pb during PREX-2. The black histogram in the top plot shows all the events. The middle plot shows the events accepted by the main detector (red histogram), and the events missed by the main detector (blue histogram). The bottom plot shows the accepted events (from the middle plot) divided by the total events (from the top plot). The elastic peak, main detector edge, and first four excited state positions (in momentum space) are shown and labeled using different colored vertical lines.  The edge of the quartz deetctor is indicated by the yellow vertical line. 
}
\label{fig:PREX_dilution_fraction}
\end{figure}

\begin{figure}
\centering
\includegraphics[scale=.30]{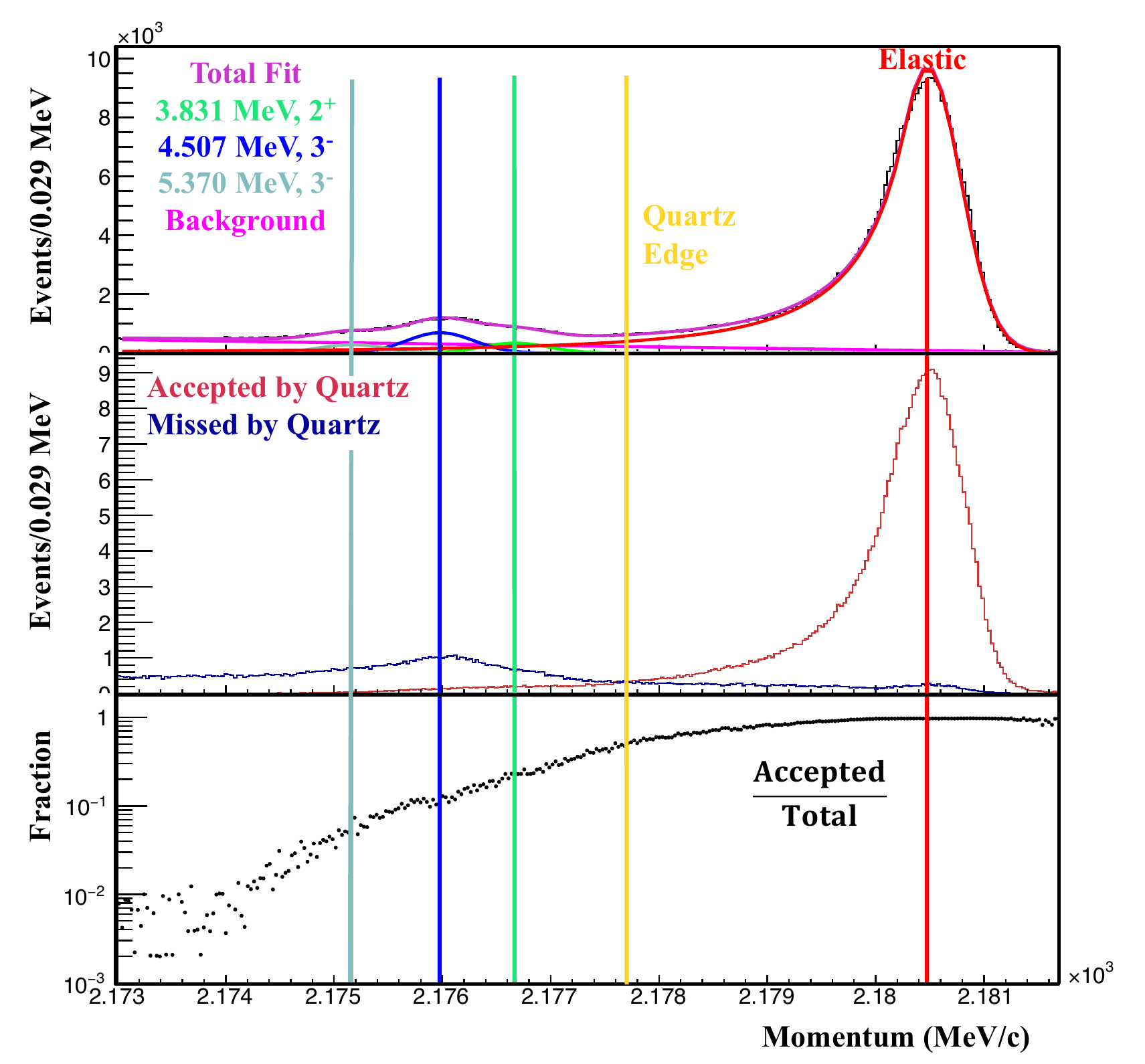}
\caption{The momentum distribution for $^{48}$Ca. The black histogram in the top plot shows the total flux. In the middle plot, the red histogram shows the flux accepted by the main detector, and the blue histogram shows the events missed by the main detector. The bottom histogram shows the accepted events divided by the total events. The relative positions of the elastic peak, main detector edge, and the three excited states are shown by different colored vertical lines. The fitted shapes from the excited states \textcolor{red}{using Crystal Ball functions}
are shown as colored histograms. 
}
\label{fig:CREX_dilution_fraction}
\end{figure}

For PREX-2 kinematics, the low-lying excited states of $^{208}$Pb had extremely small relative strength,
given the detector acceptance. The momentum spectrum of $^{208}$Pb for a typical counting mode run is shown in Fig.~\ref{fig:PREX_dilution_fraction}.  The small relative strength, coupled with the fact that the asymmetries for inelastic scattering is expected to be comparable to the elastic $A_{\rm PV}$, resulted in negligible inelastic background correction for PREX-2. A 0.02\% relative uncertainty was assigned to account for inelastic contributions. In contrast, for CREX kinematics, the low-lying excited states of $^{48}$Ca had larger relative strength. The relative cross sections for the first three excited states of $^{48}$Ca (as compared to the elastic cross section) and their corresponding acceptance probabilities in the main detector for a typical run are shown in Fig.~\ref{fig:CREX_dilution_fraction}.
The relative cross sections, acceptance probabilities, and resulting 
background fractions, averaged throughout the experiment, are presented in Table~\ref{tab:CREX_inelastic}.

\begin{table}[htbp]
    \centering 
    \caption{Inelastic background fraction for CREX. $\sigma^i_{\rm rel}$ is the differential cross section for excited state $i$, relative to the elastic cross section. ``$P_i$'' is the probability for a scattered electron from the given state to be detected by the main detector, and $f_i$ is the resulting fraction of the main detector signal from the given excited state ($f_i = \sigma^i_{\rm rel}P_i$).}
    \label{tab:CREX_inelastic}
\begin{ruledtabular}
\begin{tabular}{c c c c}
Ex. State $J^\pi$(MeV) & $\sigma^i_{\rm rel}$ (\%) & $P_i$ (\%)  & $f_i$ (\%) \\ \hline 
$2^+$ (3.831 MeV) & 2.27  & 25.92  & 0.59 \\ 
$3^-$ (4.507 MeV) & 4.51  & 14.68  & 0.66 \\ 
$2^-$ (5.370 MeV) & 1.88  & 6.44   & 0.12 \\ 
\end{tabular}
\end{ruledtabular}
\end{table}

\subsubsection{Rescattering Backgrounds}
\label{sec:rescatteringBgr}
Those inelastic and/or the radiative tail electrons whose energy was sufficiently low so that they did not transport correctly to the focal plane could scatter off the spectrometer's vacuum chamber, near the third quadrupole magnet.  These rescattered electrons could reach the main detector acceptance and contaminate the signal. Therefore, the strength of this background needed to be well understood. During the HAPPEX experiments \cite{HAPPEX_long_paper}, 
an approximate model of this background was developed and confirmed with simulations and elastic proton measurements, though the septum magnet used in the present work was different. Using this model, the strength of the accepted background signal, $B_{\rm rs}$, is given by
\begin{equation}\label{eq:BGfraction}
B_{\rm rs} = \int_{E_{\rm thr}}^{E_{\rm max}} dE P_{\rm rs}(E) R(E),
\end{equation}
where $E_{\rm thr}$ is the inelastic threshold, $E_{\rm max}$ the maximum energy loss, $P_{\rm rs}(E)$ is the rescattering function given by 
\begin{equation}\label{eq:REXfn}
P_{\rm rs}(E) = \tau \times \left(\frac{E_{\rm dep}}{E_0}\right),
\end{equation}
and $R(E)$ is the ratio of the inelastic to elastic cross-section given by
\begin{equation}\label{eq:RESratio}
R(E) = \frac{\left(\frac{d\sigma}{d\Omega dE}\right)_{\rm inelastic}}{\left(\frac{d\sigma}{d\Omega}\right)_{\rm elastic}}.
\end{equation}
$E_{\rm dep}$, $E_0$, and $\tau$ are the energy deposited by rescattered electrons on the detector, energy deposited by elastically scattered electrons, and the probability for the inelastic electrons to rescatter inside the spectrometers, respectively.  
Here $\frac{E_{\rm dep}}{E_0} \approx1$; therefore, $P_{\rm rs}(E) \approx \tau$.  

\begin{figure}
\centering
\includegraphics[scale=.30]{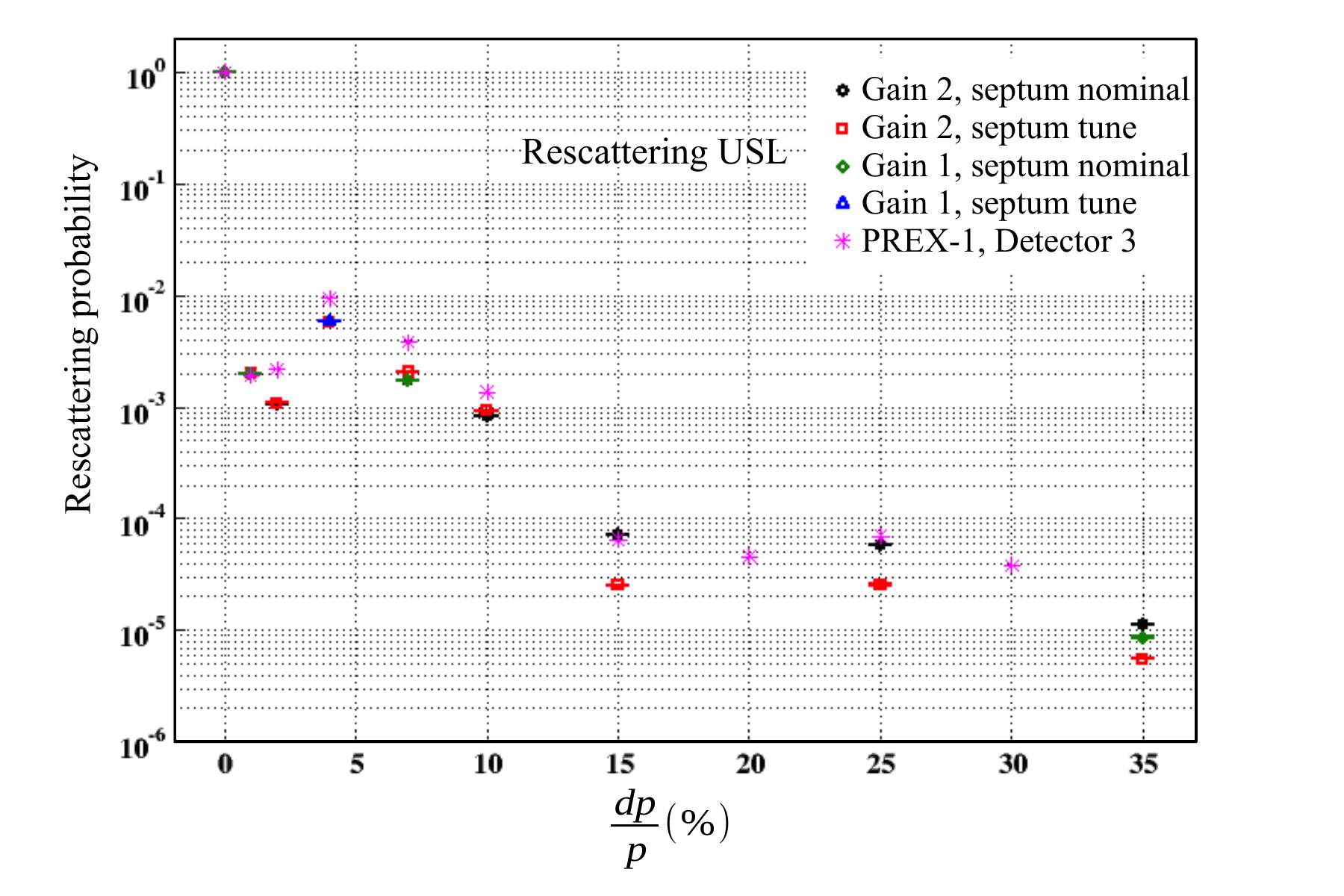}
\caption[Rescattering probability plotted against HRS dipole setting.]{Rescattering probability plotted against the precentage change in the HRS dipole setting, $\frac{dp}{p}$, in percent from the nominal production setting. Gain 1 and Gain 2 are two different gain settings of the main detector PMT.  The data are from the upstream quartz detector on the Left HRS (labeled as ``USL'').  Septum nominal means the septum field was kept at the production setting, while septum tune means the septum field was changed by the same percentage as the other magnets~. Also shown are the results of similar measurement taken during PREX-I.}
\label{fig:rescatteringUSL}
\end{figure}

Measurement of $P_{\rm rs}(E)$ was performed by measuring the energy deposited in the detectors while
increasing the fields of the HRS magnets (dipole and quadrupoles) -- forcing trajectories of the elastic electrons to follow the trajectories of the inelastic events during the production field setting (ramming most of them into the vacuum chamber wall at the dipole bend).  
Data were taken in integrating mode, both with septum magnets kept at nominal production settings, and with the septum field increased by the same fraction as the other magnets. The results were compared to check how sensitive the model was to the septum field.  We also took the runs at two different gain settings of the main detector PMTs; both sets of gain were higher than production settings in order to produce a reasonable signal level in the ADC (since rescattering rates were very low).  These tests were performed only in the left-HRS due to the right-HRS dipole's extremely slow ramp rate, but much past experience has shown the two HRSs to be highly symmetric.  The rescattering probability as a function of the fractional momentum change is shown in Fig.~\ref{fig:rescatteringUSL}.  

Since we changed the magnet $dp$ settings in discrete steps ($i$), Eq.~\ref{eq:BGfraction} can be simplified to
\begin{equation}
B_{\rm rs} = \sum\limits_i P_{\rm rs}(E)_i \times [R(E) dE]_i,
\end{equation}
where the relative rate, $R(E) dE$, was obtained from simulation of the radiative tail from elastic scattering.
To obtain the background fractions due to the rescattered electrons we need to multiply the rescattering probability by the relative cross-section for each $dp$ setting.  From this, we get $\sum\limits_i f_i = B_{\rm rs} = 1.67 \times 10^{-4}$.  
To estimate the asymmetry of these electrons we assumed the linear approximation for the $Q^2$-dependence of the asymmetry, $A_i = A_0 Q_i^2$, where $i$ represents each of the $dp$ settings and $A_0$ is a constant to be calibrated using the measured $A_{\rm PV}$. 
Since $Q^2 = 2 E_0 E^\prime (1-\cos\theta)$, $Q^2$ (and thus $A_i$) varies with the scattered energy $E^\prime$ as we change the $dp$ settings. This is because each spectrometer $dp$ setting transports a different range of $E^\prime$ events.  
The result is 
$\sum\limits_i f_i A_i = 0.0877$ ppb,  
giving a negligibly small correction to the measured asymmetry.  Therefore, we did not make any correction, but we included the systematic uncertainties from $\sum\limits_i f_i$ and $\sum\limits_i A_i$ into the total systematic error for the measurement.

Note that by changing $dp$ we were effectively scanning the HRS $E^\prime$ acceptance over the radiative tail region, where the $Q^2$ is different than at the elastic peak.
In reality, radiative effects, such as Bremsstrahlung emission/absorption, can happen before the main physics scattering vertex, at the vertex, or after the vertex.  For example, external radiation emitted before the vertex results in a decrease in $E$ and hence $Q^2$.  These types of effects would need simulation to quantify, but given the small size
of this background 
our simple linear model of $Q^2$ and the asymmetry of the radiative tail events was sufficient.  

The 
systematic uncertainty to the asymmetry from spectrometer rescattering is shown in Table~\ref{tab:ApvPREXfinal} for PREX-2, but is neglected in Table~\ref{tab:ApvCREXfinal} for CREX, since it was not significant due to the smaller radiative tail of calcium.

\subsubsection{Spectrometer Magnet Poletip rescattering}
\label{sec:poletipBgr}
The dipole and quadrupole magnets of the HRS contain magnetized iron.  The iron of the spectrometer magnets is spin-aligned and rescattering from the magnetized iron is a source of systematic uncertainty due to a polarization-dependent asymmetry in Møller scattering.  Practically, high-energy electrons do not hit the poletips while the acceptance-defining collimators stop low-energy electrons from hitting the poletips.

The correction to the measured asymmetry from poletip rescattering is
\begin{equation}
dA = f P_{e1} P_{e2} A_{\rm Fe}
\end{equation}
where $f$ is a fraction of the signal from the poletip rescattering, $P_{e1}$ is the polarization of the rescattered electron, $P_{e2}$ ($\approx 0.03$ taken from HAPPEX~\cite{HAPPEX_long_paper}) is the polarization of the atomic electron in the dipole of the spectrometer wall, and $A_{\rm Fe}$ ($\approx 0.04$) is the analyzing power of the iron.  Combining simulations and the rescattering study (see Fig.~\ref{fig:rescatteringUSL}) we estimated $f \approx 10^{-9}$ for PREX-2 and $f \approx 10^{-8}$ for CREX.  The electron polarization ($P_{e1}$) was estimated to be $\approx 0.26$ for PREX-2 and $\approx 0.55$ for CREX.  The resulting $dA$ was several orders of magnitude below 1 ppb for both experiments and therefore we did not make any corrections.

\subsection{Acceptance Function $\epsilon(\theta)$}
\label{sec:PointingMethod}
\input{crex_acceptance_function_new}

\subsection{Transverse asymmetry contamination}
\label{sec:Atmeasurement}
A beam normal single asymmetry A$_{n}$ arises when fully transversely (i.e. the polarization direction is perpendicular to the incident electron momentum) polarized electrons scatter off nuclei. In this case, the scattering amplitude develops an azimuthal modulation due to the interaction of the electron's magnetic moment and the magnetic field of the nuclei; in the hard-photon regime the modulation is due to an interference between the imaginary part of the two-photon exchange amplitude and the one-photon exchange amplitude~\cite{DERUJULA1971365,twophoton}. During PREX-2 and CREX measurements were performed with the same apparatus as the main parity-violating measurements but with the electron beam polarization aligned completely vertical~\cite{AnResults}. The results for $^{208}$Pb and $^{48}$Ca can be found in Table~\ref{tab:AnMainResults}. Of particular interest is the $^{208}$Pb result which confirms the PREX-1 measurement showing that for heavy nuclei the transverse asymmetry is highly suppressed, in stark contrast with theoretical expectations.

\begin{table}[ht]
    \centering
    \caption{Transverse asymmetry, $A_n$, measurements  for PREX-2 $^{208}$Pb and CREX $^{48}$Ca with the corresponding total uncertainties (statistical and systematic uncertainties combined in quadrature). $E_0$ is the beam energy.}

    \begin{ruledtabular}
    \begin{tabular}{@{\hspace{2em}} ccc @{\hspace{2em}}}
          $E_0$ &  \multirow{2}{*}{Target} & \multirow{2}{*}{$A_n$ (ppm)}  \\
           ${\rm (GeV)}$ & &  \\
     \hline
          0.95 & $^{208}$Pb & $\ 0.4\pm0.2$\\ 
          2.18 & $^{48}$Ca  &  $\ -9.4\pm1.1$ \\
    \end{tabular}
    \end{ruledtabular}
    \label{tab:AnMainResults} 
\end{table}

During longitudinal running, residual transverse polarization of the electron beam could lead to a false asymmetry. A polarization component in the scattering plane would produce an up-down asymmetry in the detectors. This effect was measured during the run by making use of the auxiliary $A_T$ detectors as well as monitoring of the beam energy stability.  
A polarization component perpendicular to the scattering plane would produce a left-right asymmetry. This effect largely canceled due to the use of both arms of the HRS. Checks were performed to assess the lack of symmetry between the two arms by making use of the main quartz detectors. For both PREX-2 and CREX the impact of transverse polarizations were smaller than the statistical errors so no corrections were applied (i.e. $A_T=0$ in Eq.~\ref{eq:asyCorrected}), however a systematic uncertainty was assigned based on the precision of the checks:  0.26 ppb for PREX-2, and for CREX 0.70~ppb for the vertical component and 10~ppb for the horizontal component.

\subsection{Beam Polarization}
Beam polarization is a normalization factor for the asymmetry (see Eq.~\ref{eq:asyCorrected}).   
In this section we discuss the polarization measurements using the M{\o}ller, Compton, and Mott polarimeters.

\subsubsection{M{\o}ller Polarization Results  }
\label{sec:mollerresults}
M{\o}ller polarimetry was used to determine the beam polarization to a relative precision of 0.89\% during PREX-2 and 0.85\% during CREX~\cite{NIMA2023167506}. The uncertainty was dominated by systematic uncertainties, with the largest contribution coming from our knowledge of the target polarization. The target polarization uncertainty includes both how well we know the target foil magnetization at saturation and the uncertainty in the degree of saturation; these uncertainties are estimated as 0.63\% for PREX-2 and 0.57\% for CREX. The M{\o}ller polarimetry data were taken during dedicated low-current runs, and included measurements aimed at testing our model of the spectrometer acceptance and analyzing power.

One particularly vexing issue for precision M{\o}ller polarimetry is the so-called Levchuk Effect~\cite{Levchuk:1992np}. This phenomenon modifies the effective analyzing power due to the transverse momentum imparted by the inner (and therefore unpolarized) electrons of the iron atoms.
\begin{figure}
	\includegraphics[width=0.50\textwidth]{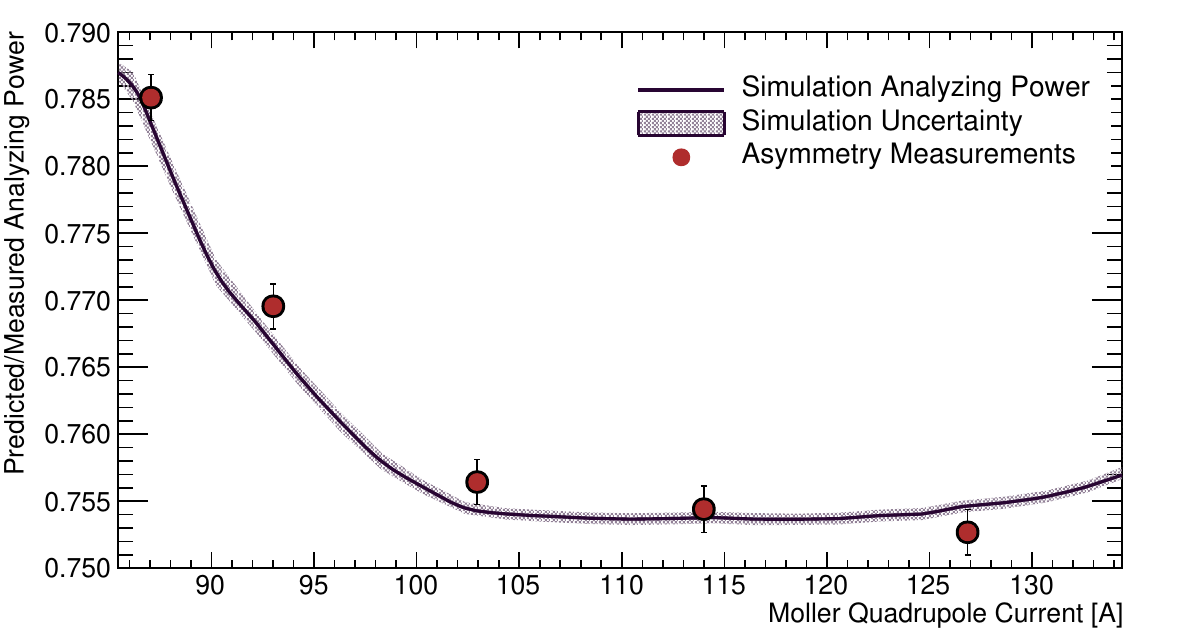}
 	\caption{An example of the data used to tune the optics of the M{\o}ller polarimeter spectrometer during CREX,  compared to expectation from simulation. 
    The acceptance-averaged analyzing power 
    $\langle A_{zz} 
    \rangle$ is plotted vs.\ the current in the first quadrupole in the polarimeter. 
    The simulation accurately accounts for the Levchuk effect by using Fermi-bound momentum distributions of target electrons from open-shell Hartree-Fock calculations for bulk iron~\cite{NIMA2023167506}.  \label{fig:MollerQuadScan}}
\end{figure}

Figure~\ref{fig:MollerQuadScan}
shows a key result from our systematic studies during CREX,  comparing the spectrometer's measured analyzing power $\langle A_{zz}\rangle$, as a function of the current in the first quadrupole of the spectrometer, with simulation. The sharp rise below 100~A is due to electrons with large transverse momentum scraping off the inner surface of the dipole, eliminating some scattering from inner target electrons. We were able to precisely simulate the effective analyzing power of the spectrometer, especially after including realistic atomic wave functions, and thus minimize the systematic uncertainty on $\langle A_{zz}\rangle$. See~\cite{NIMA2023167506} for complete details.

\begin{figure*}
	\includegraphics[width=1.0\textwidth]{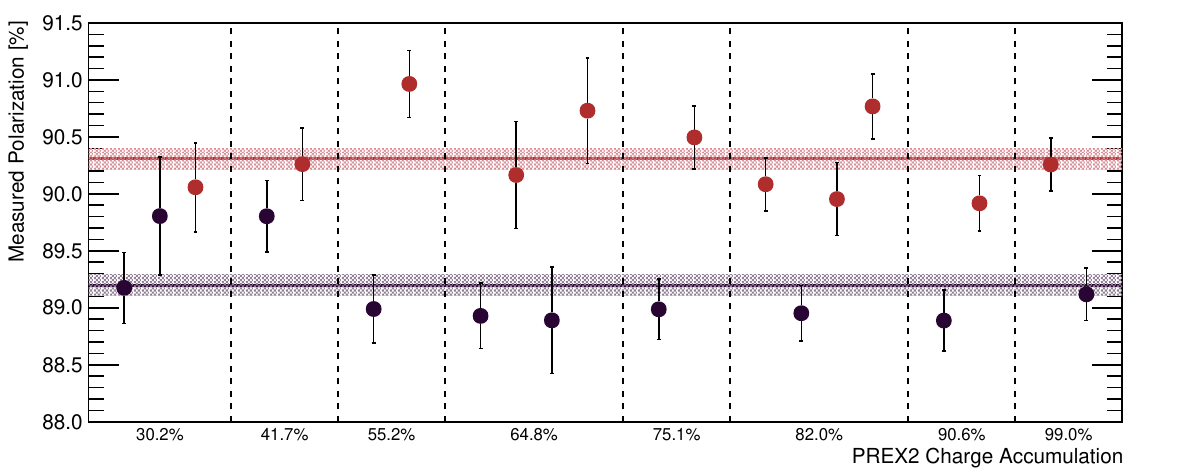}
	\caption{Beam polarization measurements using the M{\o}ller polarimeter during PREX-2, over the course of about six weeks. A clear discrepancy of $\approx1\%$ was observed between the half-wave-plate ``IN'' and ``OUT''. See text for discussion.
	\label{fig:MollerPREX}}
\end{figure*}
Figure~\ref{fig:MollerPREX} shows results from the M{\o}ller polarimeter measurements during PREX-2, with the accumlated fraction of PREX-2 data collected indicated for each measurement period. There is a clear $\approx1\%$ difference between runs taken with the insertable half-wave plate in the polarized source ``IN'' or ``OUT'' of the laser beam. This difference was later traced to be a result of the source laser system, which inadvertently created a net circular polarization difference between the two half-wave plate configurations. This configuration error was largely corrected for CREX. The net correction for beam polarization was an average over the run which accounted for the fraction of data collected in each half-wave plate configuration.

\subsubsection{Compton Polarization Results}\label{sec:comptonresults}
The technique of Compton Polarimetry was also used to determine the beam polarization for PREX and CREX. Using methods developed for the Compton polarimeter in Hall C~\cite{Narayan:2015aua}, a dramatic improvement over previous experiments in Hall A was made in the ability to determine the laser polarization within the Fabre-P\'erot optical cavity.  
While this remains the dominant uncertainty in the measurement for CREX, it has allowed a new record for electron beam polarimetry to be achieved~\cite{Zec:2024iky}.

While the Compton polarimeter was available for the full CREX run, issues with the laser system precluded use of the system until partway through the PREX-2 run period.  In addition, analysis of the PREX-2 data was complicated by an electronics issue that led to variations in the photon detector pedestal, and the interaction of the large neutron background from the Pb target caused large backgrounds in the GSO photon detector.  The combination of these factors resulted in a larger uncertainty for the PREX Compton measurements than was achieved for CREX.

Even after attempts to remove periods of unstable pedestal from the data sample for PREX, clear, significantly non-statistical behavior remained. We assigned a larger systematic error ($0.96$\%) to account for the non-statistical variation. The uncertainties for the two experiments are summarized in Table~\ref{tab:compton_systematics}.

\begin{figure}[htb]
{\includegraphics*[width=\columnwidth]{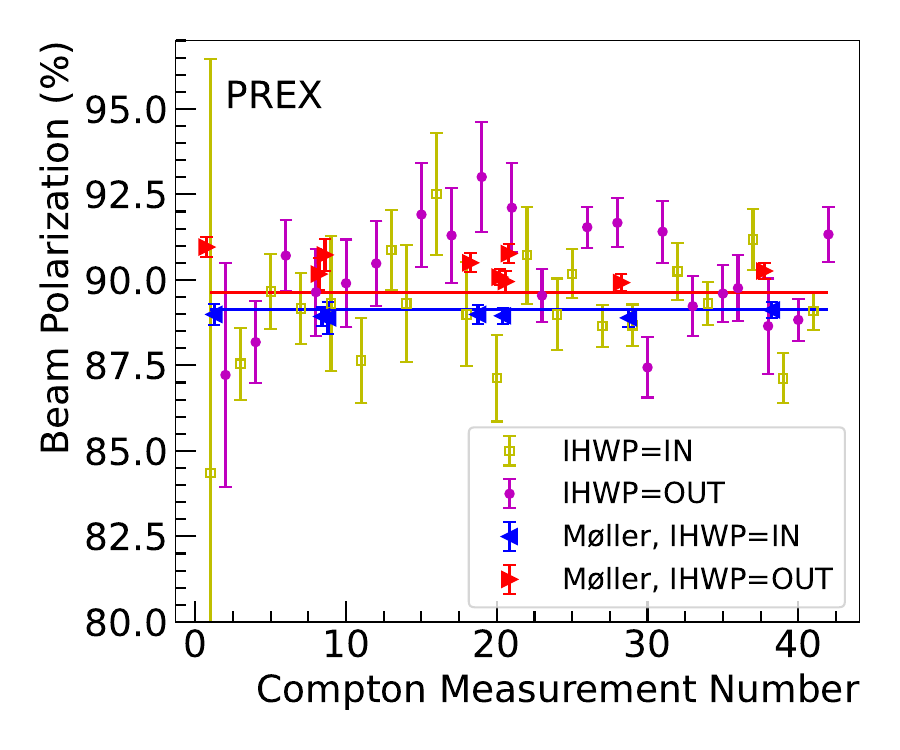}}
\caption{Measured beam polarization during the PREX-2 experiment from the Compton polarimeter, during the portion of the run that the polarimeter was available.  Uncertainties are statistical only. Polarization measurements from the M\o ller polarimeter are also shown.  Horizontal lines denote constant fits to Compton results for the PREX-2 run period.  The Compton and M\o ller results are in reasonable agreement.
\label{fig:PREX_compton_result}}
\end{figure}

\begin{figure*}[htb]
{\includegraphics*[width=\linewidth]{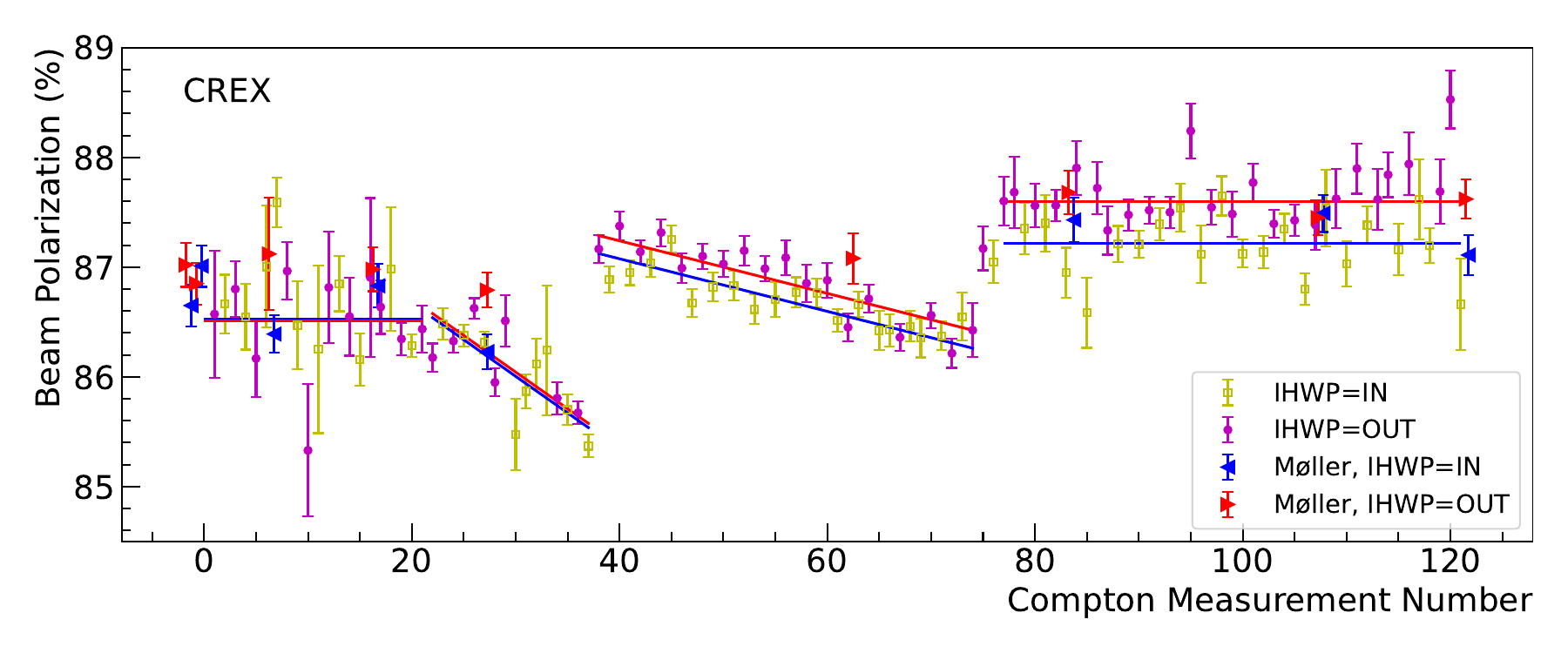}}
\caption{Measured beam polarization during the CREX experiment from the Compton polarimeter.  Uncertainties are statistical only.  The observed changes in polarization are due to changes in the quantum efficiency of the photocathode which increases when ``reactivated'' and may decay slowly with time.  The data are consistent with the fit shown (line segments) with $\chi^2$ per degree of freedom $=1.3$.  The difference in polarization between the two states of the insertable half-wave plate is due to unaccounted-for birefringence in the vacuum window in the polarized source.  Polarization measurements from the M\o ller polarimeter are also shown.  The two polarimeters are in excellent agreement.  Note that the vertical scale is 3 times smaller than in Fig.~\ref{fig:PREX_compton_result}.
\label{fig:CREX_compton_result}}
\end{figure*}

The time-dependent values of the beam polarization are shown in Figs.~\ref{fig:PREX_compton_result} and~\ref{fig:CREX_compton_result}, demonstrating excellent agreement between the two polarimeters and, in the case of CREX,  showing that the Compton polarimeter was able to track small changes due to changes in the photocathode at the beam source. The final results are averages over the polarization measurements shown in Figs.~\ref{fig:PREX_compton_result} and~\ref{fig:CREX_compton_result}, weighted by the square-inverse of the statistical uncertainty in the parity-violating asymmetry in the main measurement taken during the same time periods. The result for PREX-2 is $P_e = 89.60 \pm 0.37 \text{\,(stat)} \pm 0.99 \text{\,(syst)}\%$.  Note that this average value applies only to the latter part of the PREX run period (starting with slug 45). The result for CREX is $P_e = 86.90 \pm 0.02 \text{\,(stat)} \pm 0.31 \text{\,(syst)} \%$~\cite{Zec:2024iky}, which is the highest-precision electron beam polarimetry result we are aware of ($dP/P=0.36$\%).

\begin{figure}[htb]
	\includegraphics[width=\columnwidth]{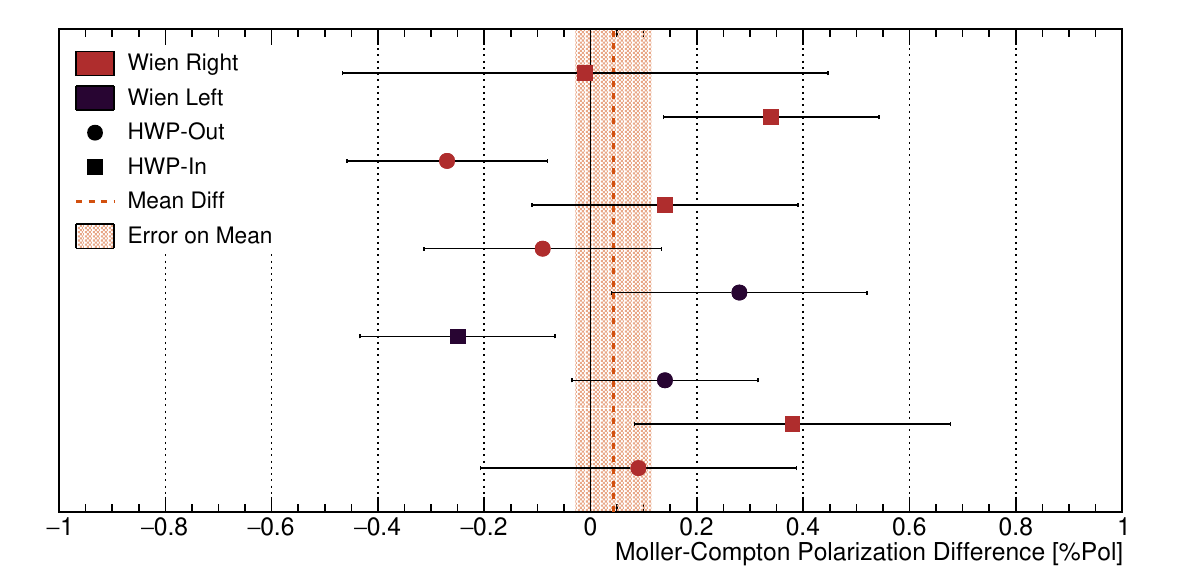}
	\caption{Comparison of beam polarization measurements between the M{\o}ller and Compton polarimeters during CREX. The Wien rotator orientation (red/black) and half-wave-plate state (square/circle) for each comparison is indicated. Error bars are statistical only. The weighted-average difference between the techniques, shown by the vertical dotted line, with uncertainty indicated by hatched region,  is seen to be statistically consistent with zero.  
	\label{fig:MollerCREX}}
\end{figure}

Contemporaneous results from the Compton and M{\o}ller polarimeters during CREX are compared in Figure~\ref{fig:MollerCREX}. Here,  M{\o}ller polarimeter results taken during dedicated low-current running are compared with Compton results taken during neighboring periods of regular high-current data taking. Figure~\ref{fig:MollerCREX} shows these 
independent 
measurements to be in excellent agreement.

\begin{table}[htb]
    \centering
    \caption{Major uncertainties for the Compton polarimetry measurements for PREX-2 and CREX~\cite{Zec:2024iky}.}
    \begin{ruledtabular}
    \begin{tabular}{l|c c} 
    & \multicolumn{2}{c}{$dP/P$(\%)}\\ 
    Source                     & PREX-2 & CREX \\ 
    \hline
    Laser polarization               & 0.38 & 0.25 \\
    Collimated spectrum distortion   & 0.30 & 0.20 \\
    Detector gain shift              & 0.22 & 0.15 \\
    Beam energy                      & 0.10 & 0.05 \\
    Helicity-state polarization difference & 0.03 & 0.03 \\
    Detector non-linearity            & 0.08 & 0.02 \\
    Pedestal fluctuations            & 0.96 & 0.00 \\
    \hline
    \textbf{Systematic}                   & \textbf{1.11} & \textbf{0.36} \\
    \textbf{Statistical}                   & \textbf{0.37} & \textbf{0.02} \\
    \hline
    \textbf{Total}                   & \textbf{1.17} & \textbf{0.36} \\
    
    \end{tabular}
    \end{ruledtabular}
    
    \label{tab:compton_systematics}
\end{table}

\subsubsection{Mott Polarization Results}

In order to measure the transverse asymmetries in various targets during PREX-2 and CREX, the beam polarization was temporarily changed from longitudinal polarization to transverse polarization. To support these transverse asymmetry measurements, the Wien spin flipper was changed from the FLIP-LEFT or FLIP-RIGHT configurations that were used to deliver longitudinally polarized beam to the Hall for the parity-violation measurements, to the VERTICAL configuration. The Mott polarimeter was used in this process. For vertical polarization, the first Wien Filter rotated the spin by 90$^\circ$ and the spin solenoid settings were set such that there is no net spin rotation, while for horizontal polarization, the spin solenoids were set for a net spin rotation of 90$^\circ$. Table~\ref{table:MottPol} summarizes the results of Mott measurements during transverse polarization running portion of the experiments.

\begin{table}[h]
    \centering
    \caption{Measurements of the horizontal ($P_x$) and vertical  ($P_y$) beam polarization components in the injector using the Mott polarimeter during 
    transverse asymmetry measurements.}
    \begin{ruledtabular}
    \begin{tabular}{@{\hspace{2em}} c  c  c @{\hspace{2em}}}
        
        Experiment  & P$_x$ (\%) & P$_y$ (\%) \\   \hline
        PREX-2  &  $0.5 \pm 0.8$ & $86.9 \pm 0.9$ \\
        CREX & $1.4 \pm 0.4$ & $87.3 \pm 0.7$ \\
        
    \end{tabular}
    \end{ruledtabular}
     \label{table:MottPol}
\end{table}

\subsection{Blinding factor}
\label{sect:blinding}    

We used a blinded analysis to avoid biases during data-taking or analysis. A numerical factor called the blinding factor, whose value is initially unknown to the experimenters, was added to all calculations of the physics asymmetry to shift it.  The maximum shift was approximately $\pm 1 \sigma$ where $\sigma$ is the anticipated statistical error of a slug, where a slug is about 1 day of running.  This shift was sufficiently small that we could still see on-line that the data were behaving in a statistical way and that the asymmetry was reversing sign with the halfwave plate reversals. The blinding factor was generated by a computer code based on a string of characters.   After all the corrections and systematic errors were evaluated and applied, the blinding factor was removed, yielding the final result for the asymmetry.  As a cross-check, the entire analysis chain was repeated without the blinding factor.

\subsection{Summary of Measurements}

\begin{figure}
	\centering
	\includegraphics[width=0.45\textwidth]{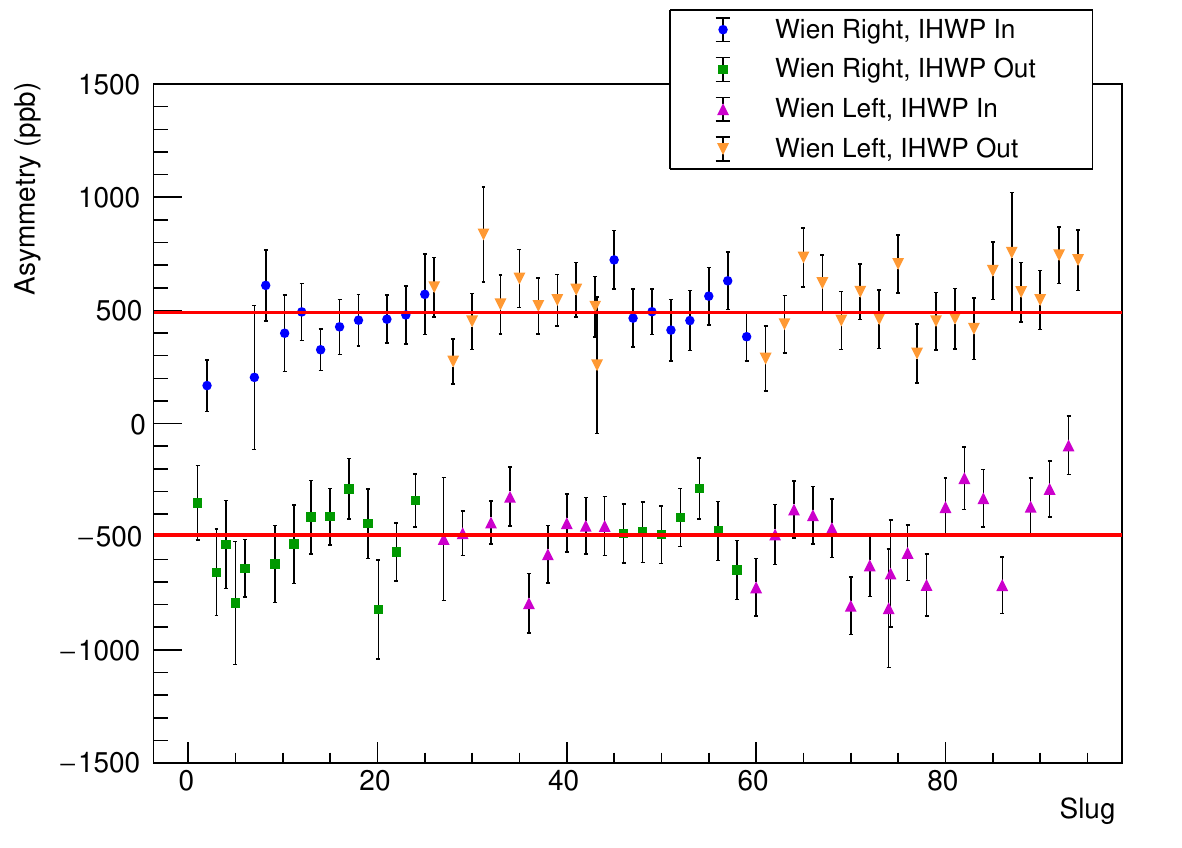}
	\caption{Unblinded slug averages of the PREX-2 slug averaged beam corrected asymmetry using Lagrange multiplier regression without sign correction, showing the effectiveness of the slow reversals.} 
	\label{fig:PREX2SlugPlot}
\end{figure}

\begin{figure}
	\centering
	\includegraphics[width=0.45\textwidth]{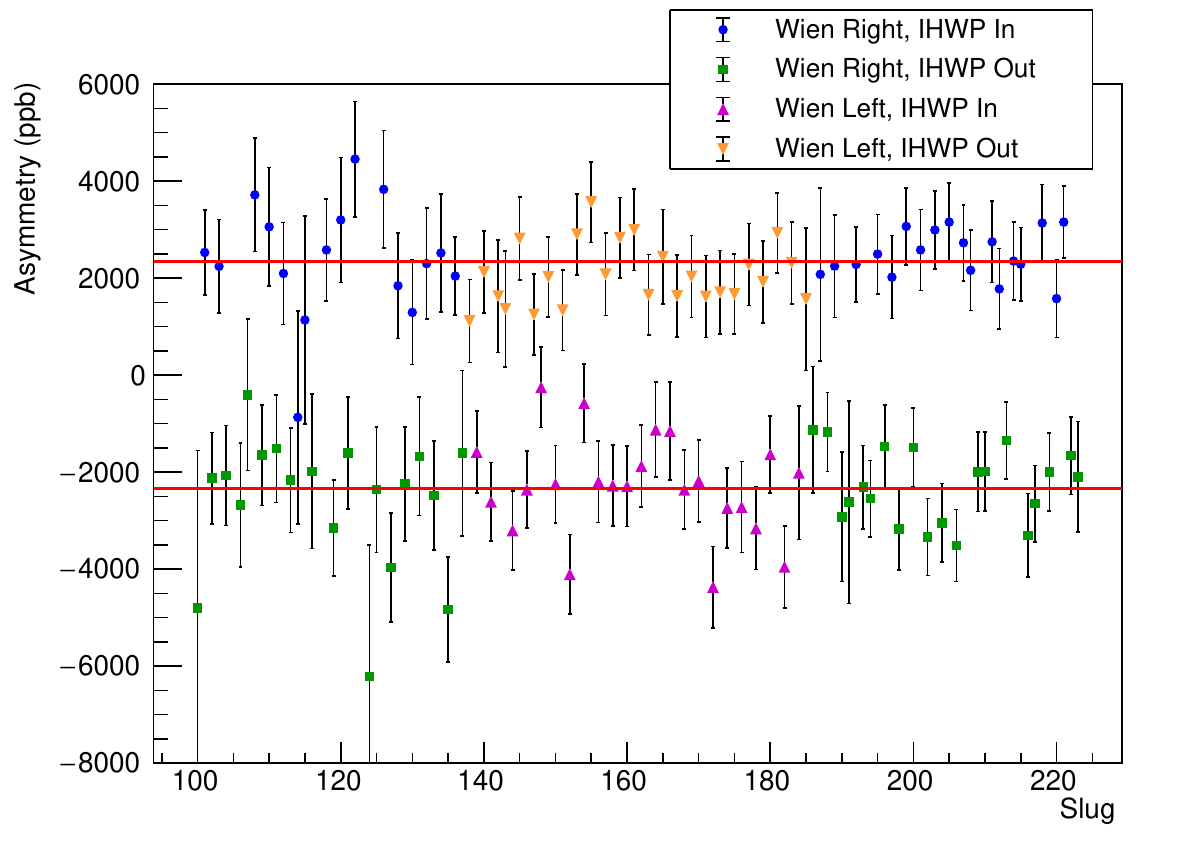}
	\caption{Unblinded slug averages of the CREX slug averaged beam corrected asymmetry using Lagrange multiplier regression without sign correction, showing the effectiveness of the slow reversals.}
	\label{fig:CREXSlugPlot}
\end{figure}

The PREX-2 and CREX final, unblinded asymmetries with beam corrections applied were computed using Eq.~\ref{eq:asyCorrected}, with $A_{\rm corr}$ computed from $A_{\rm raw}$. The asymmetries averaged over slugs are shown with the slow-control sign reversals due to the IHWP and double-Wien left uncorrected in Figs \ref{fig:PREX2SlugPlot} and \ref{fig:CREXSlugPlot}. The slug-averaging periods across each experiment for the four permutations of sign reversal are shown in Table~\ref{tab:Final_Stats_results}. PREX-2 had a $\chi^2$/NDF = 117.5/95 and CREX had  $\chi^2$/NDF = 95.2/120 when averaged this way. Given the probabilities for these $\chi^2$/NDF, the behavior of the final asymmetry data for each slow control state across the two experiments are consistent with the expectation of statistical behavior.  Furthermore, the asymmetry width explained in Sec.~\ref{sec:detectors} and Fig.~\ref{fig:prex_correction_histogram} had shown that the statistical error was largely consistent with counting statistics for the counting of electrons in the detectors.  Finally, we note that the statistical error is larger than the systematic error. Tables~\ref{tab:ApvPREXfinal} (PREX-2) and ~\ref{tab:ApvCREXfinal} (CREX) show all the significant corrections and systematic uncertainties to extract A$^{\rm meas}_{\rm PV}$ listed on the bottom row of the table. The final statistical and systematic uncertainties are also shown. It should be noted that the correction asymmetries in Table~\ref{tab:ApvPREXfinal} and ~\ref{tab:ApvCREXfinal} take polarization and background fractions into account while they are reported without normalization in ~\cite{ref:PrexII}.

\begin{table}
  \centering

  \caption{PREX-2 and CREX experiment slug-averaged quantities for each slow-reversal data set.  A slug is about one day of running. The quantities shown are the average asymmetry, the standard deviation in the asymmetry, the $\chi^2$, the number of degrees of freedom (NDF), and the probability (p-value) for this  $\chi^2$/NDF.} 
  
  \label{tab:Final_Stats_results}
  \begin{ruledtabular}
  \begin{tabular}{c c|c c c c c}
  
    Wien & IHWP  & $ \langle A \rangle$ & $ \sigma_{\langle A \rangle} $  & $\chi^2$ & NDF & p-value \\
    \hline \hline
    \multicolumn{2}{r|}{PREX-2 slugs} & 492.6 & 13.5 & 117.5 & 95 & 0.06 \\ \hline
     Right &  In  & 455.8 & 28.7 & 18.3 & 18 & 0.44 \\
     Left  &  Out & 533.5 & 24.5 & 31.6 & 28 & 0.29 \\
     Right &  Out & 481.2 & 31.7 & 16.0 & 20 & 0.72 \\
     Left  &  In  & 484.8 & 25.1 & 46.9 & 26 & 0.01 \\

    \hline 
    \multicolumn{2}{r|}{CREX slugs} & 2336.0 & 83.8 & 95.2 & 120 & 0.95 \\ \hline
     Right &  In  & 2517.5 & 160.5 & 16.5 & 33 & 0.99 \\
     Left  &  Out & 2108.6 & 174.8 & 13.2 & 24 & 0.96 \\
     Right &  Out & 2347.7 & 159.4 & 29.8 & 38 & 0.83 \\
     Left  &  In  & 2334.0 & 177.8 & 32.8 & 22 & 0.06 \\

  \end{tabular}
  \end{ruledtabular}
\end{table}

\begin{table}[h]
\centering
\caption{PREX-2: Significant corrections and systematic uncertainties to extract the asymmetry listed on the bottom row with its statistical uncertainty. The correction asymmetries are normalized to account for polarization and background fractions.}
\label{tab:ApvPREXfinal}
\begin{ruledtabular}
\begin{tabular}{c|cc}
\multicolumn{1}{c|}{Contribution}   & Absolute {[}ppb{]} & Relative {[}\%{]}  \rule{0pt}{2.4ex}  \rule[-1.2ex]{0pt}{0pt}  \\ \hline
Beam Polarization               
& $60.6 \pm 5.2 $                
& $11.0 \pm 1.0$\%           \rule{0pt}{2.4ex}  \rule[-1.2ex]{0pt}{0pt}  \\ 
Beam Trajectory and Energy                
& $-71.9 \pm 3.0$
& $13.1 \pm 0.5$\%          \rule{0pt}{2.4ex}  \rule[-1.2ex]{0pt}{0pt}   \\ 
Charge Correction           
& $24.6 \pm 0.2$                    
& $4.5 \pm 0$\%           
\rule{0pt}{2.4ex}  \rule[-1.2ex]{0pt}{0pt}   \\ 
Target Diamond Foils            
& $0.7 \pm 1.4$                
& $0.1 \pm 0.3$\%         \rule{0pt}{2.4ex}  \rule[-1.2ex]{0pt}{0pt}    \\ 
Spectrometer Rescattering       & $0.0 \pm 0.1$                & $0.0 \pm 0.0$\%         \rule{0pt}{2.4ex}  \rule[-1.2ex]{0pt}{0pt}    \\ 
Inelastic Contributions     
& $0.0 \pm 0.1$                 & $0.0 \pm 0.0$\%         \rule{0pt}{2.4ex}  \rule[-1.2ex]{0pt}{0pt}    \\ 
Transverse Asymmetry            & $0.0 \pm 0.3$                 & $0.0 \pm 0.1$\%         \rule{0pt}{2.4ex}  \rule[-1.2ex]{0pt}{0pt}    \\ 
Detector non-linearity           & $0.0 \pm 2.7$                & $0.0 \pm 0.5$\%         \rule{0pt}{2.4ex}  \rule[-1.2ex]{0pt}{0pt}    \\ 
Angle Determination      
& $0.0 \pm 3.5$                 & $0.0 \pm 0.6$\%          \rule{0pt}{2.4ex}  \rule[-1.2ex]{0pt}{0pt}   \\ 
Acceptance Function      
& $0.0 \pm 2.9$                 & $0.0 \pm 0.5$\%          \rule{0pt}{2.4ex}  \rule[-1.2ex]{0pt}{0pt}   \\ \hline
Total Systematic Uncertainty
& $\pm 8.2$                
& $\pm 1.5$\%         \rule{0pt}{2.4ex}  \rule[-1.2ex]{0pt}{0pt}    \\  
PREX-2 \hskip 0.05in ${\rm A}_{\rm PV}^{\rm meas}$ 
\hskip 0.05in with stat. err.
& $550 \pm 16$  &
$\pm 2.9$\%         \rule{0pt}{2.4ex}  \rule[-1.2ex]{0pt}{0pt}    \\ 
\end{tabular}
\end{ruledtabular}
\end{table}

\begin{table}[h]
\centering
\caption{CREX: Significant corrections and systematic uncertainties to extract the asymmetry listed on the bottom row with its statistical uncertainty. The correction asymmetries are normalized to account for polarization and background fractions.}
\label{tab:ApvCREXfinal}
\begin{ruledtabular}
\begin{tabular}{c|cc}
\multicolumn{1}{c|}{Contribution}   & Absolute {[}ppb{]} & Relative {[}\%{]}  \rule{0pt}{2.4ex}  \rule[-1.2ex]{0pt}{0pt}  \\ \hline
Beam Polarization               
& $382 \pm 13 $                
& $14.3 \pm 0.5$\%           \rule{0pt}{2.4ex}  \rule[-1.2ex]{0pt}{0pt}  \\ 
Beam Trajectory and Energy                 
& $68 \pm 7$
& $2.5 \pm 0.3$\%          \rule{0pt}{2.4ex}  \rule[-1.2ex]{0pt}{0pt}   \\ 
Charge Correction           
& $112 \pm 1$                    
& $4.2 \pm 0$\%           
\rule{0pt}{2.4ex}  \rule[-1.2ex]{0pt}{0pt}   \\ 
Isotope Purity            
& $19 \pm 3$                
& $0.7 \pm 0.1$\%         \rule{0pt}{2.4ex}  \rule[-1.2ex]{0pt}{0pt}    \\ 
3.831 MeV ($2^{+}$) inelastic       & $-35 \pm 19$ 
& $-1.3 \pm 0.7$\%         \rule{0pt}{2.4ex}  \rule[-1.2ex]{0pt}{0pt}    \\
4.507 MeV ($3^{-}$) inelastic       & $0 \pm 10$ 
& $0 \pm 0.4$\%         \rule{0pt}{2.4ex}  \rule[-1.2ex]{0pt}{0pt}    \\
5.37 MeV ($3^{-}$) inelastic       & $-2 \pm 4$ 
& $-0.1 \pm 0.1$\%         \rule{0pt}{2.4ex}  \rule[-1.2ex]{0pt}{0pt}    \\
Transverse Asymmetry            & $0.0 \pm 13$                 & $0.0 \pm 0.5$\%         \rule{0pt}{2.4ex}  \rule[-1.2ex]{0pt}{0pt}    \\ 
Detector non-linearity           & $0.0 \pm 7$                & $0.0 \pm 0.3$\%         \rule{0pt}{2.4ex}  \rule[-1.2ex]{0pt}{0pt}    \\  
Acceptance Function      
& $0.0 \pm 24$                 & $0.0 \pm 0.9$\%          \rule{0pt}{2.4ex}  \rule[-1.2ex]{0pt}{0pt}   \\ 

Radiative corrections ($Q_{\rm W})$      
& $0.0 \pm 10$                 & $0.0 \pm 0.4$\%          \rule{0pt}{2.4ex}  \rule[-1.2ex]{0pt}{0pt}   \\
\hline

Total Systematic Uncertainty
& $\pm 40$                
& $\pm 1.5$\%         \rule{0pt}{2.4ex}  \rule[-1.2ex]{0pt}{0pt}    \\
CREX \hskip 0.05in ${\rm A}_{\rm PV}^{\rm meas}$
\hskip 0.05in with stat. err.
& $2668 \pm 106$  & $\pm 4.0$\%         \rule{0pt}{2.4ex}  \rule[-1.2ex]{0pt}{0pt}    \\  
\end{tabular}
\end{ruledtabular}
\end{table}

%% file: crex_acceptance_function_new.tex
{

Scattered electrons were detected incident on the small quartz detector at the spectrometer focus. Before reaching the detectors two elements of the spectrometer provided collimation for the scattered flux: the vacuum enclosure channel of the septum magnet, and a lead collimator at the entrance to the first quadrupole magnet (Q1) of the HRS. The experimental design aimed to minimize the loss of full-energy electron flux at any location other than the lead collimator and the edge of the quartz detector in the dispersive direction in the focal plane. Scattered electrons with momenta less than about 9\% of the elastic peak are lost at the Q1 collimator, while the quartz is positioned with the detector edge corresponding to $dp/p\approx 0.2\%$ below the elastic peak. The angular acceptance at the collimator ranged approximately from $4^{\circ}$--6$^{\circ}$ in scattering angle.  

In order to guide the comparison of theoretical calculation with the experimental kinematics, an acceptance function $\epsilon({\theta})$ has been generated using a {\sc geant4} Monte Carlo simulation (see Sec.~\ref{sec:simulation}). This function provides the relative acceptance as a function of lab-frame polar scattering angle $\theta$:
\begin{equation}
    \epsilon(\theta) = \frac{1}{\epsilon_0} \frac{\frac{dN_{\text{det}}(\theta)}{d\theta}}{\frac{dN_{\text{sca}}(\theta)}{d\theta}} 
    \label{eq:acceptance_definition}
\end{equation}
with $dN_{\rm sca}/ d\theta$ the distribution of events over all generated scattering angles,  $dN_{\rm det}/ d\theta$ the distribution of detected events over scattering angle, and $\epsilon_0$ a normalizing factor such that 
\begin{equation}
    \int \epsilon(\theta) \, \sin\theta \, d\theta = 1.
    \label{eq:acceptance_norm}
\end{equation}

Theoretical modeling typically produces an analyzing power for scattering in an idealized case of an electron beam of fixed energy interacting with a nucleus to scatter with energy and angle determined by the elastic kinematics.  
The measured asymmetry includes the effects of interaction of electrons in the target pre and post scattering vertex, with multiple-scattering as well as radiative and ionization energy loss. 
Besides these effects, the simulation also accounts for internal Bremstrahlung using an effective radiation length. The acceptance function derived from simulation represents the kinematics of the scattering vertex for each accepted event. The effect of momentum spread due to radiation is extremely small due to the small momentum acceptance of the integrating detector.  The associated systematic uncertainties on $A_{PV}$ due to acceptance function calibration and modeling (see Tab.~\ref{tab:ApvPREXfinal} and ~\ref{tab:ApvCREXfinal}) are calculated using a representative model (FSUGold~\cite{Fattoyev:2010mx}) of the $^{208}$Pb or $^{48}$Ca nucleus.

The measured $A_{PV}$ can be calculated as the convolution of the theoretical analyzing power with the accepted rate distribution, averaged over the accepted rate distribution: 
\begin{equation}
    A_{\text{mea}} = \frac{\int d\theta \sin\theta A(\theta) \frac{d\sigma}{d\Omega} \epsilon(\theta)}{\int d\theta \sin\theta \frac{d\sigma}{d\Omega} \epsilon(\theta)} .
    \label{eq:acceptance_function}
\end{equation}

The acceptance function is calibrated to rate distributions measured in the tracking system of the HRS (see~\ref{sec:detectors} for details) for data collected at very low beam current ($<100$~nA) to enable reconstruction of individual electron tracks without excessive distortion due to high rates. The methods used to calibrate the spectrometer optics and determine the acceptance function are described below.

\subsubsection{Spectrometer Optics Calibration} \label{sec:opticscal}
Due the addition of the septum magnet and the optimization for excluding background and capturing elastic kinematics in the small quartz detector, the spectrometer optics tunes used for PREX-2 and CREX were significantly different than the standard HRS tune~\cite{HallA_NIM}. This optimization required a tight focus of the elastic peak in the non-dispersive focal plane direction while maintaining a large ($\approx 14$~m) dispersion to separate the elastic peak from the inelastic nuclear excitation levels. Magnetic models of the septum and HRS magnets were used for particle tracking through the spectrometer to design an initial configuration, but significant empirical tuning was needed to obtain the required optics performance due to imperfections in descriptions of the magnetic elements. 

Calibration of the spectrometer optics was performed using an insertable sieve collimator (see Fig.~\ref{sievedrawing} for RHRS sieve). The collimators were 5~mm thick Tungsten plates with a precision-machined pattern of holes, placed  1.1~m downstream of the target to create ray packets of well-defined scattering angle for optics reconstructions. Data was collected with a thin graphite foil and unrastered beam. Separate runs were collected with the beam position at different transverse locations on the target to cover the rastered beam range during production data taking, and with the spectrometer magnets set to scale all fields in three steps of 1\% to examine the momentum dependence. 

Track information of the scattered electrons, consisting of positions ($x_{fp}, ~y_{fp}$) and angles ($\theta_{fp}, ~ \phi_{fp}$) in the dispersive and non-dispersive axes of the focal plane coordinate system, was obtained using straight-line track-fitting of hits on the two drift chambers~\cite{HallA_NIM}.
These coordinates are used to calculate the parameters of the initial ray $y_{tg}, \phi_{tg}, \theta_{tg}, \delta$ corresponding to the horizontal position and angle, vertical angle, and momentum offset $\delta p/p$ relative to the central optics ray. 
Elastically scattered tracks through each hole in the sieve collimator correspond to specific values of ($y_{tg}, ~\phi_{tg}, ~\theta_{tg}, ~\delta$), which depend on the transverse beam location.  
The momentum offset was calibrated using the momentum shift between the $^{12}$C elastic and first excited state peaks.  
 
The expansion of the focal plane coordinates is performed up to fifth order, constrained by the sieve collimator calibration data. Not all elements are used in the fifth order expansion, with terms pruned to avoid overfitting~\cite{HallA_NIM}.   Figure~\ref{fig:sieveplot} shows a representative plot of the reconstructed sieve pattern (two-dimensional of $\phi_{tg}$ vs. $\theta_{tg}$).  The reconstructed center of each ray packet was compared to known relative positions of the sieve holes, and residuals used to minimize the fit.  
 
A systematic imperfection remained, in which a shift in the beam position would lead to a non-zero average of residuals in the horizontal angle $\phi_{tg}$. This corresponds to an insufficient $y_{tg}$ calibration, as follows.  For a fixed angle $\phi_{tg}$, an offset in $y_{tg}$ would correspond to a shift in the scattering vertex along the beam direction. For an ideal optics reconstruction accounting for  $y_{tg}$ dependence,  a scattered track from a vertex which might be transversely offset in the horizontal direction and longitudinally offset in the beam direction would be successfully reconstructed. The observed systematic residuals for optics data taken with the beam transversely offset implies that a target shifted along the beamline will also be imperfectly reconstructed. An offset along the beam direction between production targets and the optics target required an additional correction, which was applied during the calculation of the acceptance function as described below.   

\begin{figure}[!h]
    \includegraphics[width=0.3\linewidth]{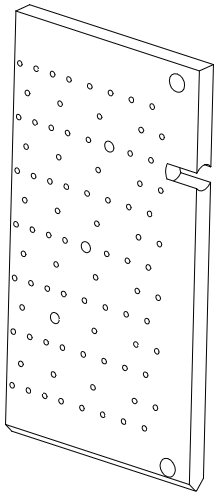}
    \caption{Sieve collimator for the RHRS. The angle between the target and the bigger hole in the middle of the sieve defines the central optics ray of the spectrometer. A similar sieve was in place for the LHRS.}
    \label{sievedrawing}
\end{figure}

\subsubsection{Central Scattering Angle Calibration}
The optics calibration provides for reconstruction of the target track parameters relative to the central optics ray. The angle of this ray was determined by using nuclear recoil in order to accurately and precisely obtain this critical calibration. A water target with stainless steel walls was employed, to allow the simultaneous measurement of elastic scattering from hydrogen and oxygen nuclei.  The sieve collimator was used so that each observed track had a well defined angle relative to the optic axis. As with the optics calibration, several different horizontal beam positions were used to vary the scattering angle through each sieve hole in a precise way and several momentum settings were used to check consistency. The calibration was performed using data collected at the beginning and end of the CREX run period, where the higher beam energy compared to PREX-2 provided a more favorable cross-section ratio between the hydrogen and oxygen elastic scattering. 
Given that both PREX-2 and CREX had the same experimental setup (i.e. the spectrometers were not moved between the two sets of measurements), the measurement at the higher energy applies for both. 
Additional data taken at the lower beam energy of PREX-2 was cross-checked for consistency.
Figure~\ref{fig:pointing}  shows a typical momentum spectrum obtained using the water target for CREX.  
Only the hydrogen and oxygen peaks were used for fitting the central scattering angle. The extracted central angles, after correction for a small difference in location along the beam axis of the water and $^{12}$C optics target,  
are in agreement with the survey values. The result 
had an uncertainty of $\pm$~0.02$^{\circ}$ which accounted primarily for the variation in results over all sieve holes, beam positions, and spectrometer momentum settings.

\begin{figure}
  \centering
  \includegraphics[width=1.0\linewidth]{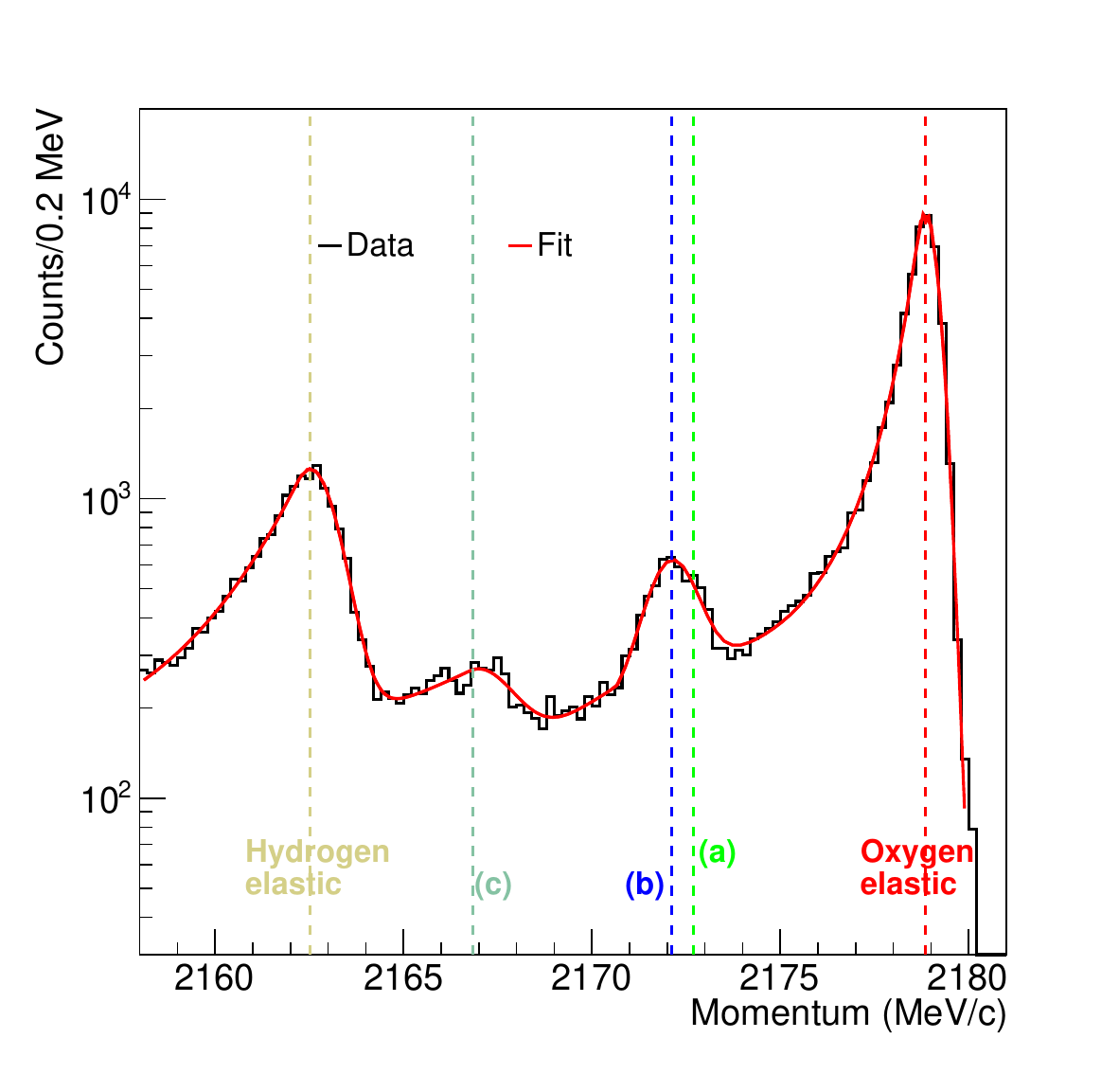}
  \caption{Momentum distribution of scattered electrons on the LHRS focal plane from the water target with E$_{\rm beam}$=2.18 GeV. The fitting was performed using a combination of Crystal Ball
  and Gaussian functions including the (a) first (6.13 MeV), (b) second (6.92 MeV) excited peaks, and (c) a peak around 11 MeV (where O$_2$ has a couple of excited states)~\cite{nndc}. \label{fig:pointing}}
\end{figure}

\begin{figure}[!h]
    \includegraphics[width=1.0\linewidth]{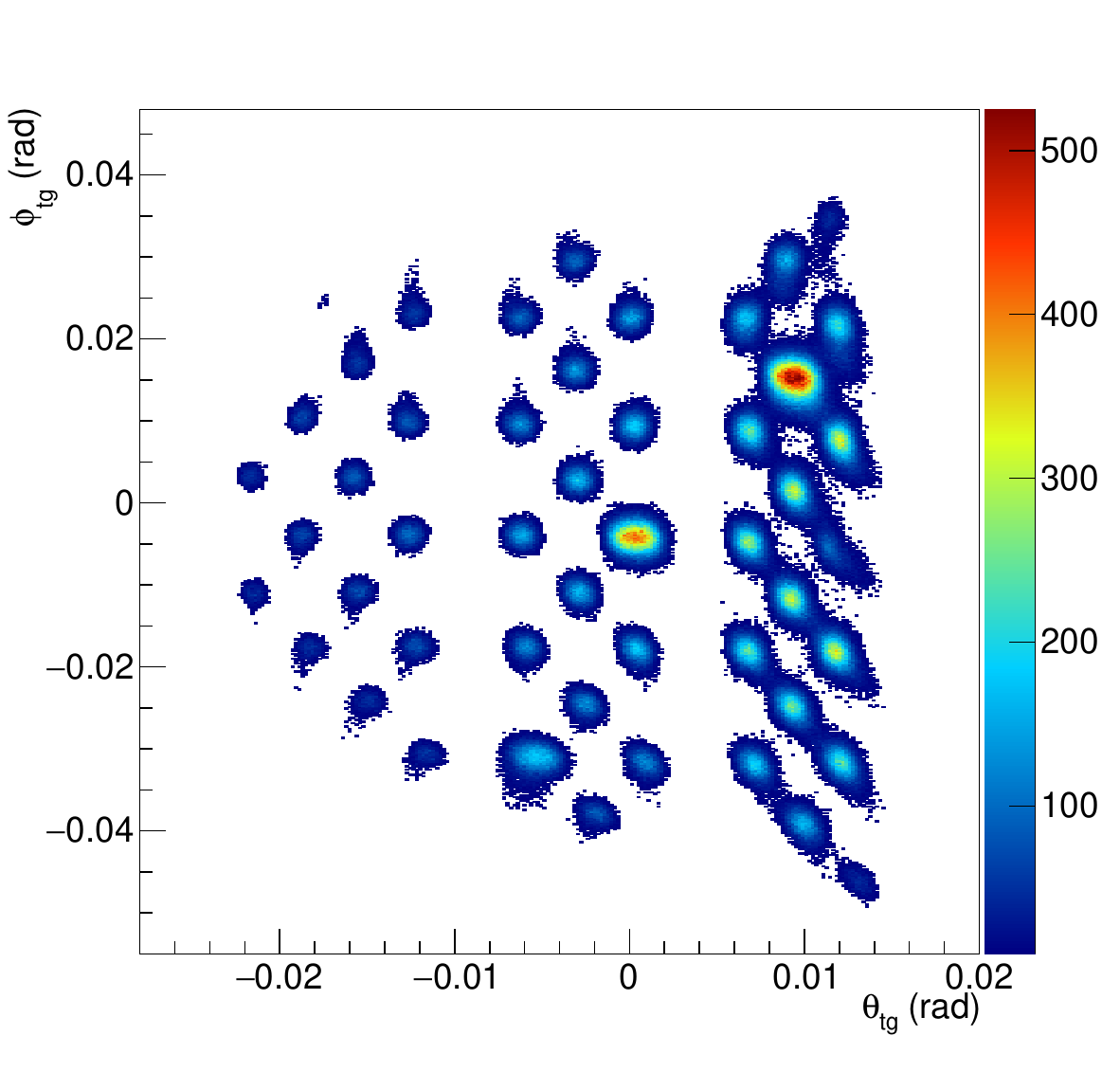}
    \caption{Representative plot of the reconstructed vertex angles of the events going through the RHRS sieve collimator during the CREX experiment. See text for more information.}
    \label{fig:sieveplot}
\end{figure}

\subsubsection{Matching Acceptance Function to Observed Distribution}

The acceptance was calibrated with data collected using production lead and calcium targets, with no sieve and very low beam current ($\approx$10-30~nA for PREX-2, 100-200~nA for CREX). 
For this calibration, the quartz detector was operated with increased gain, to allow individual pulses to be discriminated. Large trigger scintillators covered the full spectrometer acceptance, and tracks were measured in the VDCs, as described in Sec.~\ref{sec:CountingDAQ}. Measurements, about 50 in all, were made at various times throughout data collection from each target, with raster size and beam position matching production running conditions. 
These runs were taken with low currents, ensuring that the VDCs were operating without measurable rate dependence.
The reconstructed distributions, with cuts on a signal in the quartz detector, were used as the benchmark for the distribution of accepted tracks.  

These distributions were compared with results from a {\sc geant4}-based simulation package. Electrons generated in elastic kinematics were tracked through realistic magnetic geometries for both the septum and HRS magnets, taking into account initial and final state radiation losses and multiple scattering inside the target. A virtual detector representing the quartz detector in the spectrometer focal plane defined accepted tracks in the simulation. The geometry of the acceptance defining collimator was applied as a cut on simulated tracks, as were locations in the septum vacuum chamber where the chamber walls were closest to interference with accepted tracks. Scans were done on these interference positions to account for possible mis-positioning of the vacuum-loaded chamber, as well as up to few percent variations in the scale of the septum field, which accounted for the primary sources of uncertainty in the spectrometer acceptance.  The simulated accepted distributions agreed well with the reconstructed $\theta_{tg}$ and $\phi_{tg}$ distributions from the measurements with the production targets. 
The simulation examined distributions of the polar scattering angle $\theta_{lab}$ and $Q^2$, and also calculated $A_{PV}$ using the nuclear structure parameterization FSUGold~\cite{Fattoyev:2010mx}. Simulation runs in which the rate-weighted average of $\langle \theta_{lab}\rangle$ agreed within 0.2\% with the measured average from counting data were considered to match the measured distributions. 
Such matching simulations included a wide range of variations of the simulation geometry, but in each case the acceptance distribution agreed well enough that the average of $A_{PV}$ evaluated with FSU-Gold was consistent. The variations included changes larger than the construction tolerances of the different experimental collimation elements.

In order to examine the importance of a precise match of this acceptance distribution, the average rate-weighted $\langle A_{PV} \rangle$ was compared to that of a simple constant acceptance constructed such that the average and RMS of the angle distribution matched the measured acceptance. For this toy-model acceptance, the average PREX-2 $\langle A_{PV} \rangle$  matched within 0.4\%. In this way, the result is seen to be quite insensitive to the detailed shape of the acceptance function. 

The aforementioned mis-calibration of $y_{tg}$ (end of ~\ref{sec:opticscal}) requires a systematic correction to the reconstructed scattering angle. This was incorporated into the simulation, for which the systematic mis-reconstruction was added to each simulated track before comparison to the measured distribution. For CREX, the resulting correction of -0.7\% on the average scattering angle was applied. 
The same shift on average scattering angle of -0.7\% in PREX-2, evaluated using the FSUGold model, implies a shift of $\delta A_{PV}/A_{PV}\approx -0.9\%$.  
This correction, which is smaller than the quoted systematic uncertainty and significantly smaller than the statistical uncertainty $\delta A_{PV}/A_{PV} \pm 3\% ~\left( \mbox{stat} \right) \pm 1.5\% ~ \left( \mbox{syst} \right)$, has not been applied  to the published PREX-2 result.

Using the selected optimal models, the acceptance functions for each HRS were calculated using Eq.~\ref{eq:acceptance_definition}. The acceptance function for PREX-2 and CREX (averaged over the two HRSs) are shown in Fig.~\ref{fig:acceptance_function}~\cite{ref:PrexII,ref:crex}.  The acceptance functions for PREX-2 shows the greater effect of multiple scattering in passage of the lower energy beam electrons through the $^{208}$Pb targets.  A systematic uncertainty is quoted for the acceptance function for each experiment, which accounts for the possible range  of $\langle A_{PV} \rangle$ due to the precision of central angle calibration, variations in the simulation model parameters, and the range of variation in results from the acceptance calibration data collected during each run.  The FSUGold model, and a model based on FSUGold but with a 1\% increase in the neutron skin, were used to determine the sensitivity of $\langle A_{PV} \rangle$ to each of these sources of uncertainty.

\begin{figure}[!h]
    \centering
    \includegraphics[width=1.0\linewidth]{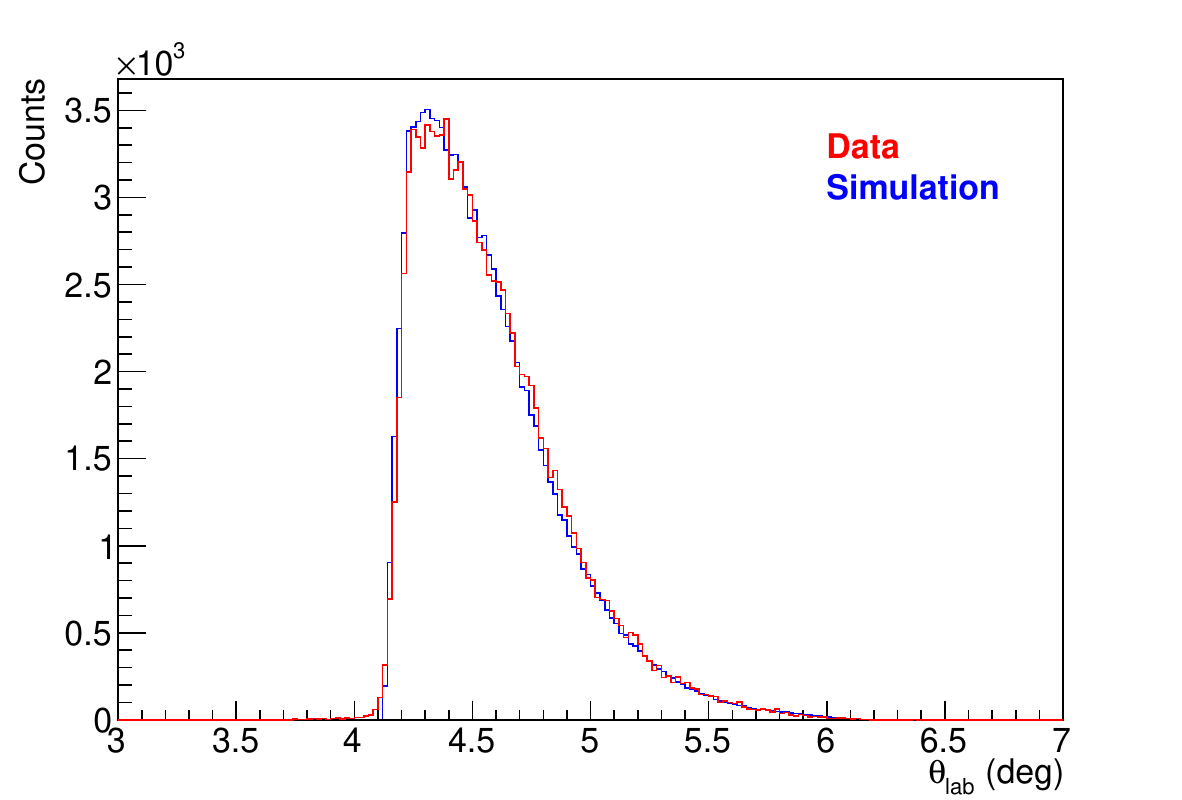}
    \caption{Comparison of $\theta_{\rm lab}$ between the optimal model and data for LHRS for the CREX kinematics. The red line is data, while the blue line is from simulation.}
    \label{fig:thetalab}
\end{figure}

\begin{figure}
\centering
    \includegraphics[width=1.0\linewidth]{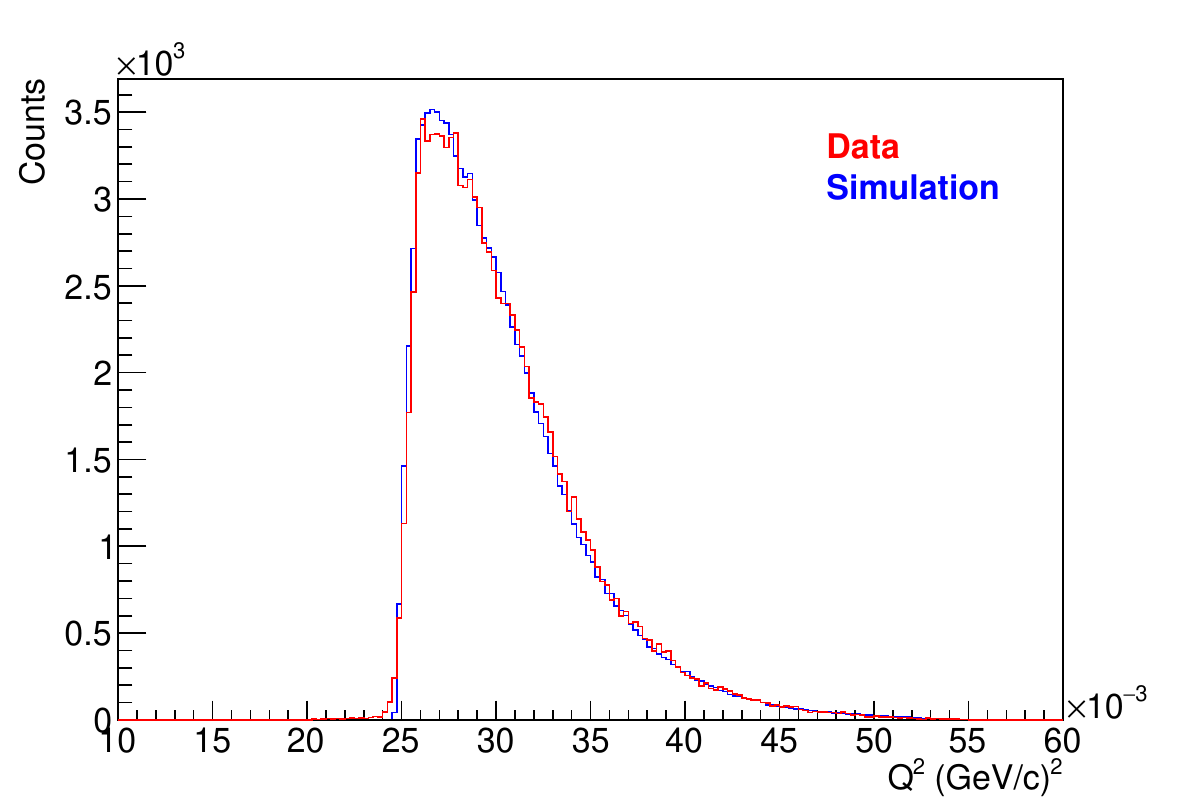}
    \caption{Same as Fig.~\ref{fig:thetalab}, but for $Q^2$.}
    \label{fig:Q2}
\end{figure}

\begin{figure}[!h]
    \centering
    \includegraphics[width=1.0\linewidth]{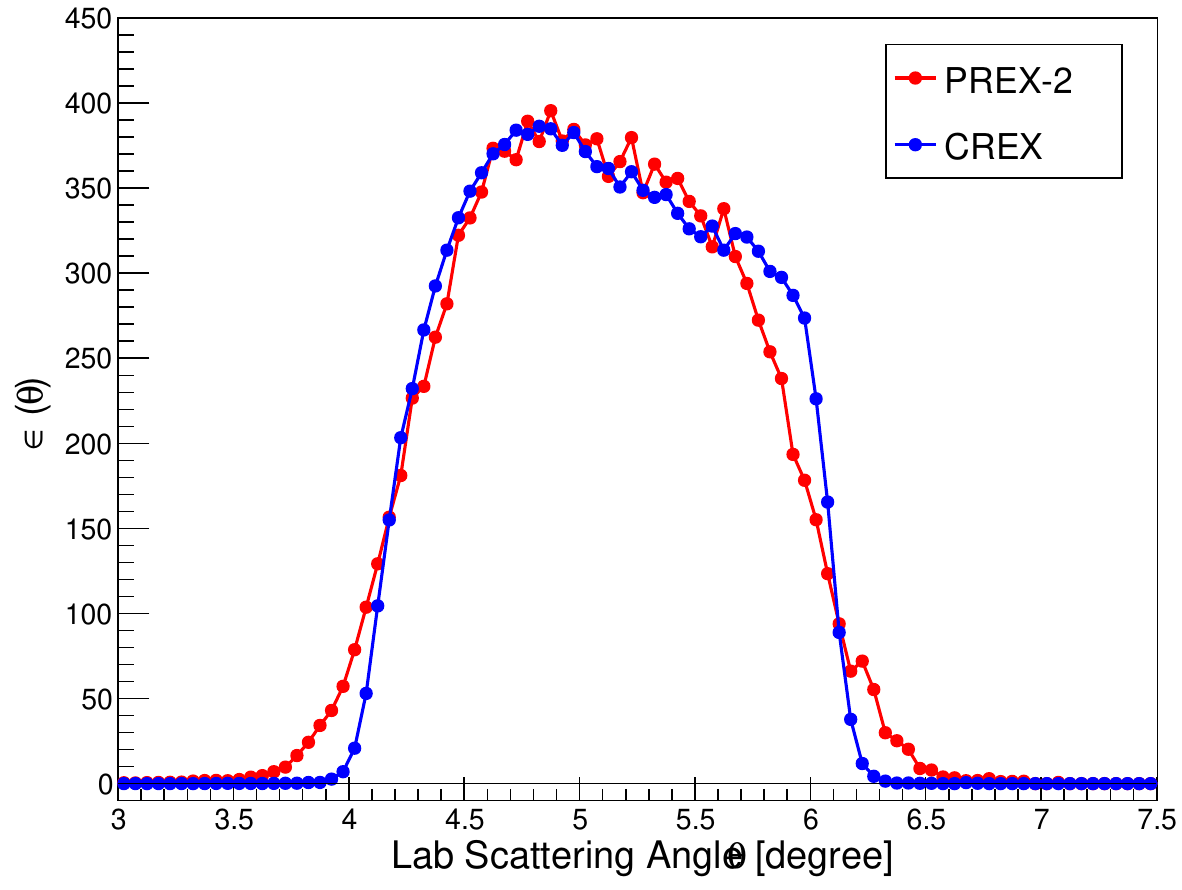}
    \caption{PREX-2 and CREX acceptance function $\epsilon(\theta)$.}
    \label{fig:acceptance_function}
\end{figure}

}

%% file: z4_resultsV2.tex
\section{RESULTS}
\label{sec:results}
The main experimental result of these experiments are the weak form factors for $^{208}$Pb and $^{48}$Ca. As discussed in Sec.~\ref{part:overview} these form factors can be extracted unambiguously from the experimentally obtained parity-violating asymmetries.

The parity-violating asymmetries for PREX-2 and CREX are provided in Table~\ref{tbl:pvasym}.  Comparisons to PREX-1 \cite{ref:prexI} are made in Sec.~\ref{sec:NeutronRadius}.  The previous sections described the experimental data selection and correction procedures, and the radiative and background corrections. 

\begin{table}[!h]
    \centering
    \caption{The parity-violating asymmetries determined in the PREX-2 and CREX experiments.}
    \begin{ruledtabular}
    \begin{tabular}{c c c c  c c c }
        
        Target & $E_0$ & $\left<Q^2\right>$  & $\left<\theta\right>$  &  $A_{PV}$ $\pm$ stat. $\pm$ sys. \\ 
         & (MeV) & (GeV$^2/c^2$) & (degrees) & (ppb) \\
        \hline
        $^{208}$Pb & 953 & 0.00616 & 4.7 & 550 $\pm$ 16 $\pm$ 8 \\
        $^{48}$Ca  & 2118 & 0.0297 & 4.5 & 2668 $\pm$  106 $\pm$ 40 \\
        
    \end{tabular}
    \end{ruledtabular}
    \label{tbl:pvasym}
    \end{table}

\begin{table}[]
    \centering
    \begin{ruledtabular}
    \caption{Weak charge values used in the calculation of $F_W$ for each nucleus. We fix the value of $q_p=0.0713$~\cite{Qweak:NIM} for both nuclei.}
    \begin{tabular}{@{\hspace{2em}} ccc@{\hspace{2em}} }
       Nucleus & $Q_W$ & $q_n$ \\ \hline
        $^{48}{\rm Ca}$ & $-26.0\pm0.1$ & $-0.9795$ \\
        $^{208}{\rm Pb}$ & $-117.9\pm0.3$ & $-0.9821$
    \end{tabular}
    \end{ruledtabular}
    \label{tab:qweak}
\end{table}

As detailed in Sec.~\ref{part:overview}, the weak form factor $F_W$ for both $^{208}$Pb and $^{48}$Ca can be obtained with virtually no model dependence. 
The procedure makes use of the well-known electromagnetic charge form factor and the total weak charge of the respective nucleus (see Table~\ref{tab:qweak}). Model predictions of $A_{\rm PV}$ are calculated using the methods described in \cite{Horowitz:1998vv} using the electromagnetic charge form factor at large momentum transfer
based on the experimental results of Emrich {\it et al.}~\cite{Emrich} for $^{48}{\rm Ca}$ 
and the work of De Vries {\it et al.}~\cite{DEVRIES1987495} for $^{208}{\rm Pb}$. 
To accurately compare to the experimental result, the calculated asymmetry is folded with the acceptance function (see Sec.~\ref{sec:PointingMethod}) to get a single value of $\langle A_{\rm PV} \rangle$ for each model. By using multiple models (see Tables \ref{tab:models1} and \ref{tab:models2}) to calculate the value of $F_W$ at the experimental average $Q^2$ a clear linear dependence can be seen between the calculated weak form factor and the parity-violating asymmetry~\cite{Mammei:2023kdf}. 

The model-independent linear relationship between $\langle A_{\rm PV} \rangle$ and  $F_W$ ensures that the quoted form factor uncertainties in the results presented in Table~\ref{tab:ffacs} are entirely due to experimental uncertainties. It should be further emphasized here that our experimental asymmetry measurement uncertainties are dominated by the fluctuations of counting statistics for both $^{48}{\rm Ca}$ and $^{208}{\rm Pb}$.

\begin{table}[htb]
    \centering
    \caption{Weak form factor $F_W$ extracted at the momentum transfer $Q$ indicated.}
    \begin{ruledtabular}
    \begin{tabular}{@{\hspace{2em}} ccc@{\hspace{2em}} }
        Nucleus & Q [fm$^{-1}$] & $F_W(Q^2)$ \\ \hline
        $^{48}{\rm Ca}$ & 0.8733 & $0.1304\pm0.0055$ \\
        $^{208}{\rm Pb}$ & 0.3977 & $0.368\pm0.013$
    \end{tabular}
    \end{ruledtabular}
\label{tab:ffacs}
\end{table}

\subsection{Extraction of Form Factor Differences}
\begin{figure}[htb]
    \centering
    \includegraphics[width=0.50\textwidth]{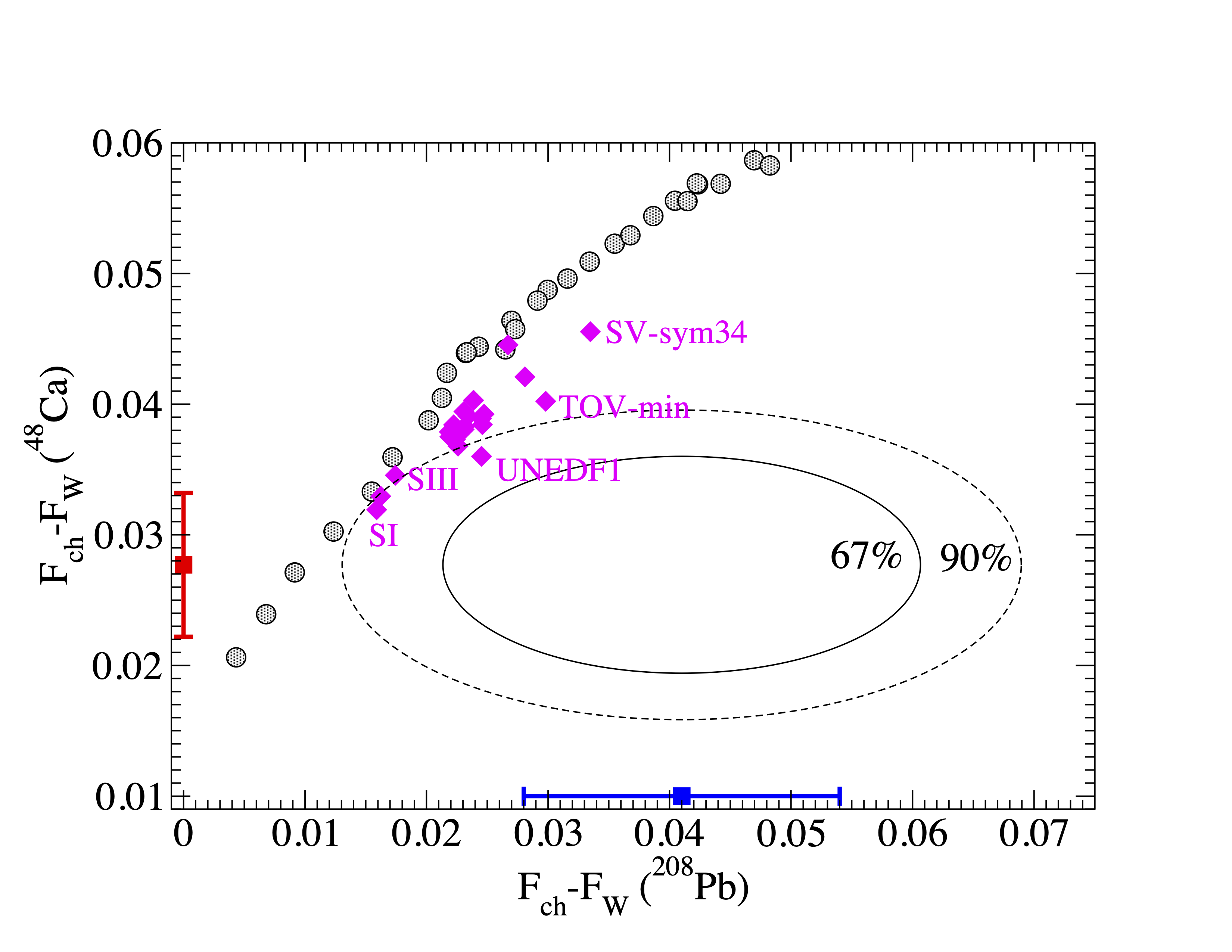}
    \caption{Difference between the charge and weak
    form factors of $^{48}$Ca (CREX) versus that of $^{208}$Pb (PREX-2) at their respective momentum transfers.  The blue (red) data point shows the PREX-2 (CREX) measurements. The ellipses are joint PREX-2 and CREX 67\% and 90\% probability contours.  The gray circles (magenta diamonds) are a range of relativistic (non-relativistic) density functionals.  For clarity only some of these functionals are labeled.  The complete list of functionals used here are found in Table~\ref{tab:models1} (relativistic) and Table~\ref{tab:models2} (non-relativistic).}
    \label{PREXVsCREX}
\end{figure}

The cleanest (least model-dependent) way to demonstrate the distribution of neutrons extending beyond that of protons in the two nuclei is by evaluating difference between the charge and the weak form factors: the ``form factor skin'', $F_{\rm ch} - F_W$.  The experimental results for the form factor skins are presented in Table~\ref{tab:ffSkin}. Using the extracted weak form factors for the two experiments one can subtract the well known charge form factors, both taken at the average $Q^2$, for a variety of models (see the full list in Tables \ref{tab:models1} and \ref{tab:models2}). We plot the predictions for each model in Fig.~\ref{PREXVsCREX} along with the 67\% confidence intervals resulting from each experiment. What is shown is that the entire model set we included, which was chosen to be a representative sample of popular models/formalisms used in other works, falls outside of the 67\% ellipse. Furthermore, only a select few models fall slightly within the 90\% ellipse. 

\begin{table}[htb]
    \centering
    \caption{Form factor differences (``skins'') at the momentum transfer $Q$ indicated.}
    \begin{ruledtabular}
    \begin{tabular}{@{\hspace{2em}} ccc@{\hspace{2em}} }
        Nucleus & Q [fm$^{-1}$] & $F_{\rm ch} - F_W$ \\ \hline
        $^{48}{\rm Ca}$ & 0.8733 & $0.0277\pm0.0055$ \\
        $^{208}{\rm Pb}$ & 0.3977 & $0.041\pm0.013$
    \end{tabular}
    \end{ruledtabular}
\label{tab:ffSkin}
\end{table}

The figure points to the apparent inability of nuclear models to describe the weak form factors for a relatively small nucleus like $^{48}{\rm Ca}$ and the large nucleus of $^{208}{\rm Pb}$ simultaneously. We point out again that the ellipses describing the experimental results in the 2-D parameter space of the two nuclear form factor differences have very little theoretical uncertainty.

\section{INTERPRETATION}
\label{sec:interpretation}
The results detailed above play an important role in refining our knowledge of neutron-rich nuclear matter and the corresponding equation of state, and constitute the optimal way to demonstrate the power and impact of the experimental results. In the following, we combine these results with models to obtain quantities such as the weak radius and neutron skin. These latter quantities are more often presented
in the literature studying the nuclear equation of state. However, these extractions will introduce model uncertainties.
In particular, the model uncertainties play an important role for $^{48}{\rm Ca}$, while those for the extractions for $^{208}{\rm Pb}$ are assessed to be well below the experimental uncertainties. 

\subsection{Weak Radius}
Starting with the cleanest extraction, the case of $^{208}{\rm Pb}$, we note that for a heavy nucleus the weak density is well approximated by a two-parameter Fermi function~\cite{Piekarewicz:2016}
\begin{eqnarray}
    \rho_W(r,c,a) = \rho_0 \frac{\sinh{(c/a)}}{\cosh{(r/a)}+\cosh{(c/a)}}\\
    \nonumber \rho_0 = \frac{3Q_W}{4\pi c(c^2+\pi^2a^2)} ,
\end{eqnarray}
where $a$ is the surface thickness and $c$ is a radius parameter. The density is normalized so that $\int d^3r\rho_W=Q_W$. 

The model dependency arises in the surface thickness term which we constrain using a wide range of density functionals, as detailed in~\cite{Reed:2020fdf}.  In the case of 
$^{208}{\rm Pb}$, 
there is very little model error in the surface thickness, which makes the extraction of the weak radius a nearly 
model-independent measurement. 

In contract, the extraction for $^{48}$Ca has more model dependence due to the surface thickness playing a much larger role for the $^{48}$Ca nucleus. As a result, the above extraction method does not work well. To get around this, we make use of the predictions of the form factor skin 
$F_{\rm ch}-F_W$ and the weak skin $R_W-R_{\rm ch}$ from several relativistic and non-relativistic energy density functionals (EDFs). We note that the inclusion of spin-orbit currents to the weak and charge densities has a larger effect on the non-relativistic EDFs than on the relativistic EDFs, causing additional model dependency. We follow the method described in detail in the CREX paper~\cite{ref:crex}. The weak skin ($R_W^{48}-R_{\rm ch}^{48}$) shows a linear relationship with the form factor skin over the range of different models and we extract this dependence using a fit.
We assign a model uncertainty that encapsulates all the model predictions, making it a conservative estimate.  

The calculated weak skins in $^{48}$Ca and $^{208}$Pb (from PREX-2) are 
\begin{eqnarray}
    &&\nonumber R_W^{48}-R_{\rm ch}^{48}=0.159\pm0.026\,(\mathrm{exp})\pm0.023\,(\mathrm{model}) \,\mathrm{fm}\\\nonumber
    &&R_W^{208}-R_{\rm ch}^{208}=0.292\pm0.082\,(\mathrm{exp})\pm0.013\,(\mathrm{model}) \, .\mathrm{fm}
\end{eqnarray}

\subsection{Neutron Radius }
\label{sec:NeutronRadius}

Much like the weak skin and form factor skin, the neutron skin and weak skin share a similar linear trend for relativistic models, albeit with some spread in the non-relativistic models. We can use the same procedure as above to obtain the neutron skin in both nuclei by fitting a line to the relativistic models to obtain the slope and then adjusting the intercepts to obtain a conservative model uncertainty. As a result, we obtain the following for the neutron skins in $^{48}$Ca and $^{208}$Pb:
\begin{eqnarray}
    \nonumber R_\mathrm{skin}^{48} = 0.121\pm0.026\,(\mathrm{exp})\pm0.024\,(\mathrm{model}) \,\mathrm{fm}\\\nonumber
    R_\mathrm{skin}^{208} = 0.278\pm0.078\,(\mathrm{exp})\pm0.012\,(\mathrm{model}) \,\mathrm{fm} .
\end{eqnarray}

The $^{208}$Pb neutron skin result for PREX-2 is  consistent with the result from PREX-1 \cite{ref:prexI}, see Table~\ref{tab:LeadSkins}.  PREX-1 ran nine years prior to PREX-2 with a similar apparatus and a different beam energy of 1.06 GeV.  The analysis methods described in this paper for PREX-2 and CREX were similar to the ones used for PREX-1. 
In the following, results involving the neutron skin of $^{208}$Pb use the combined result of PREX-1 and PREX-2, while results involving the weak form factor or weak radius are only shown for PREX-2 since the two experiments ran at different $Q^2$, requiring interpolation, and since PREX-1 did not add significantly to the overall accuracy. 

\begin{table}[htb]
    \centering
    \caption{Comparison of PREX-1 and PREX-2 neutron skin results and the grand average.}
    \begin{ruledtabular}
    \begin{tabular}{@{\hspace{3em}} cc@{\hspace{3em}} }
        Experiment &  $R_n - R_p$ (fm) \\ \hline
        PREX-1 &  $0.33^{+ 0.16}_{ - 0.18}$ \\
        PREX-2 & $0.278\pm0.079$\\
        Combined PREX &
        $0.283\pm0.071$
    \end{tabular}
    \end{ruledtabular}

    \label{tab:LeadSkins}
\end{table}

\subsection{Comparisons to Models}

\begin{figure}[htb]
    \centering
    \includegraphics[width=0.50\textwidth]{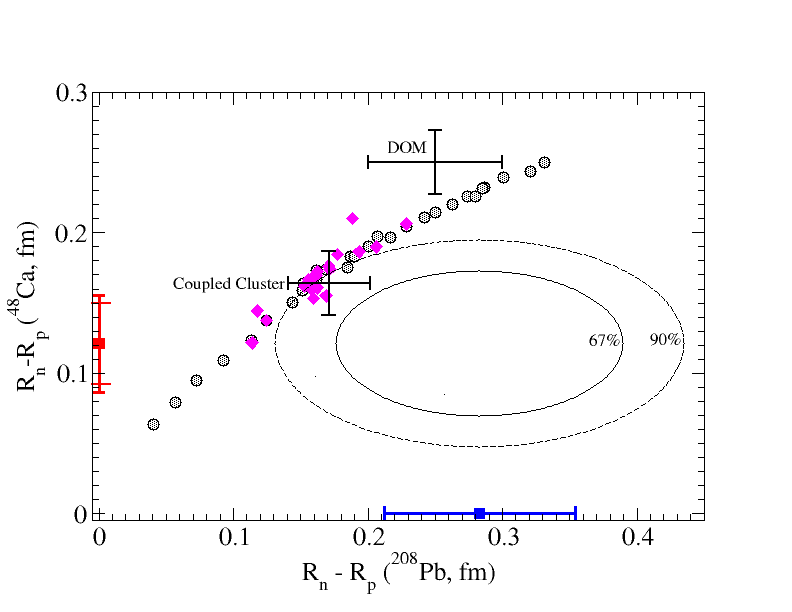}
    \caption{$^{48}$Ca neutron minus proton radius versus that for $^{208}$Pb.  The PREX-2+PREX-1 experimental result is shown as a blue square, while that for CREX is shown as a red square with the inner error bars indicating the experimental error and the outer error bars including the model error.  
    The gray circles (magenta diamonds) show a variety of relativistic (non-relativistic) density functionals.  Coupled cluster \cite{Hu:2021} and dispersive optical model (DOM) predictions \cite{atkinson:2020} are also shown.}
    \label{RnpCREX_PREX:fig}
\end{figure}

Similar to  Fig.~\ref{PREXVsCREX}, in 
Fig.~\ref{RnpCREX_PREX:fig} we show the prediction of several models' neutron skin $R_n-R_p$ for each experiment compared to the results of the experiments. 
As before, there is an apparent tension between the ability of available nuclear models to simultaneously predict the experimental results for the neutron skins of both experiments and the data; we do find that more models fall within the 90\% ellipse. 
In addition, we show the results of two microscopic calculations that utilize coupled cluster \cite{Hu:2021} and dispersive optical model \cite{atkinson:2020} techniques. We find that the coupled cluster results are closer to the experimental results for the neutron skins.

\subsection{Implications for the Nuclear Equation of State}

\begin{figure}[ht]
 \centering
 \includegraphics[width=0.4\textwidth]{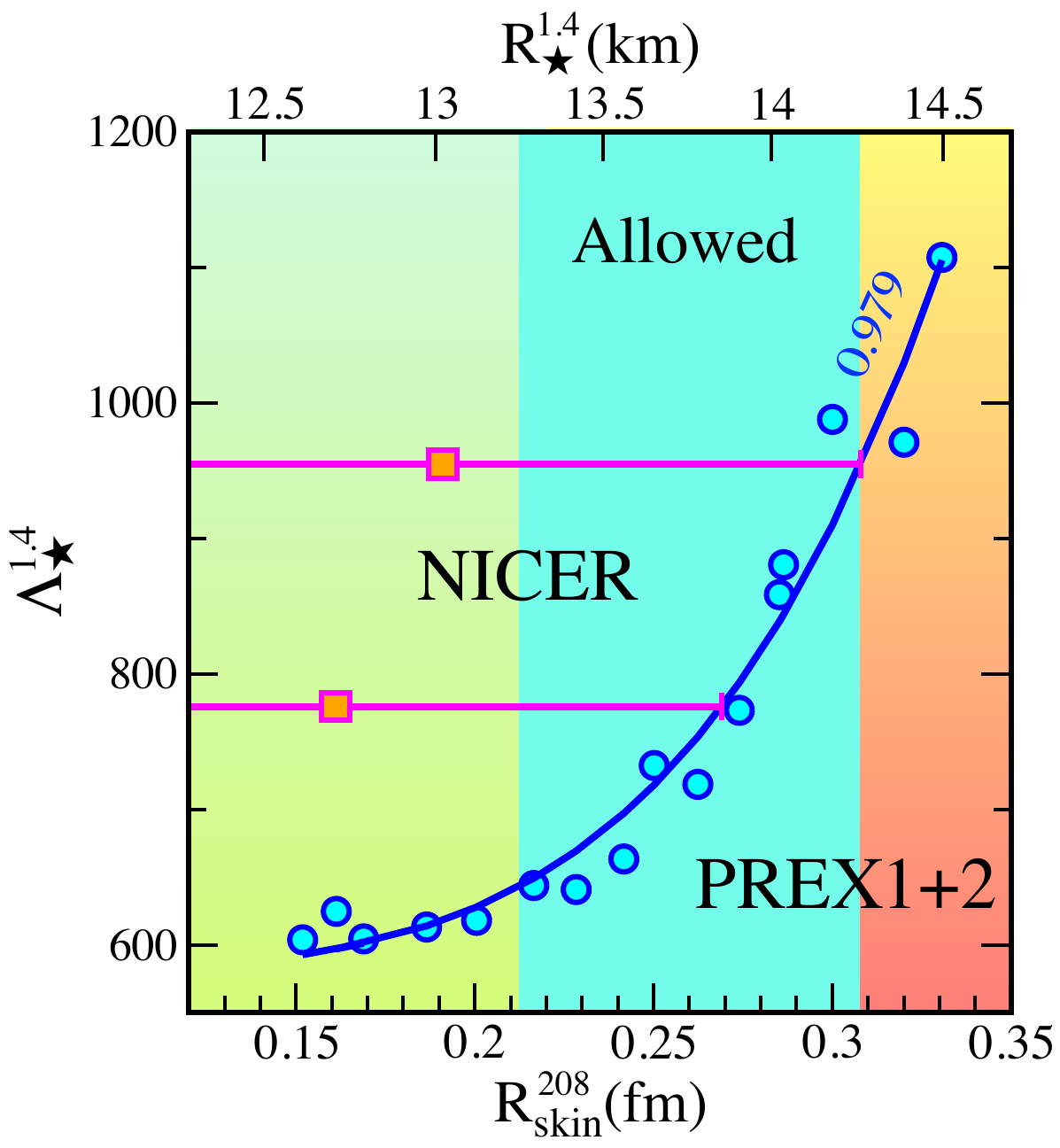}
 \caption{Neutron star observables compared with the neutron skin of $Pb$ as measured by PREX. Here we overplot the measurements of PSR J0030+0451's radius obtained with the NICER telescope \cite{Riley:2019yda} (green band) and the neutron skin of $^{208}$Pb determined from PREX (orange). The region where the two overlap is colored cyan. The blue curve shows the behavior of $\Lambda_{1.4}$ as described by several relativistic models shown as blue circles. Figure adapted from \cite{Reed:2021nqk}.}
\label{fig:prex_ns}
\end{figure}

The measurement of isovector skins in nuclei is important for the determination of the density dependence of the symmetry energy. In particular, the slope of the symmetry energy at saturation density ($L$) is known to be highly correlated with the thickness of neutron skins in neutron-rich nuclei, particularly $^{208}{\rm Pb}$. As a result of PREX-2, many works have arisen which indicate favoring large values for $L$ (e.g.~\cite{Reed:2021nqk}) which is in contrast to what is favored by other methodologies and nuclear observations. 
The thick neutron skin of $^{208}$Pb implies that $L\sim110\pm40$ MeV~\cite{Reed:2021nqk}. In stark contrast to the PREX result for $L$, the thinner neutron skin in $^{48}$Ca implies a very small value: $L\sim20\pm20$ MeV~\cite{Reed:2023}. 
The different values of $L$ from PREX and CREX suggest an incomplete understanding
of nuclear structure and/or dense nuclear matter. We emphasize again that the PREX/CREX experimental uncertainties are dominated by counting statistics and that nuclear theory predictions are separately consistent with the two results at approximately the 90\%\ confidence level.

These PREX and CREX results also have a connection to the high density EOS. By determining the neutron skins
of $^{48}$Ca and $^{208}$Pb one may find important constraints on astrophysical systems, in particular the radius of neutron stars.
The pressure of extra neutrons in neutron-rich nuclei causes a neutron-rich skin to form and comes from the same mechanism which governs the radius of 
neutron stars.
The connection with the neutron skin arises because the pressure of neutron matter is well-approximated by $P_{\rm PNM}\sim\frac{1}{3}\rho_0L$ where $\rho_0$ is nuclear saturation. From this connection to $L$, a well-known relationship between the neutron skin of a large nucleus such as $^{208}$Pb and the radius of a $1.4M_\odot$ 
neutron star
\cite{Horowitz:1999fk, FSUGold} can be established. It is noted that extracting connections to the high density EOS with the smaller $^{48}$Ca nucleus could have more corrections arising from its more important surface region. 

The tidal deformability $\Lambda$ of a 1.4 $M_\odot$ neutron star was constrained by the gravitational wave event GW170817 \cite{Abbott:2018} and provides an additional constraint on the radius of a neutron star. The neutron skin measurement of PREX and astrophysical observations of neutron star radii and $\Lambda$ are shown in Fig.~\ref{fig:prex_ns}.  There is mild tension between the tidal deformability, which suggests a somewhat smaller pressure of dense matter, and the neutron skin of $^{208}$Pb, suggesting a larger pressure.  Future GW observations and  parity-violating electron scattering experiments (such as MREX~\cite{Mammei:2023kdf,ref:MESA}) could sharpen this comparison.

\section{Conclusions}
\label{sec:Conclusions}

The parity-violating electron scattering experiments PREX and CREX have determined weak form factors, weak radii, neutron radii, and neutron skins of $^{48}$Ca and $^{208}$Pb. We have described the experimental method
and the data analyses in detail and shown that the experimental errors are dominated by counting statistics. These results have implications for nuclear structure, neutrino-nucleus elastic scattering, atomic parity violation, the equation of state of dense matter, and X-ray and gravitational wave observations of neutron stars.

%% file: z5_App.tex
\begin{table}[htb] 
\caption{List of all relativistic energy density functional models used in the CREX and PREX-2 analysis. Note there are several FSUGold2 and IU-$\delta$ models with different values of the slope parameter of the symmetry energy $L$. Also listed are $R_n-R_p$ and $F_{\rm{skin}}=F_{\rm ch}-F_W$ for $^{48}$Ca and $^{208}$Pb.}
    \centering
    \begin{ruledtabular}
    \begin{tabular}{l| c  c  c  c  c}
       Model & $L$ & $F_{skin}^{48}$ & $R_n-R_p^{48}$ & $F_{skin}^{208}$ & $R_n-R_p^{208}$  \\
        & [MeV] & & [fm] & & [fm] \\\hline
    FSUGarnet\cite{Chen:2015} & 50.96 & 0.0439 & 0.1665 & 0.0232 & 0.1614 \\
    FSUGold\cite{FSUGold} & 60.44 & 0.0488 & 0.1974 & 0.0300 & 0.2073 \\
    FSUGold2\cite{Reed:2021nqk} & 47.00 & 0.0424 & 0.1641 & 0.0217 & 0.1520 \\
    FSUGold2\cite{Reed:2021nqk} & 50.00 & 0.0444 & 0.1736 & 0.0243 & 0.1691 \\
    FSUGold2\cite{Reed:2021nqk} & 54.00 & 0.0464 & 0.1830 & 0.0270 & 0.1867 \\
    FSUGold2\cite{Reed:2021nqk} & 58.00 & 0.0479 & 0.1902 & 0.0291 & 0.2007 \\
    FSUGold2\cite{Reed:2021nqk} & 69.00 & 0.0509 & 0.2041 & 0.0334 & 0.2287 \\
    FSUGold2\cite{Reed:2021nqk} & 76.00 & 0.0523 & 0.2105 & 0.0355 & 0.2420 \\
    FSUGold2\cite{Reed:2021nqk} & 90.00 & 0.0544 & 0.2204 & 0.0387 & 0.2626 \\
    FSUGold2\cite{Reed:2021nqk} & 100.00 & 0.0556 & 0.2260 & 0.0405 & 0.2743 \\
    FSUGold2\cite{FSUGold2} & 112.68 & 0.0568 & 0.2319 & 0.0424 & 0.2866 \\
    IUFSU\cite{Fattoyev:2010mx} & 47.21 & 0.0440 & 0.1731 & 0.0233 & 0.1618 \\
    NL3\cite{Lalazissis:1996rd} & 118.19 & 0.0555 & 0.2258 & 0.0415 & 0.2802 \\
    RMF022\cite{Chen:2015} & 63.52 & 0.0496 & 0.1967 & 0.0316 & 0.2168 \\
    RMF028\cite{Chen:2015} & 112.64 & 0.0569 & 0.2315 & 0.0422 & 0.2854 \\
    RMF032\cite{Chen:2015} & 125.63 & 0.0587 & 0.2439 & 0.0470 & 0.3206 \\
    TFa\cite{Fattoyev:2013yaa} & 82.50 & 0.0529 & 0.2145 & 0.0368 & 0.2505 \\
    TFb\cite{Fattoyev:2013yaa} & 122.53 & 0.0569 & 0.2395 & 0.0442 & 0.3006 \\
    TFc\cite{Fattoyev:2013yaa} & 135.24 & 0.0583 & 0.2502 & 0.0483 & 0.3313 \\
    IU-$\delta$ & -40.00 & 0.0206 & 0.0637 & 0.0043 & 0.0408 \\
    IU-$\delta$ & -30.00 & 0.0239 & 0.0795 & 0.0068 & 0.0568 \\
    IU-$\delta$ & -20.00 & 0.0271 & 0.0950 & 0.0092 & 0.0722 \\
    IU-$\delta$ & -10.00 & 0.0302 & 0.1090 & 0.0123 & 0.0928 \\
    IU-$\delta$ & 0.00 & 0.0333 & 0.1234 & 0.0155 & 0.1133 \\
IU-$\delta$ & 10.00 & 0.0359 & 0.1373 & 0.0172 & 0.1247 \\
IU-$\delta$ & 20.00 & 0.0388 & 0.1501 & 0.0202 & 0.1438 \\
IU-$\delta$ & 30.00 & 0.0405 & 0.1589 & 0.0213 & 0.1510 \\
IU-$\delta$ & 40.00 & 0.0442 & 0.1750 & 0.0265 & 0.1846 \\
\end{tabular}
\end{ruledtabular}
    \label{tab:models1}
\end{table}

\begin{table}[htb]
\caption{List of all non-relativistic EDF models used in the CREX and PREX-2 analysis.}
    \centering
    \begin{ruledtabular}
    \begin{tabular}{l | c  c  c  c  c}
        Model & $L$ & $F_{skin}^{48}$& $R_n-R_p^{48}$ & $F_{skin}^{208}$ & $R_n-R_p^{208}$  \\
        & [MeV] & & [fm] & & [fm] \\\hline
    SI\cite{Skyrme:1958} & 1.22 & 0.0319 & 0.1216 & 0.0159 & 0.1138 \\
    SIII\cite{SIII} & 9.91 & 0.0345 & 0.1374 & 0.0174 & 0.1246 \\
    SKM*\cite{skms} & 45.76 & 0.0392 & 0.1551 & 0.0247 & 0.1688 \\
    SLy4\cite{SLy4} & 45.96 & 0.0391 & 0.1535 & 0.0234 & 0.1596 \\
    SLy5\cite{SLy4} & 48.14 & 0.0403 & 0.1608 & 0.0239 & 0.1622 \\
    SLy7\cite{SLy4} & 47.22 & 0.0394 & 0.1598 & 0.0231 & 0.1583 \\
    SV-K218\cite{Klupfel:2009} & 34.62 & 0.0381 & 0.1725 & 0.0231 & 0.1622 \\
    SV-K226\cite{Klupfel:2009} & 34.09 & 0.0379 & 0.1693 & 0.0228 & 0.1599 \\
    SV-K241\cite{Klupfel:2009} & 30.95 & 0.0379 & 0.1617 & 0.0219 & 0.1527 \\
    SV-bas\cite{Klupfel:2009} & 32.36 & 0.0379 & 0.1651 & 0.0223 & 0.1559 \\
    SV-kap00\cite{Klupfel:2009} & 39.44 & 0.0368 & 0.1627 & 0.0226 & 0.1580 \\
    SV-kap02\cite{Klupfel:2009} & 35.54 & 0.0373 & 0.1635 & 0.0224 & 0.1565 \\
    SV-kap06\cite{Klupfel:2009} & 29.33 & 0.0384 & 0.1669 & 0.0222 & 0.1555 \\
    SV-mas07\cite{Klupfel:2009} & 52.15 & 0.0389 & 0.1764 & 0.0245 & 0.1708 \\
    SV-mas08\cite{Klupfel:2009} & 40.15 & 0.0383 & 0.1694 & 0.0232 & 0.1616 \\
    SV-mas10\cite{Klupfel:2009} & 28.03 & 0.0375 & 0.1633 & 0.0219 & 0.1536 \\
    SV-sym28\cite{Klupfel:2009} & 7.21 & 0.0330 & 0.1444 & 0.0162 & 0.1178 \\
    SV-sym32\cite{Klupfel:2009} & 57.07 & 0.0421 & 0.1864 & 0.0281 & 0.1933 \\
    SV-sym34\cite{Klupfel:2009} & 80.95 & 0.0455 & 0.2063 & 0.0335 & 0.2287 \\
    SV-Min\cite{Klupfel:2009} & 44.81 & 0.0384 & 0.1739 & 0.0246 & 0.1716 \\
    TOV-Min\cite{TOVmin} & 76.23 & 0.0402 & 0.1900 & 0.0298 & 0.2064 \\
    UNEDF0\cite{UNEDF0} & 45.08 & 0.0445 & 0.2104 & 0.0267 & 0.1882 \\
    UNEDF1\cite{UNEDF1} & 40.00 & 0.0360 & 0.1845 & 0.0245 & 0.1770 \\
    \end{tabular}
    \end{ruledtabular}
    \label{tab:models2}
\end{table}